\documentclass[useAMS,usenatbib]{mnras}
\usepackage{graphicx}
\usepackage{amsmath}
\usepackage{amssymb}
\usepackage[TABTOPCAP]{subfigure}
\usepackage{url}
\usepackage{multirow}
\usepackage{grffile}

\newcommand{\m}{\rm\thinspace m}
\newcommand{\cm}{\rm\thinspace cm}

\newcommand{\cmsq}{\hbox{$\cm^2\,$}}
\newcommand{\msq}{\hbox{$\m^2\,$}}

\newcommand{\s}{\rm\thinspace s}

\newcommand{\Ms}{\rm\thinspace Ms}

\newcommand{\Hz}{\rm\thinspace Hz}

\newcommand{\Msun}{\hbox{$\rm\thinspace M_{\odot}$}}

\newcommand{\keV}{\rm\thinspace keV}

\newcommand{\erg}{\rm\thinspace erg}

\newcommand{\ergps}{\hbox{$\erg\s^{-1}\,$}}

\newcommand{\ergcmps}{\hbox{$\erg\cm\ps\,$}}

\newcommand{\pcmcu}{\hbox{$\cm^{-3}\,$}}

\newcommand{\ps}{\hbox{$\s^{-1}\,$}}

\newcommand{\rg}{\rm\thinspace $r_\mathrm{g}$}

\title[X-ray emission from black hole plunging regions]{Venturing beyond the ISCO: Detecting X-ray emission from the plunging regions around black holes}
\author[Wilkins, Reynolds \& Fabian]{D. R. Wilkins$^{1}$\thanks{E-mail: dan.wilkins@stanford.edu}\thanks{Einstein Fellow}, C. S. Reynolds$^{2}$ and A. C. Fabian$^{2}$\\
$^{1}$Kavli Institute for Particle Astrophysics and Cosmology, Stanford University, 452 Lomita Mall, Stanford, CA 94305, USA \\
$^{2}$Institute of Astronomy, University of Cambridge, Madingley Road, Cambridge. CB3 0HA, UK \\
}
\begin{document}

\date{Accepted 2020 February 28. Received 2020 February 19; in original form 2019 July 14}

\pagerange{\pageref{firstpage}--\pageref{lastpage}} \pubyear{2020}

\maketitle

\label{firstpage}

\begin{abstract}
We explore how X-ray reverberation around black holes may reveal the presence of the innermost stable circular orbit (ISCO), predicted by General Relativity, and probe the dynamics of the plunging region between the ISCO and the event horizon. Being able to directly detect the presence of the ISCO and probe the dynamics of material plunging through the event horizon represents a unique test of general relativity in the strong field regime. X-ray reverberation off of the accretion disc and material in the plunging region is modelled using general relativistic ray tracing simulations. X-ray reverberation from the plunging region has a minimal effect on the time-averaged X-ray spectrum and the overall lag-energy spectrum, but is manifested in the lag in the highest frequency Fourier components, above $0.01\,c^{3}\,(GM)^{-1}$ (scaled for the mass of the black hole) in the 2-4\keV\ energy band for a non-spinning black hole or the 1-2\keV\ energy band for a maximally spinning black hole. The plunging region is distinguished from disc emission not just by the energy shifts characteristic of plunging orbits, but by the rapid increase in ionisation of material through the plunging region. Detection requires measurement of time lags to an accuracy of 20 per cent at these frequencies. Improving accuracy to 12 per cent will enable constraints to be placed on the dynamics of material in the plunging region and distinguish plunging orbits from material remaining on stable circular orbits, confirming the existence of the ISCO, a prime discovery space for future X-ray missions.

\end{abstract}


\section{Introduction}
Black holes represent some of the most extreme environments in the Universe, around which the strong gravitational field, attributed to the curvature of spacetime, is described by the theory of General Relativity.

Material spiralling into black holes powers some of the most luminous objects we see in the Universe. In particular, matter accreting onto supermassive black holes in the centres of galaxies produces active galactic nuclei, or AGN, and stellar mass black holes accreting material from a companion star produce bright X-ray binaries. Energy released as matter falls in through the strong gravitational potential powers intense electromagnetic radiation, from optical to ultraviolet and X-ray wavelengths, and can launch jets of particles close to the speed of light that span vast distances from the black hole. The material spiralling into a black hole forms a flattened accretion disc within which it maintains stable circular orbits. Energy is liberated from the accretion flow by viscous forces between differentially rotating gas at different radii that transfer angular momentum outwards through the disc, causing material to progress gradually to orbits at smaller radii \citep{shaksun, pringle_81}.

General relativity predicts that there exists a radius outside the event horizon of a black hole within which stable circular orbits can no longer be maintained; the \textit{innermost stable circular orbit (ISCO)}. Once material reaches this orbit within the accretion flow and continues to lose angular momentum, it will plunge rapidly into the black hole. Indeed, a sharp transition to a plunging flow at the ISCO is observed in turbulent magnetohydrodynamic models of accretion discs \citep{reynolds_fabian2008}.

Around a non-spinning black hole, around which the spacetime geometry is described by the Schwarzschild metric, the ISCO is located at a radius of 6\rg\ from the singularity, outside the event horizon at 2\rg. As the spin of the black hole increases, the geometry is described by the Kerr metric \citep{kerr}. Relativistic frame-dragging acts to stabilise circular orbits that are prograde to the black hole spin and for a rapidly spinning black hole with spin parameter $a = J/Mc = 0.998\,GMc^{-2}$, the ISCO moves inward to a radius of 1.235\rg\ (where the gravitational radius, $r_\mathrm{g} = GM/c^2$ in Boyer-Lindquist co-ordinates, and represents the characteristic scale-length in the gravitational field around a black hole that can be scaled with black hole mass, $M$).

If the presence of an innermost stable orbit could be detected in the accretion flow around a black hole and its radius could be measured, it would represent a unique test of a specific prediction of general relativity in the strongest field regime that is only found in the immediate vicinity of black holes. Moreover, much remains unknown the physical processes (likely magnetohydrodynamic in nature, \citealt{balbus+91,balbus+98,krolik+05}) that generate the viscous stresses that drive the accretion flow and that are responsible for the acceleration of particles into luminous X-ray emitting coron\ae\ and large-scale jets. If the dynamics of matter in the innermost regions of the accretion flow can be probed, around the ISCO and within the plunging region, it would yield an understanding of the interplay between the accretion flow and magnetic fields around the black hole that determines the behaviour of the plasma and generates the intense energy output. 

Lastly, the ability to measure the spins of black holes is important both in understanding the energy source that underlies the powerful jets borne from radio galaxies \citep{blandford_znajek} and also the accretion history that led to the growth of supermassive black holes in the centres of galaxies \citep{reynolds_spin}. The spin of a black hole is commonly measured by determining the innermost radius from which either thermal \citep{mcclintock_spin} or line emission from the disc \citep{brenneman_reynolds} can be detected, assuming that the disc extends down to the innermost stable orbit and that no emission is seen from within this radius. It is important to understand how any emission from material in the plunging region might be detected and whether this emission could bias the measured values of the black hole spin parameter.

Great advances have been made in probing the structure and geometry of the accretion flows around black holes from the reflection and reverberation of continuum X-ray emission off of the infalling material. When an X-ray continuum illuminates the accretion flow, a characteristic reflection spectrum is produced containing a number of emission lines among other features \citep{ross_fabian, garcia+2010}. The most prominent of these is the iron K$\alpha$ fluorescence line at 6.4\keV\ in the rest frame of the emitting material \citep{george_fabian}. This line emission is subject to Doppler shifts from the orbital motion of the accretion flow as well as gravitational redshifts in the strong gravitational potential that vary as a function of position on the disc. The combination of these broadens a narrow emission line into a characteristic shape with a blueshifted peak and extended redshifted wing. The extent of the redshifted wing is a function of how deep the accretion flow extends into the gravitational potential and, hence, can be used to measure the spin of the black hole \citep{brenneman_reynolds}, while the precise profile of the line encodes the pattern of illumination of the disc by the primary X-ray source \citep{1h0707_emis_paper} and can be used to measure the location and geometry of the corona \citep{understanding_emis_paper} and of the accretion disc \citep{taylor_reynolds}.

Most recently, the advent of X-ray spectral-timing studies has added a further dimension to the picture that has been emerging of the inner accretion disc and corona with the detection of X-ray reverberation. The continuum emission from the corona is extremely variable and time lags are observed between correlated variations in energy bands that are dominated by the directly-observed continuum and by the emission that reverberates from the disc. These time lags correspond to the additional light travel time between the primary continuum source in the corona and the reprocessing accretion disc \citep{fabian+09,reverb_review}. Differential light travel times can be measured to the inner and outer parts of the accretion disc, producing respectively the redshifted wing and the 6.4\keV\ core of the iron K$\alpha$ line \citep{zoghbi+2012,kara_global} and measuring the variation in time lag as a function of energy not only places constraints on the extent of the corona \citep{lag_spectra_paper}, but reveals structures within the corona and the manner in which luminosity fluctuations propagate through is extent \citep{propagating_lag_paper}. Evidence is emerging that the X-ray coron\ae\ of radio quiet Seyfert galaxies contain a compact, collimated core, akin to the base of a failed jet, that dominates the rapid X-ray variability, embedded within a more slowly varying corona associated with the inner parts of the accretion disc \citep{1zw1_corona_paper}.

In this paper we explore how the plunging region, inside the innermost stable orbits around black holes, may be detected through X-ray reverberation. The production of relativistically broadened line emission from the plunging region around a non-spinning black hole was studied by \citet{reynolds+97}, while \citet{reynolds+99} incorporated the emission from material within the plunging region into the time- and energy-resolved reverberation response of the iron K$\alpha$ lines from the accretion flows onto spinning black holes following a single flare of emission from the corona. We here extend these studies, computing the full spectral response of the accretion discs and plunging regions around spinning black holes to variations in continuum luminosity, and frame studies of the plunging region in the context of the present state-of-the-art X-ray spectral timing techniques. We specifically seek to address whether emission from the plunging region can be detected and distinguished from the stably orbiting accretion disc, and whether the dynamics of the material in the plunging region can be probed with the enhanced capabilities promised by forthcoming X-ray observatories.

\section{Structure of the Disc and Plunging Region}

Material in the accretion disc, outside of the innermost stable circular orbit (ISCO), is assumed to follow stable, circular orbits with angular velocity
\begin{equation}
\label{angvel.equ}
\Omega = \frac{d\varphi}{dt} = \frac{1}{a^2 + r^\frac{3}{2}}
\end{equation}
The constants of motion, $k$ and $h$, corresponding to the energy and angular momentum of material in a circular orbit in the equatorial plane of the Kerr spacetime can be written
\begin{align}
k &= \frac{1 - 2u - au^\frac{3}{2}}{ \sqrt{1 - 3u + 2au^\frac{3}{2}}} \\
h &= \frac{1 + a^2 u^2 - 2au^\frac{3}{2}}{ u^2 \sqrt{1 - 3u + 2au^\frac{3}{2}}}
\end{align}
where $u = 1/r$.

Material experiences viscous stresses in the disc causing more rapidly orbiting material at inner radii to lose angular momentum to slower material orbiting just outside, leading to the inspiral of the material towards the black hole. Once the material reaches the innermost stable orbit and loses a small amount of angular momentum, it will transition to an unstable plunging orbit, upon which it will plunge rapidly towards the event horizon. Since mass must be conserved, the rapid increase in velocity on the plunging orbit will lead to a sharp drop in density. As a first approximation to the plunging region, we therefore assume that the accreting material experiences no torque from viscous stresses once it crosses the ISCO and therefore retains the energy and angular momentum (and hence values of $k$ and $h$) from the ISCO.

Given the value of $k$ and $h$ for a given orbit, a massive particle's 4-velocity, $\mathbf{v} = \left(\dot{t}, \dot{r}, \dot{\theta}, \dot{\varphi}\right)$, where dots represent the derivatives of the co-ordinates with respect to the particle's proper time, can be calculated. For a massive particle constrained to the equatorial plane, choosing the negative square root to calculate $\dot{r}$ for the infalling solution;
\begin{align}
\dot{t} &= \frac{1}{\Delta} \left[ \left(r^2 + a^2 + \frac{2a^2}{r} \right)k - \frac{2ah}{r} \right] \label{tdot.equ}\\
\dot{\varphi} &= \frac{1}{\Delta} \left[ \frac{2ak}{r} + \left(1 - \frac{2}{r}\right)h \right] \label{phidot.equ} \\
\dot{r} &= -\sqrt{ k^2 - 1 + \frac{2}{r} + \frac{a^2(k^2 - 1) - h^2}{r^2} + \frac{2(h-ak)^2}{r^3}} \label{rdot.equ}
\end{align}

\subsection{The stably orbiting disc}

\citet{novthorne} derive analytic expressions for the structure of a stably-orbiting thin accretion disc in the Kerr spacetime. The accretion disc around a black hole is divided into three regions, depending on the mass accretion rate, each characterised by the dominant contributions to the pressure and opacity. The accretion flow transitions from an outer region where the pressure is dominated by gas pressure and the opacity is dominated by free-free scattering, through a middle region where electron scattering dominates the opacity (but gas pressure still dominates) to an inner region where radiation pressure dominates over the gas pressure and the opacity is dominated by electron scattering. The aspect ratio, $(h / r)$ of the disc depends upon the vertical pressure balance and, thus, mass continuity dictates the functional dependence of the surface density.

Over both the middle and outer regions of the disc, the aspect ratio, $(h / r)$, is constant and the surface density, $\Sigma \propto r^{-\frac{1}{2}}$. Dividing the surface density by the height of the disc, we can approximate the average density through a vertical slice of the disc by a power law; $\rho(r)\propto r^{-\frac{3}{2}}$ over the outer and middle regions. In the inner disc, where radiation pressure dominates, the disc height, $h$, is constant and the surface density and volume density will approximately follow $r^\frac{3}{2}$. In these simulations, we shall approximate the stably orbiting accretion disc to be gas-pressure dominated, with density profile following $\rho(r)\propto r^{-\frac{3}{2}}$, though We will derive a self-consistent prescription for the density profile in the plunging region which will not depend on the density profile of the outer accretion disc. 

\subsection{The plunging region}

Mass is conserved as it flows inwards and crosses the innermost stable circular orbit (assuming there is negligible mass loss in outflows and winds from the disc). Hence, the continuity equation for the energy-momentum tensor $T^{\mu\nu}$ is obeyed:
\begin{equation}
\nabla_\mu T^{\mu\nu} = 0
\end{equation}
In the limit of a non-relativistic fluid with $p \ll \rho c^2$, in a frame co-rotating with the accretion flow, this reduces to
\begin{equation}
\partial_\mu (\rho u^\mu) = 0
\end{equation}
where $\rho$ is the proper density of the fluid, measured in the co-rotating frame, and $u^\mu$ is its 4-velocity. For a steady mass accretion rate, $\dot{m}$, through an axisymmetric thin disc with comoving column density $\Sigma$, the continuity equation can be written
\begin{equation}
\label{continuity.equ}
r \Sigma u^r = -\frac{\dot{m}}{2\pi}
\end{equation}

If the disc has constant aspect ratio $(h / r)$ and has density $\rho_0$ at some radius $r_0$, we may obtain a relation for the density as a function of radius
\begin{align}
\rho r^2 u^r &= \mathrm{const.} \\
\rho(r) &= \frac{\rho_0 r_0^2 u^r(r_0)}{r^2 u^r(r)}
\end{align}
Where $u^r(r)$ is the radial (contravariant) component of the 4-velocity at radius $r$ given by Equation~\ref{rdot.equ}, using the values of the constants $k$, $h$ and $Q$ at the ISCO.

We examine two extremal cases; the accretion flow around a non-spinning black hole (dimensionless spin parameter $a = 0$) illuminated by an isotropically-emitting point source at a height $h = 5$\rg\ above the singularity on the polar axis, and a maximally spinning black hole ($a=0.998$) with its accretion flow illuminated by a point source at height $h=2$\rg. Measurements of the illumination profiles of accretion discs around rapidly spinning black holes and the corresponding X-ray reverberation timescales suggest that the primary X-ray source is located within these short distances of the black hole, while in the non-spinning case, it is necessary to place the source sufficiently above the event horizon (now at 2\rg\ instead of approximately 1\rg\ for the rapidly spinning case) for sufficient rays to illuminate the disc and not be lost within the event horizon. Note also that if the X-ray source is produced by magnetic fields or other processes anchored to the accretion disc, now with the innermost stable orbit at 6\rg, it is likely that coron\ae\ associated with non-spinning black holes will be located at a greater distance from the black hole than in the maximal-spin case.

\begin{figure*}
\centering
\subfigure[Density profile] {
\includegraphics[width=55mm]{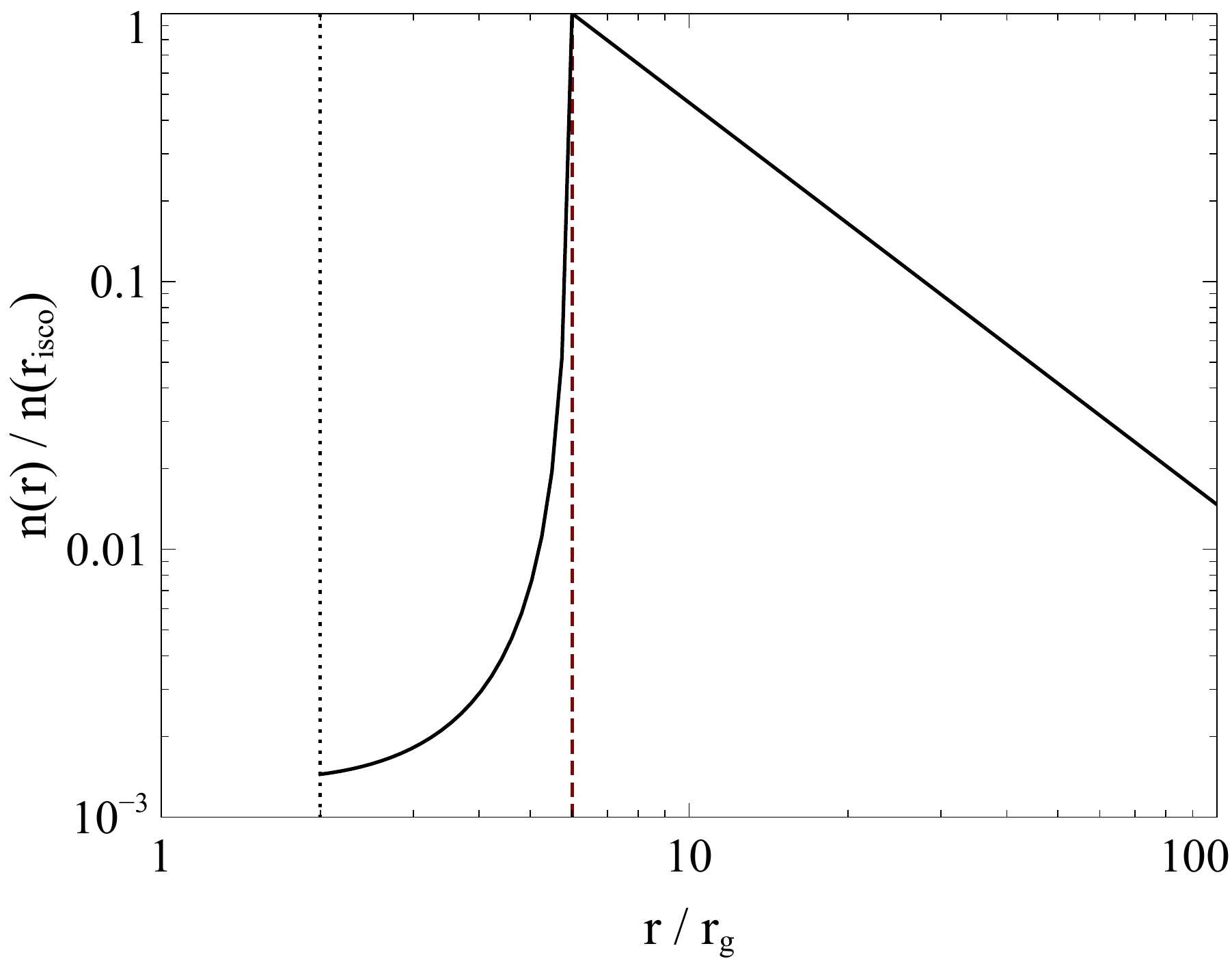}
}
\subfigure[Illumination profile] {
\includegraphics[width=55mm]{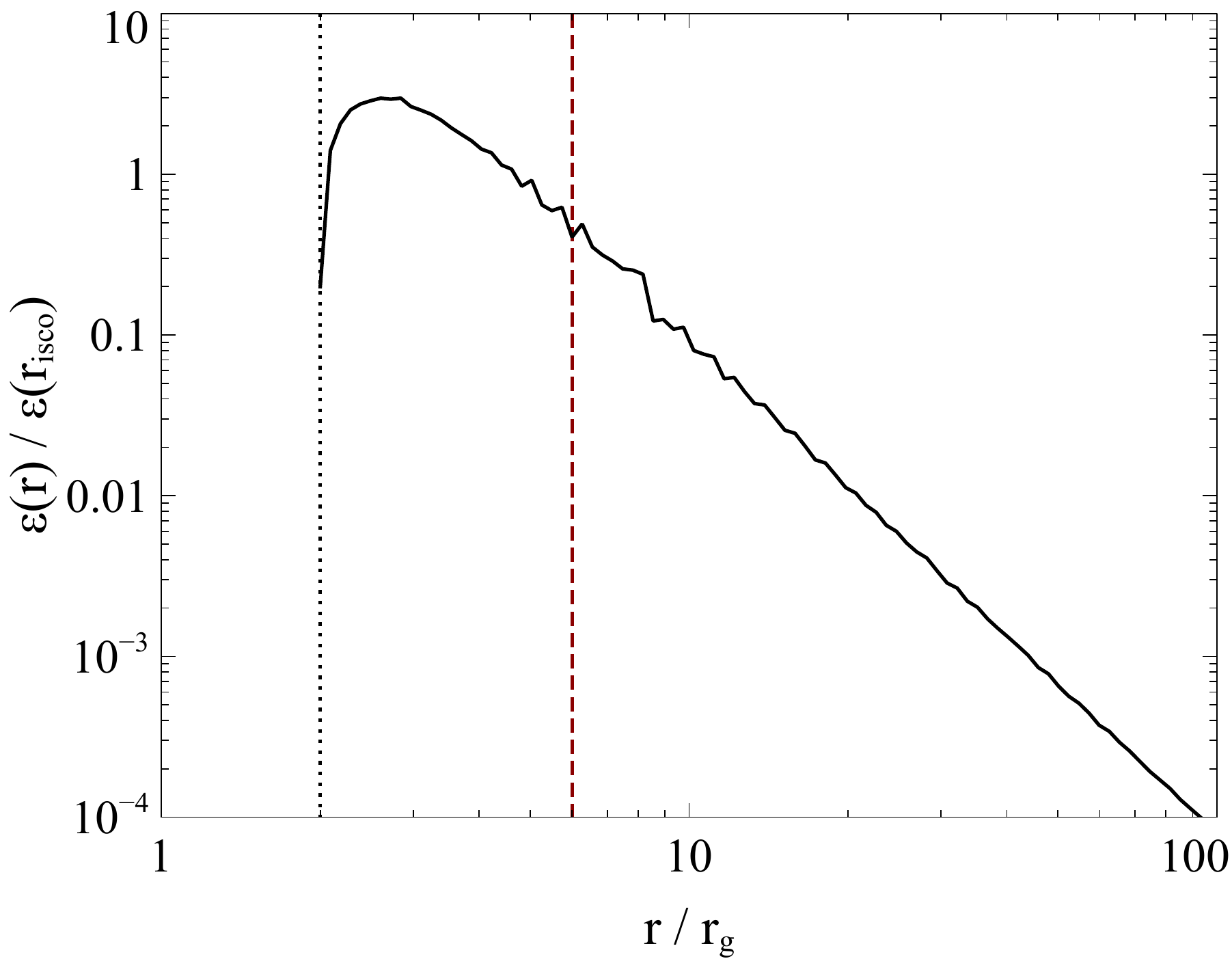}
}
\subfigure[Ionisation profile] {
\includegraphics[width=55mm]{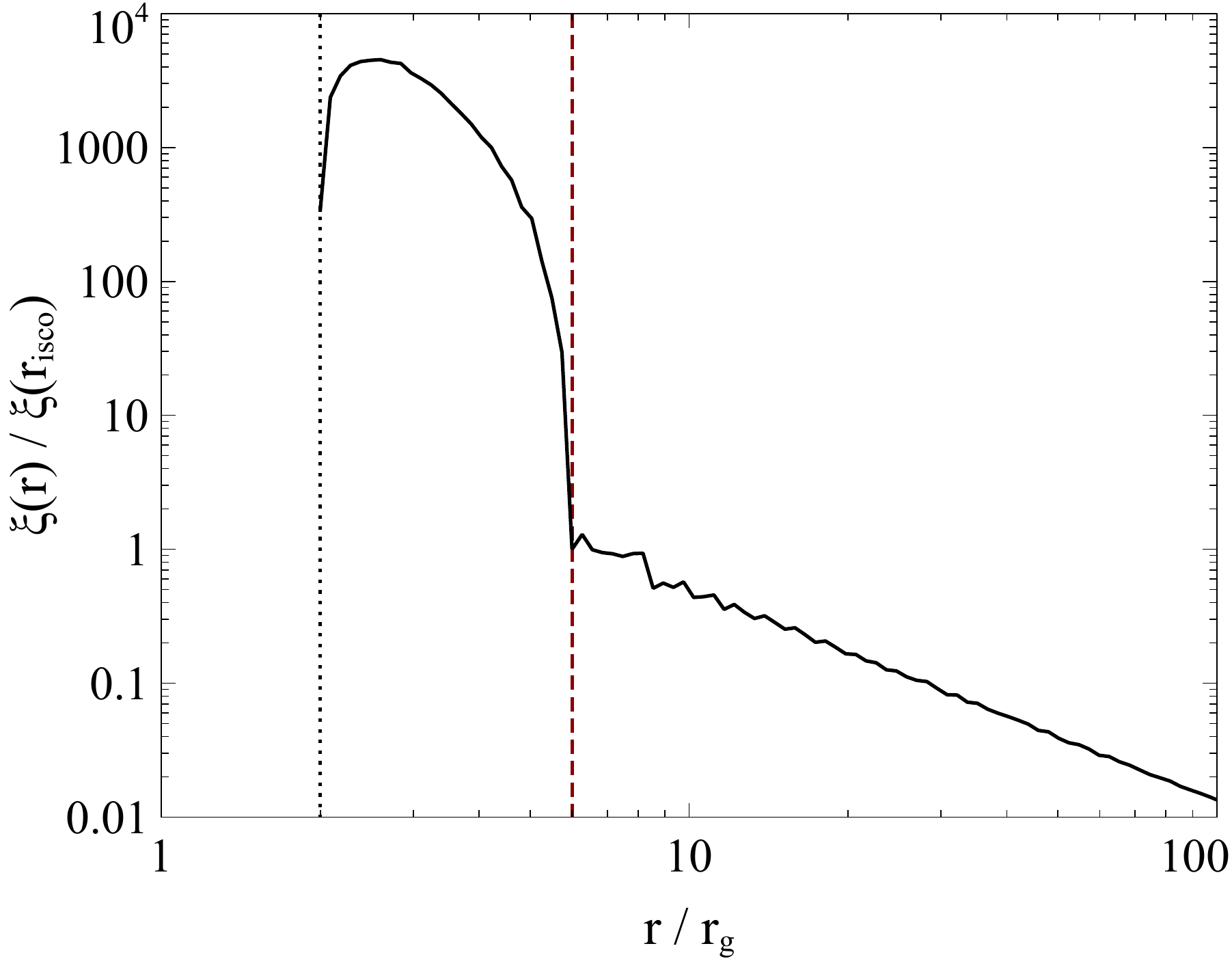}
}
\subfigure {
\includegraphics[width=55mm]{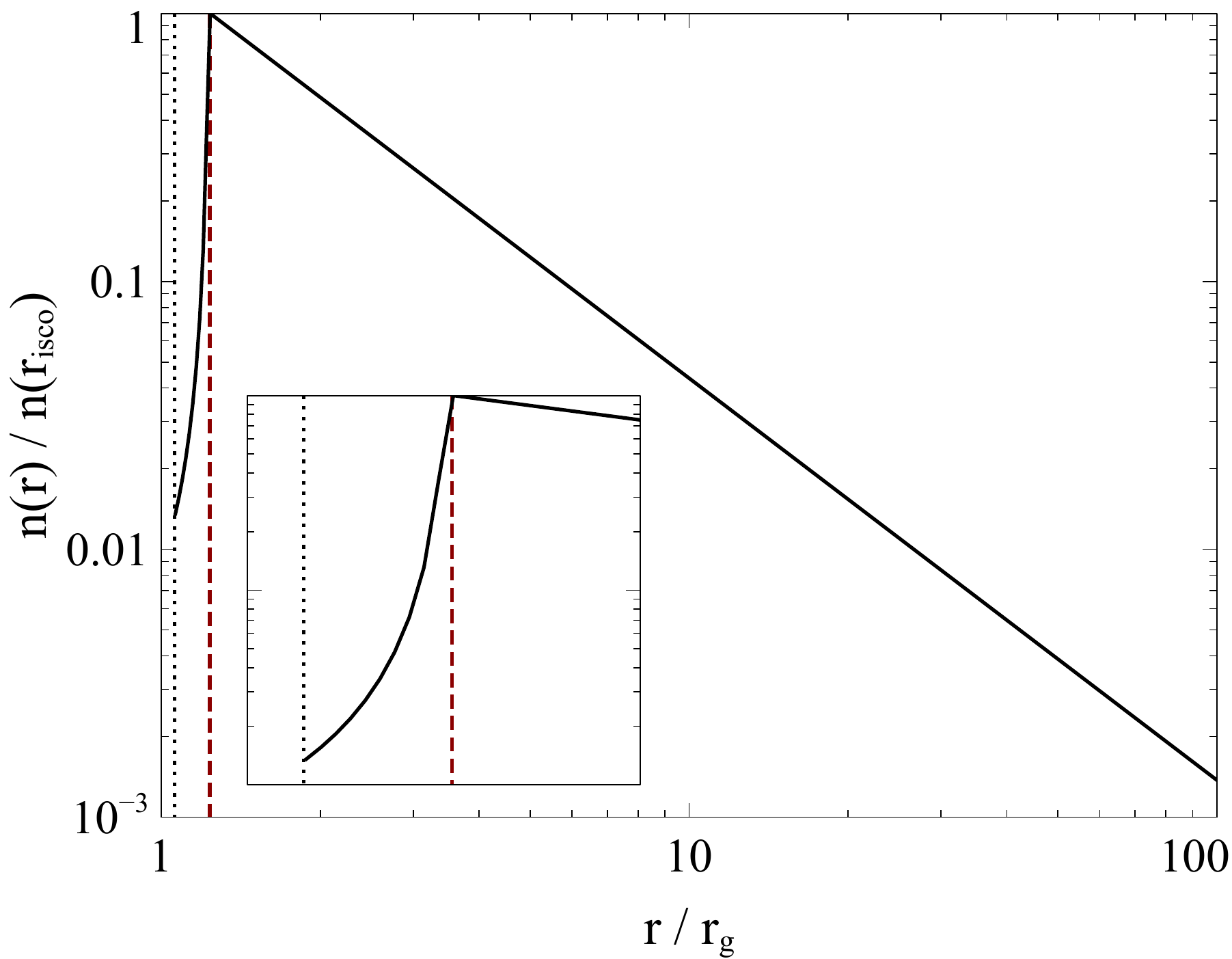}
}
\subfigure {
\includegraphics[width=55mm]{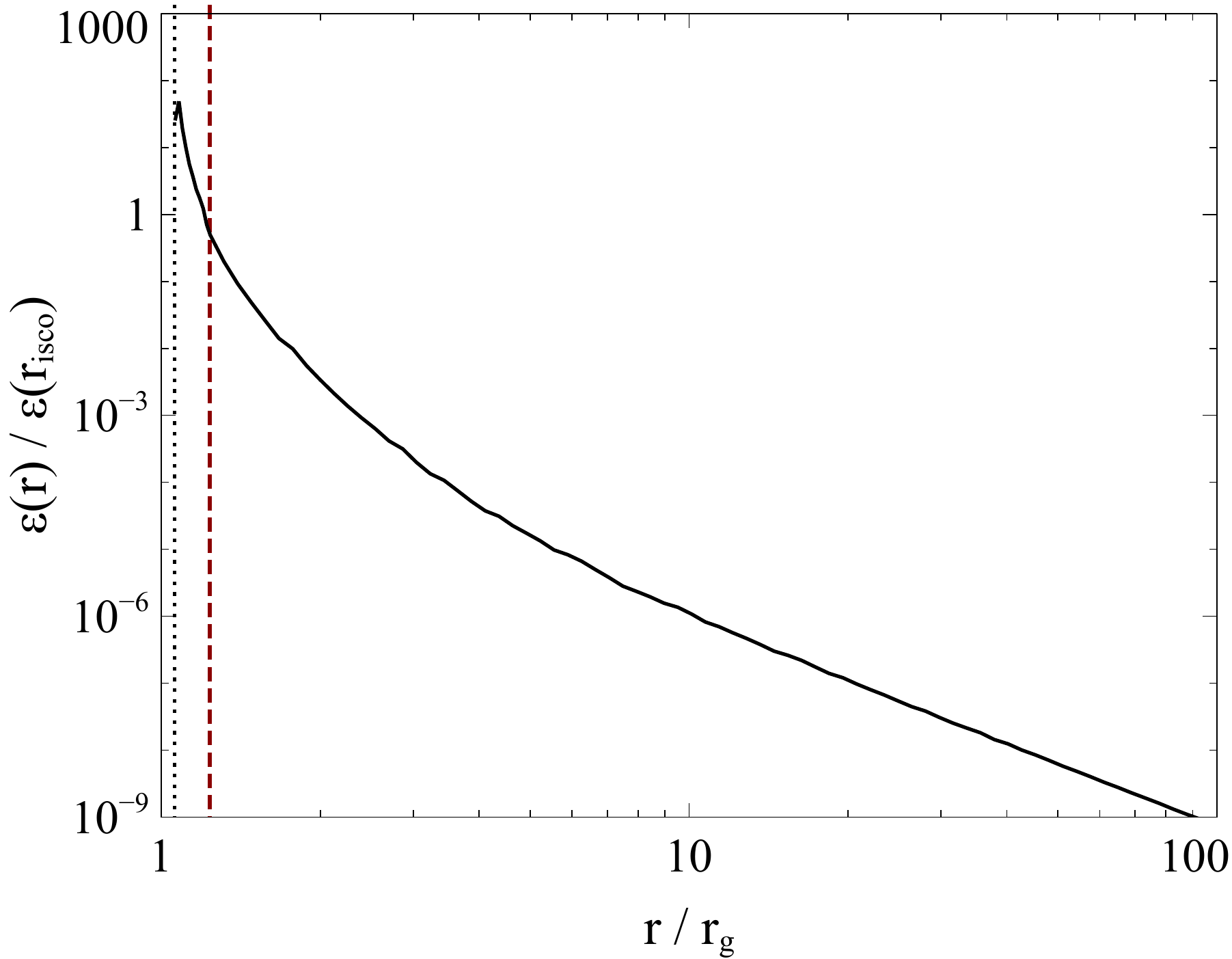}
}
\subfigure {
\includegraphics[width=55mm]{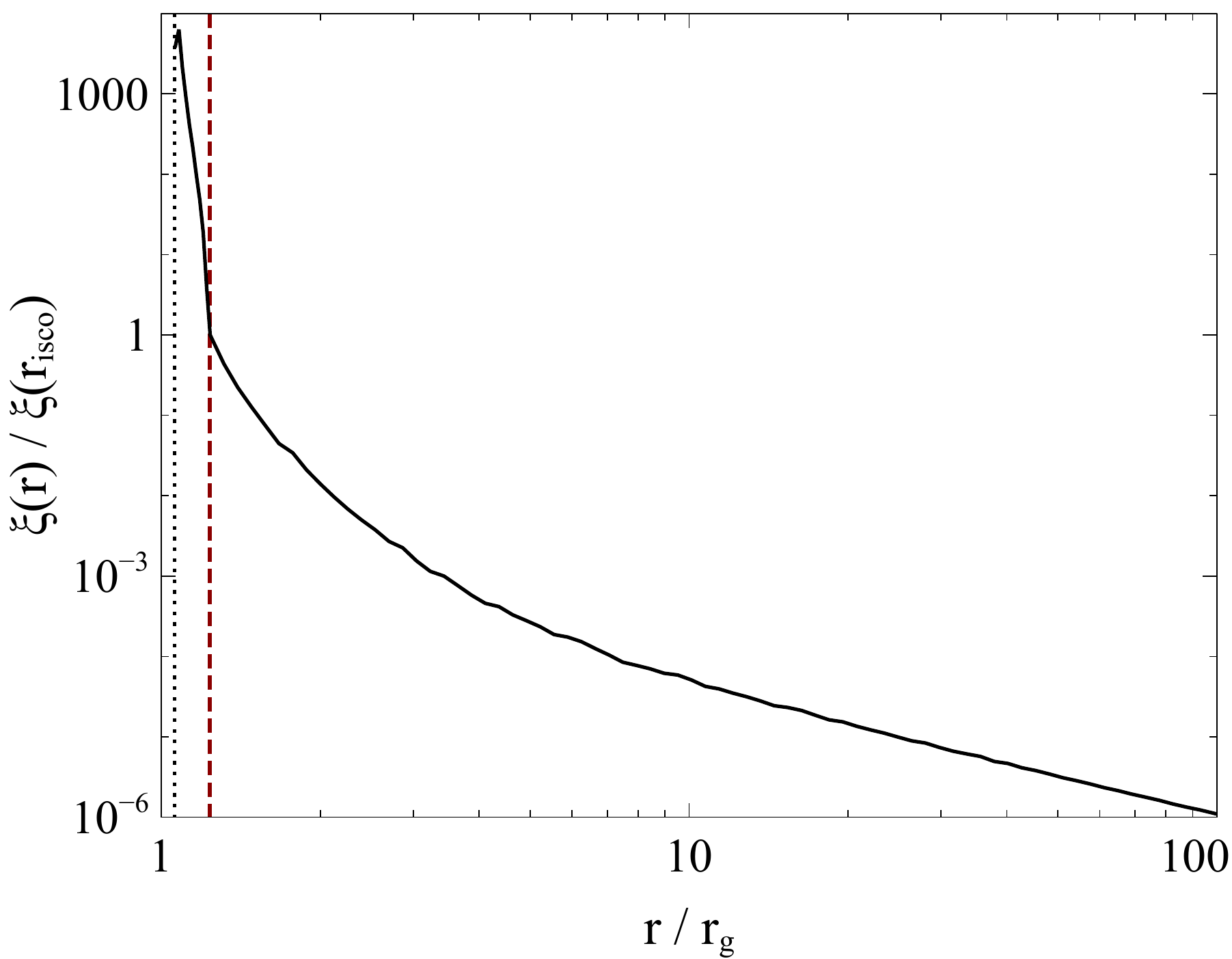}
}
\caption[]{Structure of the accretion flow around a non-spinning black hole (\textit{top}) and maximally spinning ($a=0.998$) black hole (\textit{bottom}). The left panels show the model density profile of the flow as a function of radius. Middle panels show the illumination profile; the flux received from the primary source per unit proper area. The right panels show the ionisation parameter computed from the density and illumination profiles. The red dashed line shows the location of the innermost stable circular orbit (ISCO) and the black dot-dash line shows the event horizon.}
\label{disc_profiles.fig}
\end{figure*}

\subsubsection{Optical depth through the plunging region}
\citet{reynolds+97} demonstrate that for reasonable mass accretion rates onto non-spinning black holes, the plunging region remains optically thick all the way down to the event horizon. Noting that the optical depth to electron scattering through the flow, as measured in the comoving frame, is $\tau_\mathrm{e} = \Sigma \sigma_\mathrm{T} / m_\mathrm{p}$, and expressing the mass accretion rate as a fraction of the Eddington accretion rate, Equation~\ref{continuity.equ} can be recast in terms of the optical depth.
\begin{equation}
 \tau_\mathrm{e} = \frac{2}{\eta r u^r}\left(\frac{\dot{m}}{\dot{m}_\mathrm{Edd}}\right)
\end{equation}
Where $\eta$ is the radiative efficiency, such that the bolometric luminosity is $L = \eta\dot{m}c^2$, the radius, $r$ is measured in units of \rg\ and the radial component of the velocity is measured as a fraction of $c$. Given the assumption material follows ballistic orbits and retain their angular momentum from the ISCO, the optical depth throughout the plunging region depends only on the mass accretion rate and the radiative efficiency of the accretion flow.

Assuming a radiative efficiency $\eta = 0.06$ for the accretion flow around a non-spinning black hole (assuming that the energy that is liberated is equal to the energy loss between infinity and the ISCO), $\tau_\mathrm{e} > 1$ at all radii between the ISCO and event horizon so long as $\dot{m} > 0.05\,\dot{m}_\mathrm{Edd}$. For a maximally spinning black hole, assuming $\eta \sim 0.4$, the plunging region remains optically thick so long as $\dot{m} > 0.01\,\dot{m}_\mathrm{Edd}$.

Since the plunging region remains optically thick to the event horizon, we do not expect bright emission to be detected from the photon ring around a supermassive black hole with $\dot{m} > 0.01\,\dot{m}_\mathrm{Edd}$, \textit{c.f.} in the case of the optically thin accretion flow in M87 \citep{eht1}. Bright emission from the photon ring is detected when light paths are able to orbit the black hole multiple times close to the photon orbit, accumulating emission. In this case, such light paths would be intercepted by the accretion flow.

\section{Simulating X-ray Reverberation}
The energy- and time-dependent reverberation of X-rays from the accretion flow is studied via the \textit{impulse response function} which encodes the flux received by the observer as a function of energy and time following a delta-function flash of emission from the primary source \citep{reynolds+99,cackett_ngc4151,propagating_lag_paper}. Integrating the response function along the time axis yields the spectrum of the reprocessed emission, while convolving with the time series describing the variability of the primary source (or for the case of an extended primary source, the underlying source of variability that is fed into the corona, as in \citealt{propagating_lag_paper}) produces the light curve that is observed in that energy band, enabling lag \textit{vs.} Fourier frequency and lag \textit{vs.} energy spectra to be predicted.

The illumination of the accretion flow by the primary X-ray source is computed by ray-tracing calculations in the Kerr spacetime. In the simplest case, the flow is illuminated by an isotropic point source located on the polar axis above the black hole. Rays are started at equal intervals in solid angle in the rest frame of the source (a total of 1.3 billion rays over the full $4\pi$ solid angle). The momentum vectors of the rays are transformed into the global Boyer-Lindquist co-ordinate system to obtain the constants of motion for the rays, then the paths of the rays are computed by integrating the geodesic equations until they reach the equatorial plane where the accretion flow is presumed to lie, following the method of \citet{understanding_emis_paper}, although \citet{taylor_reynolds} discuss the implications of the disc geometry and the disc having a finite scale height on the reverberation signatures. Between 750 million and 850 million rays reach the accretion flow in each simulation (depending on the black hole spin) to yield high-resolution sampling of the innermost regions.

When the rays reach the disc, the $t$ co-ordinate is recorded and the redshift is calculated by taking the inner product of the ray 4-momentum with the 4-velocity of both the emitter ($\mathcal{E}$) and observer ($\mathcal{O}$) to obtain the ratio of the energy measured by each:
\begin{equation}
	\label{redshift.equ}
g = \frac{E_\mathcal{O}}{E_\mathcal{E}} = \frac{g_{\mu\nu} p^\mu(\mathcal{O}) u_\mathcal{O}^\nu}{g_{\mu\nu} p^\mu(\mathcal{E}) u_\mathcal{E}^\nu}
\end{equation}
For rays landing on the disc, the 4-velocity of the disc element is given by $[u_\mathcal{O}^\mu] = \dot{t}(1, 0, 0, \Omega)$, while in the plunging region, $[u_\mathcal{O}^\mu] = (\dot{t}, \dot{r}, 0, \dot{\varphi})$ from Equations~\ref{tdot.equ}, \ref{rdot.equ} and \ref{phidot.equ}.

To compute the density, illumination and ionisation profiles, the accretion flow is divided into radial bins. There are 100 logarithmically spaced bins between $r_\mathrm{ISCO}$ and $r=500$\rg\ (the stably orbiting disc) and a further 25 logarithmically spaced bins between the event horizon and $r_\mathrm{ISCO}$ for the $a=0$ case and 10 in the $a=0.998$ case (where the ISCO is closer to the event horizon).

 The relative flux incident on each radius of the disc is calculated by counting the number of rays landing in each bin, each weighted by the square of the redshift factor of that ray. One factor of redshift accounts for the shift in energy of each photon travelling along the ray path and the other accounts for the photon emission or arrival rate measured by an observer at the emitting and receiving ends of the ray according to their own proper times, following the calculation of accretion disc emissivity profiles by \citet{understanding_emis_paper}. Note that the solid angle aberration between the source and observers on the disc is accounted for by the number of rays (started at equal interval in solid angle) counted in each bin. The incident energy is converted into the incident flux by dividing by the proper area of the radial bin, that is the area of a patch of disc as measured by a co-rotating observer. The calculation of the proper area of a patch on a general part of the accretion flow with 4-velocity components $u^\mu$ is outlined in Appendix~\ref{area.app}.

Once the density and illumination profiles of the accretion flow have been calculated, including the plunging region, the ionisation parameter as a function of radius can be computed, describing the ionisation state of the disc and determining the details of the reprocessed X-ray spectrum. The ionisation parameter is given by $\xi(r) = 4\pi F(r)/n(r)$, directly using the disc density in place of the hydrogen number density, $n$. As for the density profile, the ionisation parameter is defined at a specific radius, $r_0$, to be $\xi_0$ and given the relative density and relative illumination, the ionisation parameter is computed for all radii. While \citet{reynolds+97} compute the ionisation at every point directly from the illuminating flux implied by a given the mass accretion rate, defining the density and ionisation at the ISCO allows for a more direct comparison to observations without having to make assumptions about the relationship between the accretion rate, X-ray luminosity (as opposed to bolometric luminosity) and disc irradiation. The density, illumination and ionisation profiles of the disc are shown in Fig.~\ref{disc_profiles.fig}.

For each ray incident upon the disc at co-ordinates $(r, \varphi)$, the reprocessed X-ray spectrum from this position is obtained from the \textsc{xillverd} model of \citet{xillver_density}. This gives the reprocessed spectrum, as measured in the rest frame of the accreting material with variable hydrogen number density, $n$ and ionisation parameter, $\xi$. The rest-frame spectrum is varied across the disc according to the calculated variation in density and ionisation parameter.

The reprocessed rays are then transferred to an observer at infinity to obtain the spectrum that would be recorded by a telescope. The redshift of each point on the disc as seen by the observer is computed by back-tracing rays that pass perpendicular through a square image plane 10,000\rg\ from the singularity, centred at $\theta$ co-ordinate corresponding to the inclination at which the system is observed (here taken to be 60\,deg). This image plane represents the patch of sky observed by the telescope, with the parallel rays propagating perpendicular to the plane appearing to come from infinity. Due to gravitational light bending, these rays will bend in the region of space close to the black hole. When these rays hit the disc, their time is recorded and their redshift calculated using Equation~\ref{redshift.equ}. The time and redshift for each point on the accretion flow (including the plunging region) is stored in a lookup table that is referenced during the ray tracing calculation between the primary source and the flow.

Each primary ray that hits the disc produces the full reprocessed spectrum with total normalisation determined by the source-to-disc redshift. This reprocessed spectrum is shifted in energy according to the redshift between the patch of the disc and the observer and the total travel time of the ray from source to disc to observer is summed. Upon arrival at the observer, rays are weighted by the cube of the redshift between the patch on the disc and the observer according to Liouville's theorem. This accounts for time dilation between the observer and source altering the apparent arrival rate of photons along the ray as well as the solid angle into which the source can emit rays that land in unit area on the image plane, assuming the reprocessed X-rays are emitted isotropically in the rest frame of the reprocessor. The number of photons is then counted as a function of arrival time and energy at the observer to produce the impulse response function of the disc as in \citet{propagating_lag_paper}.

\section{Reverberation Response of the Plunging Region}
\subsection{Line Response}
We begin by examining the emission line response function of the accretion flow; that is the flux received by the observer as a function of time since the continuum emission from a point source is observed and energy shift (from the rest frame emission wavelength) if the photons that are produced in single emission line, here taken to be the iron K$\alpha$ fluorescence line emitted at 6.4\keV\ in the rest frame of the reprocessing material. The response at a given time interval is convolved with the rest frame emission spectrum in order to obtain the total observed spectrum at that time, while convolving the response across a particular energy band with the time series describing the variability of the primary continuum emission yields the light curve observed in that energy band.

\begin{figure*}
\centering
\subfigure[Total response] {
\includegraphics[width=55mm]{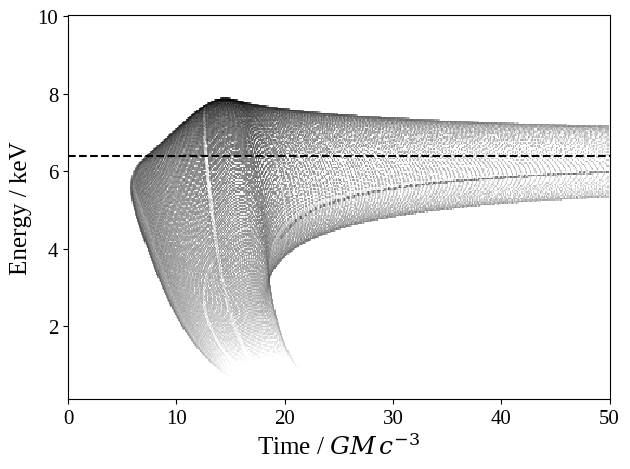}
\label{avg_arrival_propin.fig:c}
}
\subfigure[Disc response] {
\includegraphics[width=55mm]{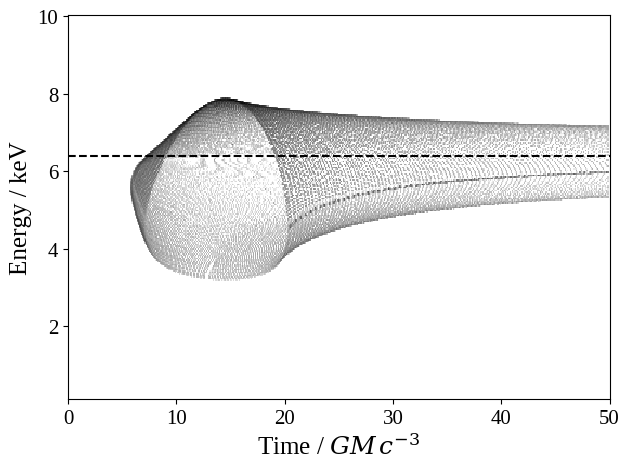}
\label{avg_arrival_propin.fig:0.01c}
}
\subfigure[Plunging region response] {
\includegraphics[width=55mm]{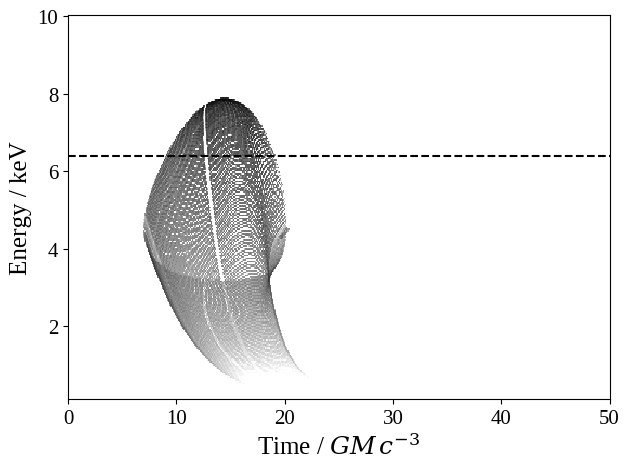}
}
\subfigure {
\includegraphics[width=55mm]{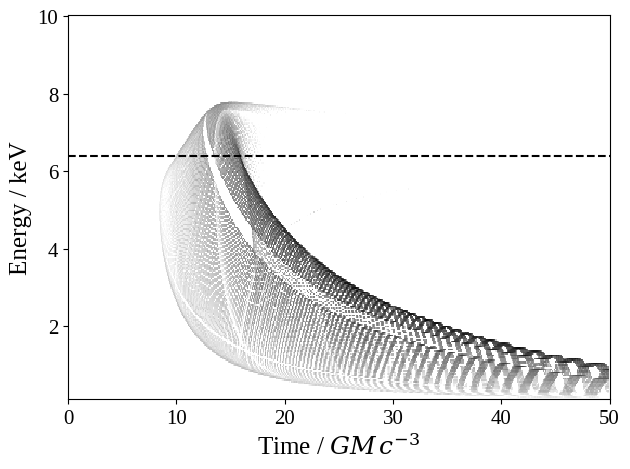}
\label{avg_arrival_propin.fig:0.1c}
}
\subfigure {
\includegraphics[width=55mm]{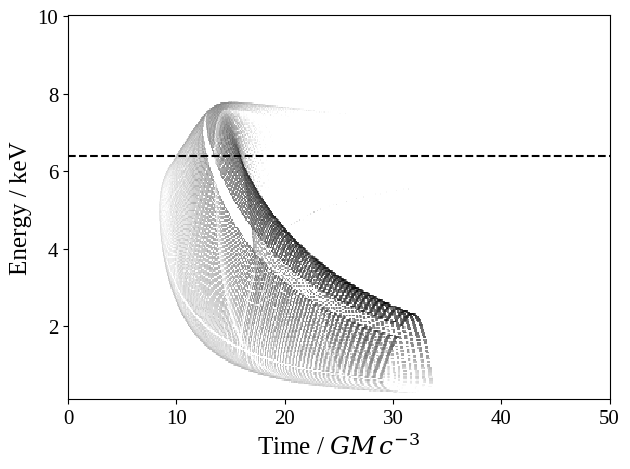}
}
\subfigure {
\includegraphics[width=55mm]{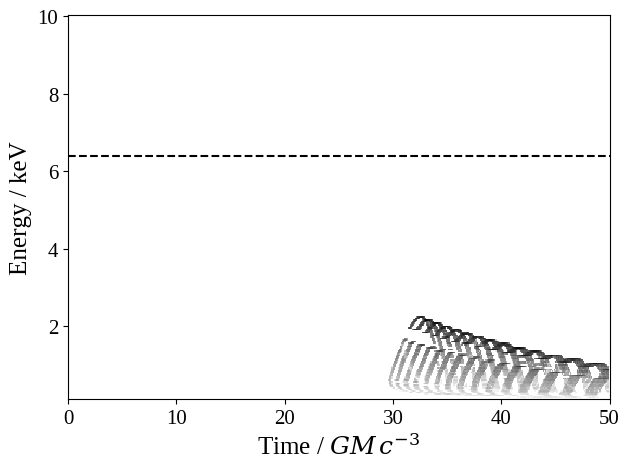}
}
\caption[]{The response function of the iron K$\alpha$ fluorescence line (emitted at 6.4\keV\ in the rest frame) from the accretion flow following a single variation in the emission from the point-like X-ray source. Shading corresponds to the photon count rate received by the observer as a function of X-ray energy and time after the detection of the continuum variation. The accretion flow is observed at an inclination of 60\,deg to the normal. Top panels show the response functions for a non-spinning black hole with the X-ray source located at a height $h=5$\rg\ on the polar axis and bottom panels show the response for a maximally spinning black hole ($a=0.998$) with the source located at a height $h=2$\rg. The left panels show the total response. The middle panels show the response from the stably-orbiting disc while the right panels show the response from the plunging region. The dashed lines show the rest frame energy of the emission line.}
\label{line_response.fig}
\end{figure*}

\begin{figure*}
\centering
\subfigure[$a = 0$] {
\includegraphics[width=85mm]{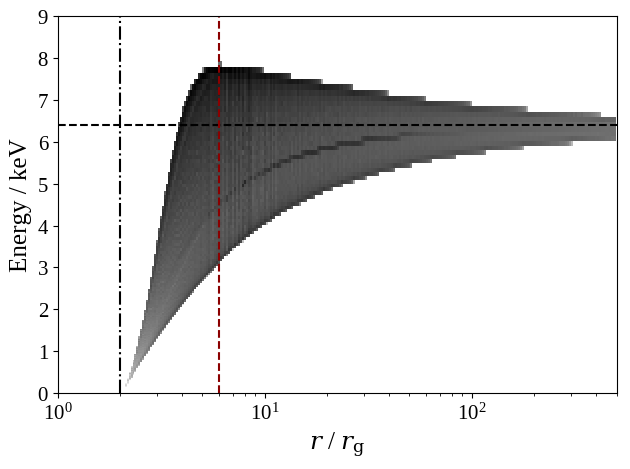}
\label{avg_arrival_propin.fig:c}
}
\subfigure[$a = 0.998$] {
\includegraphics[width=85mm]{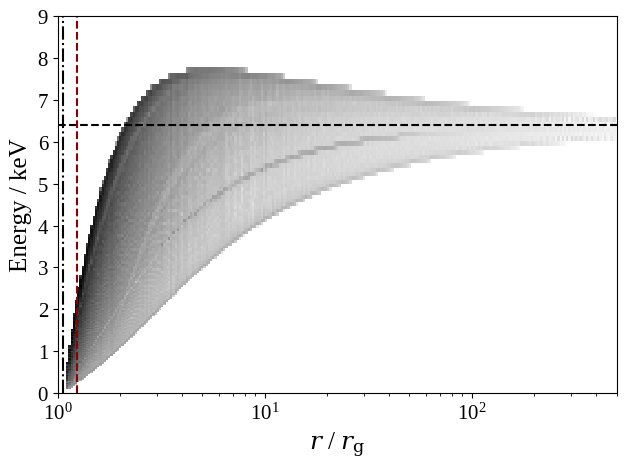}
\label{avg_arrival_propin.fig:0.1c}
}
\caption[]{The X-ray count rate from the iron K$\alpha$ fluorescence line emitted from the accretion flow as a function of radius and X-ray energy. The red dashed line shows the location of the innermost stable circular orbit (ISCO) and the black dot-dash line shows the event horizon.}
\label{enr_response.fig}
\end{figure*}

Fig.~\ref{line_response.fig} shows the response functions of the iron K$\alpha$ emission line. The left panels show the total response, while the middle and right panels show, respectively, the response functions from each the stably-orbiting portion of the accretion disc and the plunging region. Fig.~\ref{enr_response.fig} shows the observed redshifts as a function of radius.

The form of this response function has been extensively explored by \citet{reynolds+99,cackett_ngc4151,propagating_lag_paper}. The earliest response is seen from energies (\textit{i.e.} the energy shift resulting from the combined effects of Doppler shift and gravitational redshift as a function of radius) corresponding to the shortest total light travel time from source to disc to observer. For a point source at a height of 5\rg, this is the response at 5.5\keV, while lowering the source to 2\rg\ moves this point to a lower energy. Some time later, the response of more extremely redshifted and blueshifted photons from, respectively, the receding and approaching sides of the inner parts of the disc is seen. These photons must travel through the strong gravitational potential close to the black hole and hence their passage is delayed. After the initial appearance of redshifted and blueshifted emission from the near side of the inner disc, a second loop is seen in the response function as delayed, progressively more redshifted emission is seen. These delayed photons are the emission that is reprocessed by the back-side of the disc that would classically be hidden behind the black hole shadow. Gravitational lensing bends (and magnifies) these rays into our line of sight. The late-time response is seen from the outermost parts of the disc, producing a double-peaked line profile with blue- and redshifted emission from the approaching and receding sides, converging on the rest frame energy of the line as the orbital velocity decreases.

In the case of a non-spinning black hole, the plunging region, within the innermost stable circular orbit, spans a relatively large range in radius, from the event horizon at 2\rg\ to the ISCO at 6\rg. This means that with the continuum source at a height of 5\rg\ above the singularity, as much as 30 per cent of the X-rays reprocessed from the accretion flow come from inside the ISCO. The accreting material on the ISCO has sufficient angular momentum that when it first enters the plunging orbit, it still orbits the black hole several times before entering a rapid plunge close to the horizon. As a result, the reverberation response from the plunging region exhibits both red-and blueshifted emission from the receding and approaching sides. The plunging region also extends the redshifted tail below 2\keV, redshifts typically associated with more rapidly spinning black holes, now that reverberation is seen from more rapidly orbiting (plunging) material deeper within the gravitational potential. Relativistic beaming of this emission along the trajectory of the plunging material, however, means that this redshifted emission is significantly weaker than that seen from accretion flows around rapidly spinning black holes.

In the case of the maximally spinning black hole, all X-ray emission from the plunging region is observed to be redshifted. The ISCO and plunging region lie within the region sufficiently close to the singularity that the gravitational redshift is strong enough to overcome any Doppler blueshifting. This can be seen in Fig.~\ref{enr_response.fig} which shows the observed line flux as a function of radius and energy shift. Blueshifted emission around an maximally spinning black hole is seen from radii outside approximately 2\rg. Around a maximally spinning black hole, the plunging region adds a late-time response to the line at energies below 2\keV\ (for a 6.4\keV\ line). In this case, the significantly smaller plunging region, extending to the ISCO at only 1.235\rg\ means that only 2 per cent of the total reprocessed emission comes from within the ISCO.

\subsection{Full Spectral Response}

\begin{figure*}
\centering
\subfigure[Total response] {
\includegraphics[width=55mm]{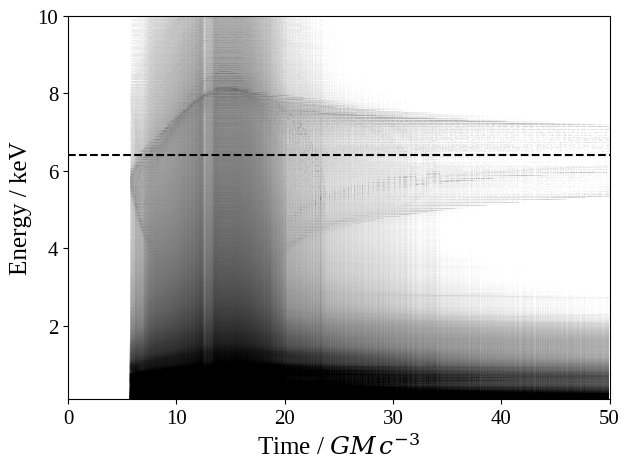}
\label{avg_arrival_propin.fig:c}
}
\subfigure[Disc response] {
\includegraphics[width=55mm]{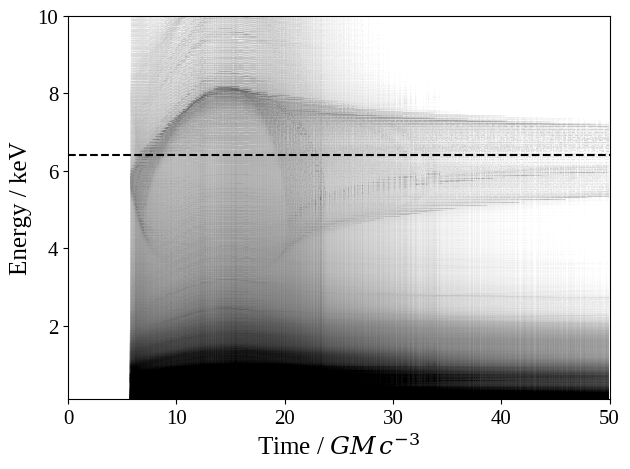}
\label{avg_arrival_propin.fig:0.01c}
}
\subfigure[Plunging region response] {
\includegraphics[width=55mm]{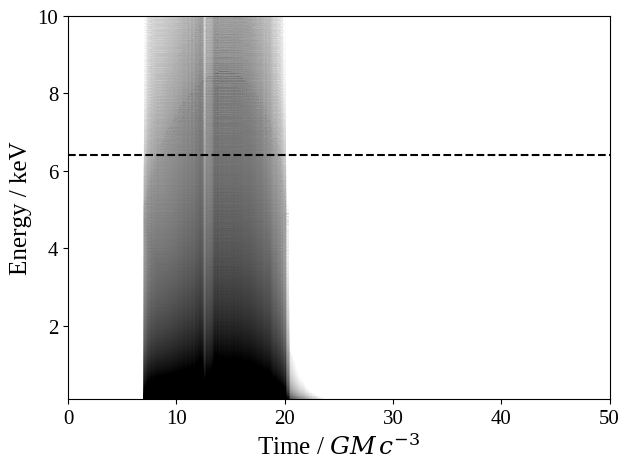}
}
\subfigure {
\includegraphics[width=55mm]{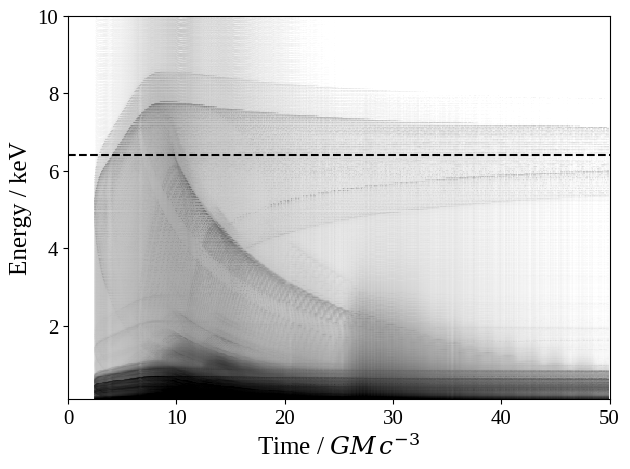}
\label{avg_arrival_propin.fig:0.1c}
}
\subfigure {
\includegraphics[width=55mm]{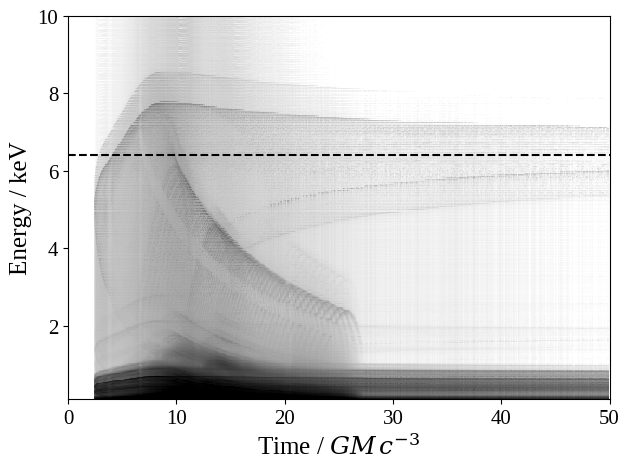}
}
\subfigure {
\includegraphics[width=55mm]{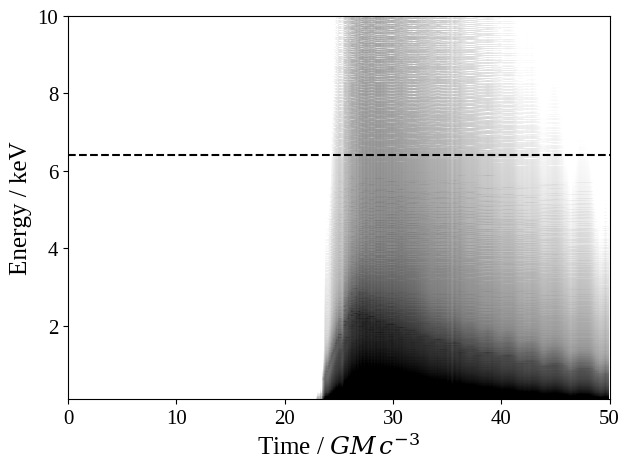}
}
\caption[]{The full spectral response function of the accretion flow following a single variation in the emission from the point-like X-ray source, computed by convolving the line response function with the rest frame spectral model \textsc{xillverd} with varying density and ionisation parameter as a function of radius on the disc. Top panels show the response functions for a non-spinning black hole with the X-ray source located at a height $h=5$\rg\ on the polar axis and bottom panels show the response for a maximally spinning black hole ($a=0.998$) with the source located at a height $h=2$\rg. The left panels show the total response. The middle panels show the response from the stably-orbiting disc while the right panels show the response from the plunging region.}
\label{spectral_response.fig}
\end{figure*}

The full spectral reverberation responses are shown in Fig.~\ref{spectral_response.fig}. The line responses, separated by radius in the accretion flow, were convolved with a rest-frame spectrum from the \textsc{xillverd} model with variable density and ionisation parameter set for that radius. Following recent findings that accretion discs in AGN may reach densities as high as $10^{19}$\pcmcu \citep{jiang_iras}, we set the hydrogen number density of the accretion disc to $10^{19}$\pcmcu, falling off as $r^{-\frac{3}{2}}$ over the outer disc to approximate the Novikov-Thorne density profile. In the cases considered here, the ionisation parameter is set to $\xi_0 = 1000$\ergcmps\ at the ISCO and its variation across the disc was computed from the density and illumination profiles. While the normalisation of the density is not important for the dynamics of the plunging region, it does affect the shape of the X-ray spectrum produced at each radius in the accretion flow. The reprocessed spectrum is produced by an incident power law continuum with photon index $\Gamma = 2.5$ and the iron abundance is taken to be eight times the Solar value, representative of the AGN in which strong reverberation signatures are seen from the accretion disc \citep[\textit{e.g.}][]{fabian+09,zoghbi+09,parker_mrk335,mrk335_corona_paper,jiang_iras}.

The most prominent feature of the spectral response is, of course, the iron K$\alpha$ fluorescence line which is clearly visible around 6.4\keV. The key difference, however, is that due to the variation in ionisation as a function of radius across the accretion flow, the rest frame energy of the iron fluorescence line is not constant. Weakly ionised material produces K$\alpha$ emission from neutral iron at 6.4\keV, though once the ionisation parameter increases to between 100 and 500\ergcmps, there are sufficient Fe\,\textsc{xvii} to Fe\,\textsc{xxiii} ions with an L-shell vacancy to absorb the emitted K$\alpha$ photons, weakening the observed line. Once the material becomes highly ionised, with $\xi > 500$\ergcmps, iron is found in the helium-like and hydrogen-like Fe\,\textsc{xxv} and Fe\,\textsc{xxvi} states which produce K$\alpha$ lines at 6.67 and 6.97\keV.

The variation in line energy through the plunging region was modelled by \citet{reynolds+97} by defining four ranges of ionisation parameter. Where $\xi < 100$\ergcmps, a narrow line is produced at 6.4\keV. Where the ionisation parameter increased to between 100 and 500\ergcmps, no line was produced due to resonant Auger destruction of the line photons by the ionised species. Where $\xi > 500$\ergcmps, the narrow line emission was shifted to 6.97\keV\ corresponding to the emission from hydrogenic iron and for $\xi > 5000$\ergcmps, no line was produced.

Here, however, the rest-frame reflection spectrum as a function of ionisation is self-consistently computed by the \textsc{xillverd} model. This produces the correct ratio of neutral, helium-like and hydrogen-like emission lines based upon the ionisation balance in the plasma and accounts for other scattering processes that are acting, most notably here the Compton scattering of the line photons produced in the most highly ionised parts of the flow by free electrons. This means that in the most highly ionised regions, the spectrum in the rest frame does not contain a strong, sharp line, rather the line is broadened. weakening the appearance of the emission line above the continuum. The \textsc{xillver} models account for Auger decay of the line by computing the branching ratios for Auger \textit{vs.} fluorescent emission but do not account for resonant Auger scattering. \citet{loisel+17}, however, suggest through laboratory experiments that in the conditions expected within the accreting plasma, resonant Auger destruction of fluorescent lines is not as important as originally anticipated.

\begin{figure*}
\centering
\subfigure[$a = 0$] {
\includegraphics[width=85mm]{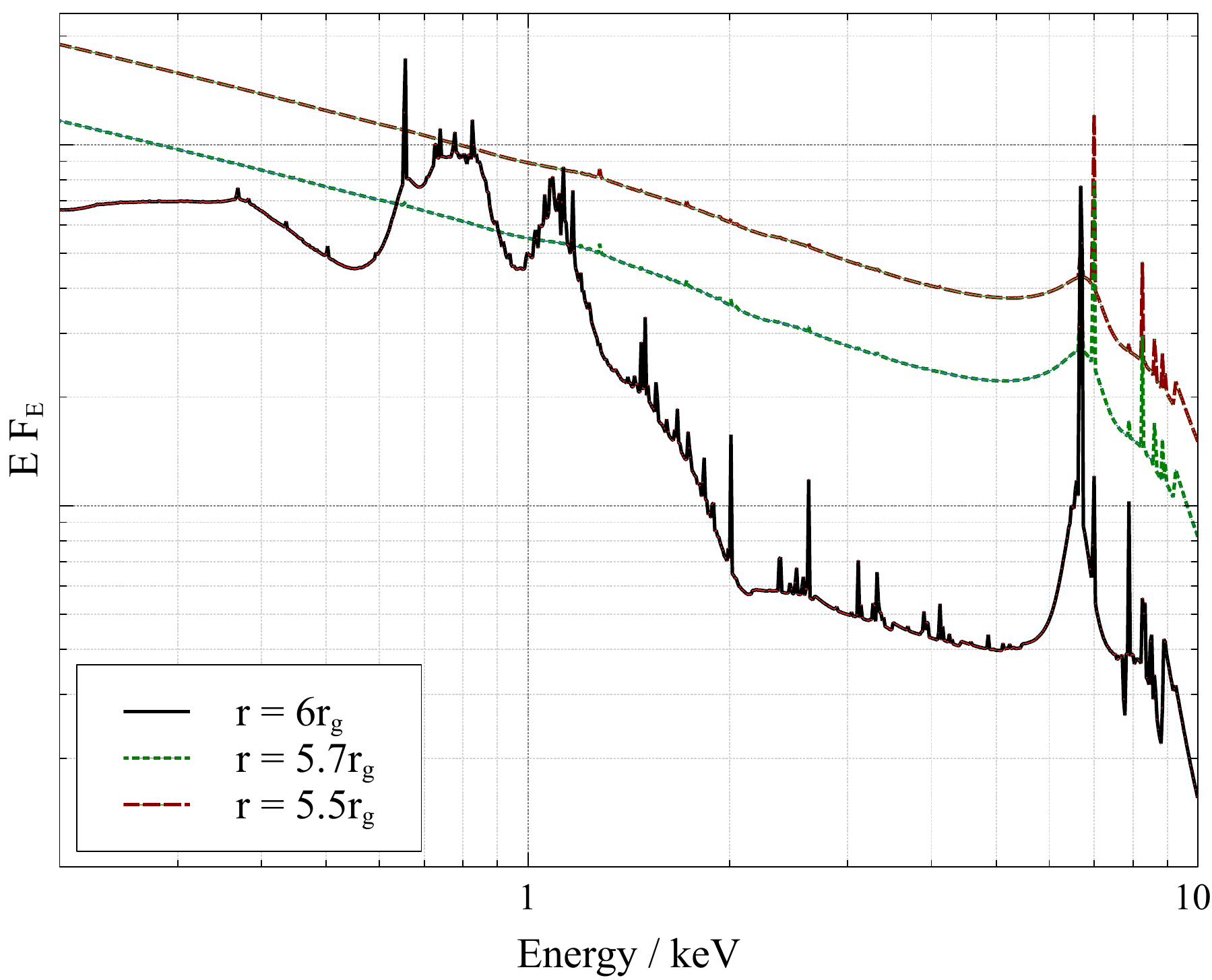}
\label{avg_arrival_propin.fig:c}
}
\subfigure[$a = 0.998$] {
\includegraphics[width=85mm]{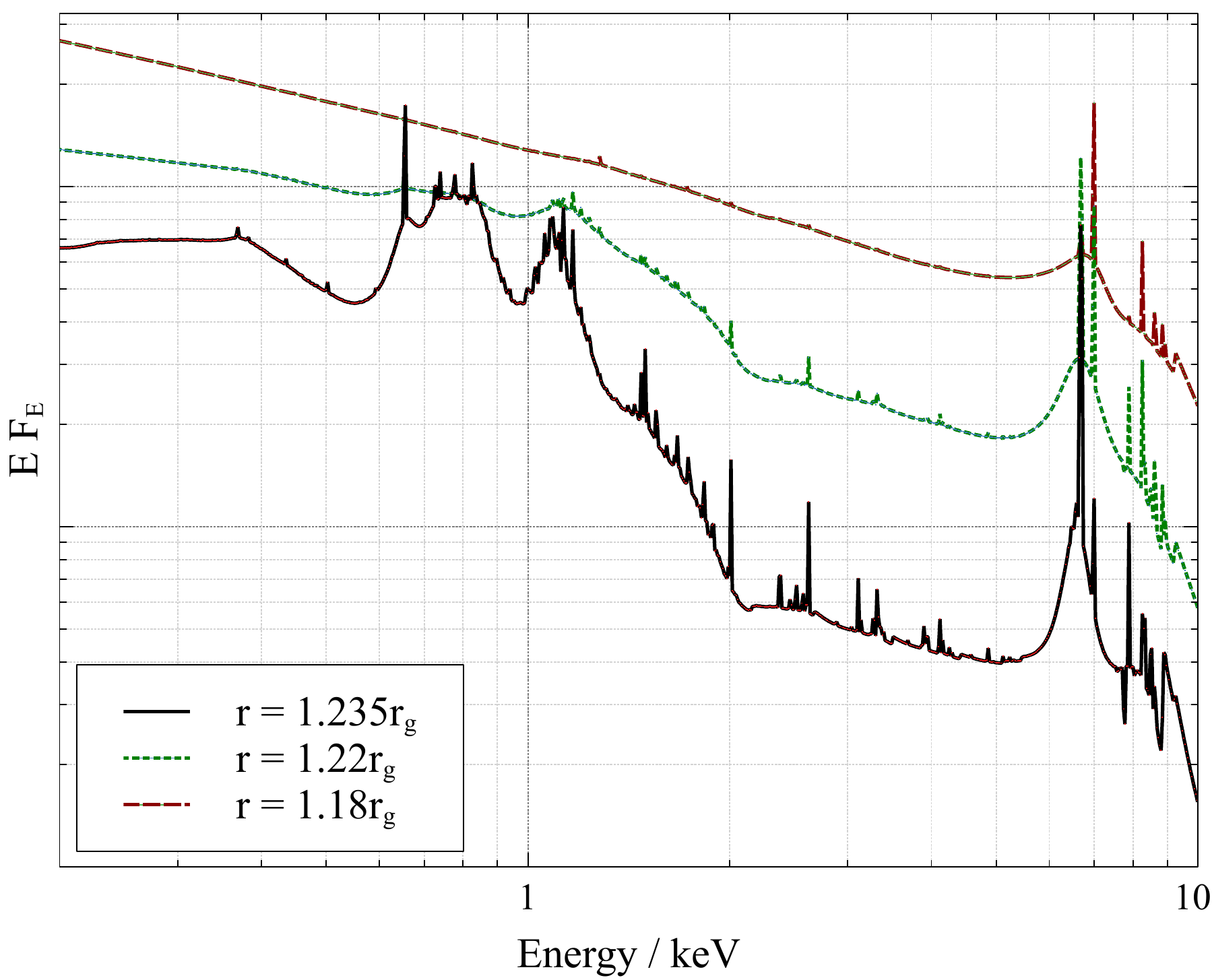}
\label{avg_arrival_propin.fig:0.1c}
}
\caption[]{The spectrum of the reprocessed X-rays, observed in the co-rotating rest frame of the disc, from radii just outside and just inside the ISCO. The ionisation of the plasma rises rapidly as the transition to plunging orbits leads to an abrupt drop in density of the strongly irradiated accretion flow.}
\label{restframe_spec.fig}
\end{figure*}

In the full spectral response function for the non-spinning black hole, the broad range of red- and blueshifts in the plunging region result in the reflected emission being smeared out into a smooth continuum-like spectrum. A broad line-like response is faintly visible in the response function, showing the blueshifted loop and tail towards low energies. This is predominantly from hydrogenic iron, emitted at 6.97\keV\ in the rest frame, though significantly weaker than the blueshifted hydrogenic line response reported by \citet{reynolds+97} and \citet{reynolds+99} due to the broadening of the line in the rest frame by Compton scattering. This variation in the rest-frame spectrum across the ISCO is shown in Fig.~\ref{restframe_spec.fig}.

In the case of the maximally-spinning black hole, the response from the plunging region is seen entirely in the most redshifted part of the iron K$\alpha$ line, extending the tail of the response to low energies (below 2\keV) at late times. The increasing ionisation, particularly the sharp increase in ionisation parameter as the density drops sharply across the ISCO, can be seen as a smearing in the highest energy part of this tail due to the combination and broadening of emission lines from different species.

Below 1\keV, the response from the soft emission lines (most notably iron L and those from oxygen and nitrogen) are blended together. The response of each of these lines will follow the shape of the iron K$\alpha$ response. The effect of the density gradient in the disc can also be seen. As discussed by \citet{xillver_density}, at high densities, the disc produces additional thermal emission by bremsstrahlung due to suppressed cooling which can extend up to 2\keV. In the disc response, this can be seen to fade away at late times as the response moves to the outer disc where the density is lower.

\section{Detecting Emission from the Plunging Region}
Having computed the X-ray reverberation response functions for the accretion disc and the plunging region between the innermost stable circular orbit and the event horizon, we now consider how X-ray emission from the plunging region may be detected in X-ray observations of accreting black holes. In the case of a non-spinning black hole, the X-rays reverberating from the plunging region form a smooth continuum-like component, responding before the reverberation from the accretion disc but after the primary continuum. Reverberation from the plunging region around a maximally spinning black hole produces a late-time response in the extremely redshifted tails of emission lines.

\subsection{The X-ray Spectrum}
\begin{figure*}
\centering
\subfigure[$a = 0$] {
\includegraphics[width=85mm]{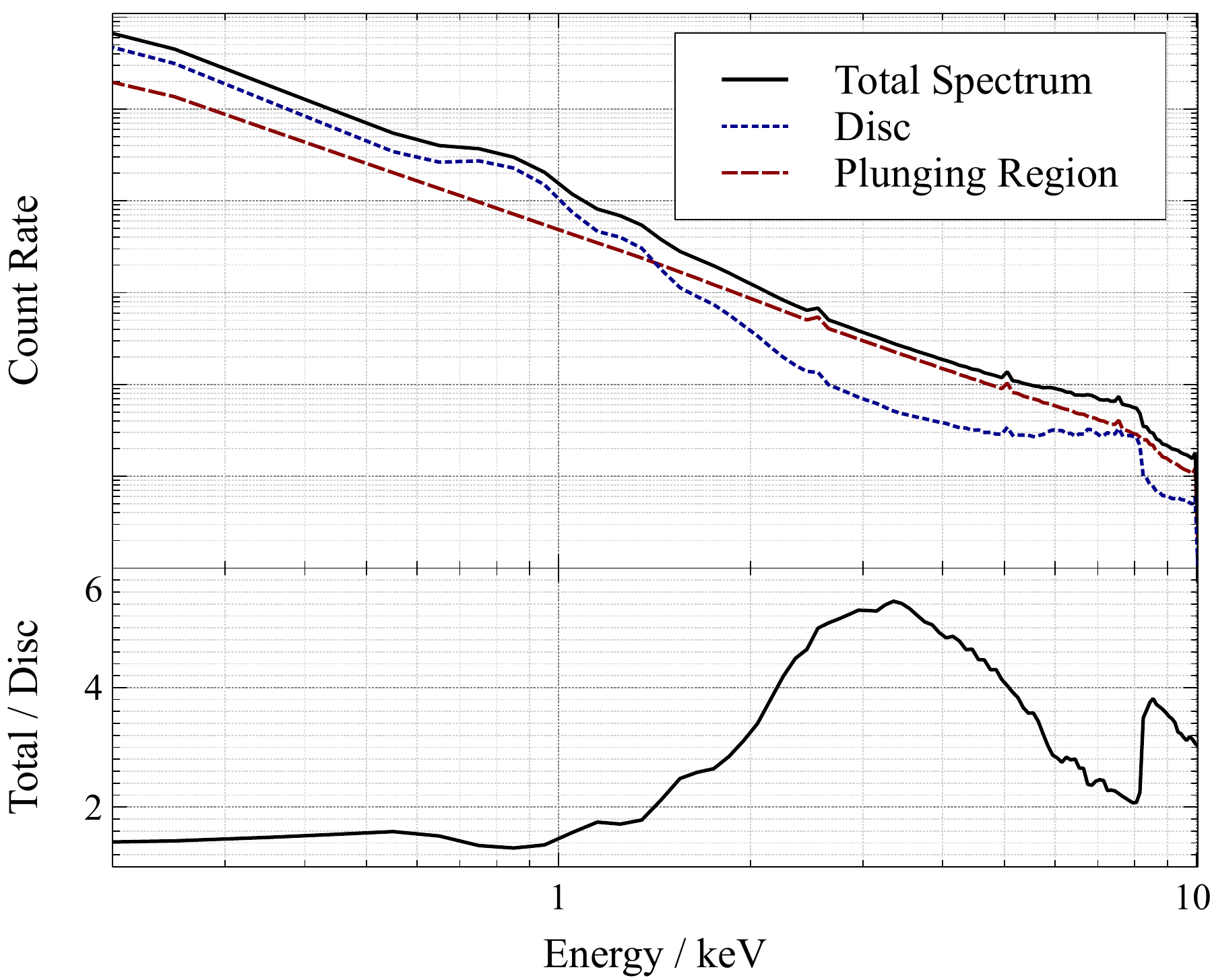}
\label{avg_arrival_propin.fig:c}
}
\subfigure[$a = 0.998$] {
\includegraphics[width=85mm]{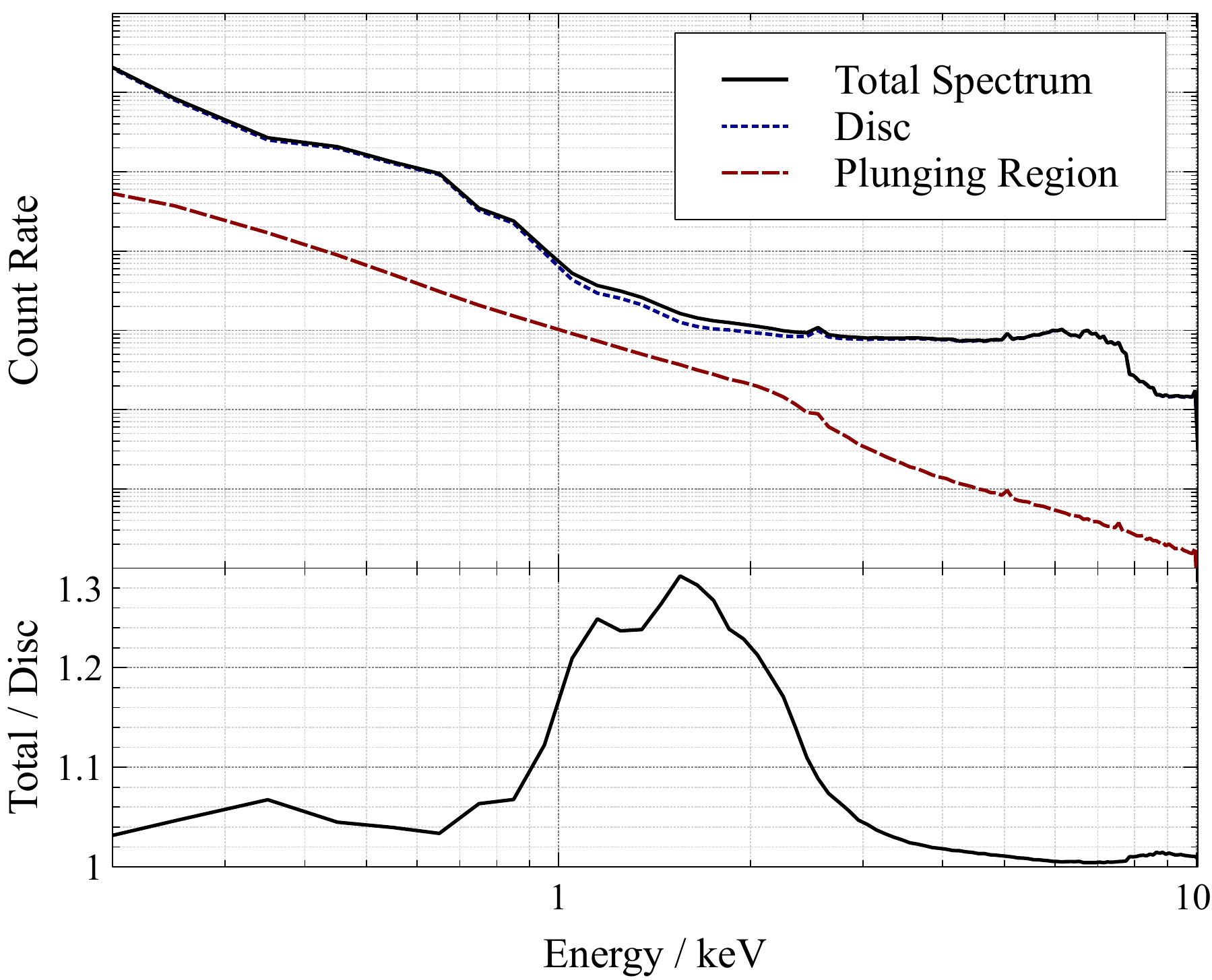}
\label{avg_arrival_propin.fig:0.1c}
}
\caption[]{The time-averaged spectrum of the X-ray emission reprocessed/reflected from the accretion flow. The total reprocessed spectrum is shown along with the contributions from the stably-orbiting disc and the plunging region. The lower panels show the ratio of the total spectrum to the spectrum arising from only the disc if no emission were seen from the plunging region.}
\label{spectrum.fig}
\end{figure*}

We first consider the time-averaged spectrum of the X-ray emission that is reprocessed from the accretion flow. Fig.~\ref{spectrum.fig} shows the observed X-ray spectrum from the accretion disc for both the non-spinning and maximally-spinning cases, showing the total spectrum as well as the contributions from the stably orbiting disc and the plunging region. Also shown is the ratio of the total spectrum to just the disc spectrum to show the difference made to observations by the plunging region. The spectrum is computed by summing the response function in each energy band across all time bins.

The extreme redshifts and blueshifts from the plunging region in the case of the non-spinning black hole cause the line emission to become smeared out, meaning that the emission closely resembles a smooth power-law continuum, albeit with a slight edge corresponding to the blueshifted edge of the iron line. Excess emission above that expected from just the disc (with no emission from the plunging region) is seen between 2 and 6\keV\ as additional redshifted emission is produced from the line photons. This excess emission would be difficult to detect as part of the reprocessed emission, however, since its smooth nature would mean it would be fit in any spectral model by adding to the directly observed continuum component. Its flux is significant compared to that reverberating from the disc since the plunging region spans a large area of space (out to 6\rg) close to the primary X-ray source.

In the case of the maximally spinning black hole, the smaller plunging region means that its emission is well below the level of the emission from the disc at all energies (and at least an order of magnitude below at most energies). The emission from the plunging region gives the biggest relative contribution between 1 and 2\keV, where a slight excess above the reprocessed continuum is produced by extremely redshifted iron K$\alpha$ line photons. The additional contribution to the time-averaged spectrum is 30 per cent at these energies, though it would likely be difficult to detect as anything other than additional primary continuum emission.

\subsection{Detection through X-ray Timing}
Though the X-ray emission from the plunging region may be difficult to detect in the time-averaged X-ray spectrum, with the reprocessed emission from the plunging region being largely indistinguishable from additional continuum emission, this emission from the accretion flow is delayed with respect to the arrival of the primary continuum at the observer. We therefore explore whether this emission can be detected using X-ray spectral timing techniques.

\subsubsection{The Lag-Energy Spectrum}

The first calculation to consider is the average response time as a function of X-ray energy, or the \textit{lag-energy} spectrum, which shows the relative times at which different energy bands respond to a variation in the continuum. It is computed by calculating the weighted average of the response function along the time axis at each energy. The continuum emission is included in the calculation as a delta function in time at $t=0$. For a point source, there is precisely one ray that reaches an observer at infinity, hence all photons from a given flash of emission will arrive at the same time. The continuum emission has a power law spectrum and is normalised relative to the sum of all reprocessed emission across the response function to produce a realistic reflection fraction (the ratio of reflected flux from the disc to directly observed continuum flux) of $R=2$. Values close to this are commonly measured in the time-average spectra of Seyfert galaxies.

\begin{figure*}
\centering
\subfigure[$a = 0$] {
\includegraphics[width=85mm]{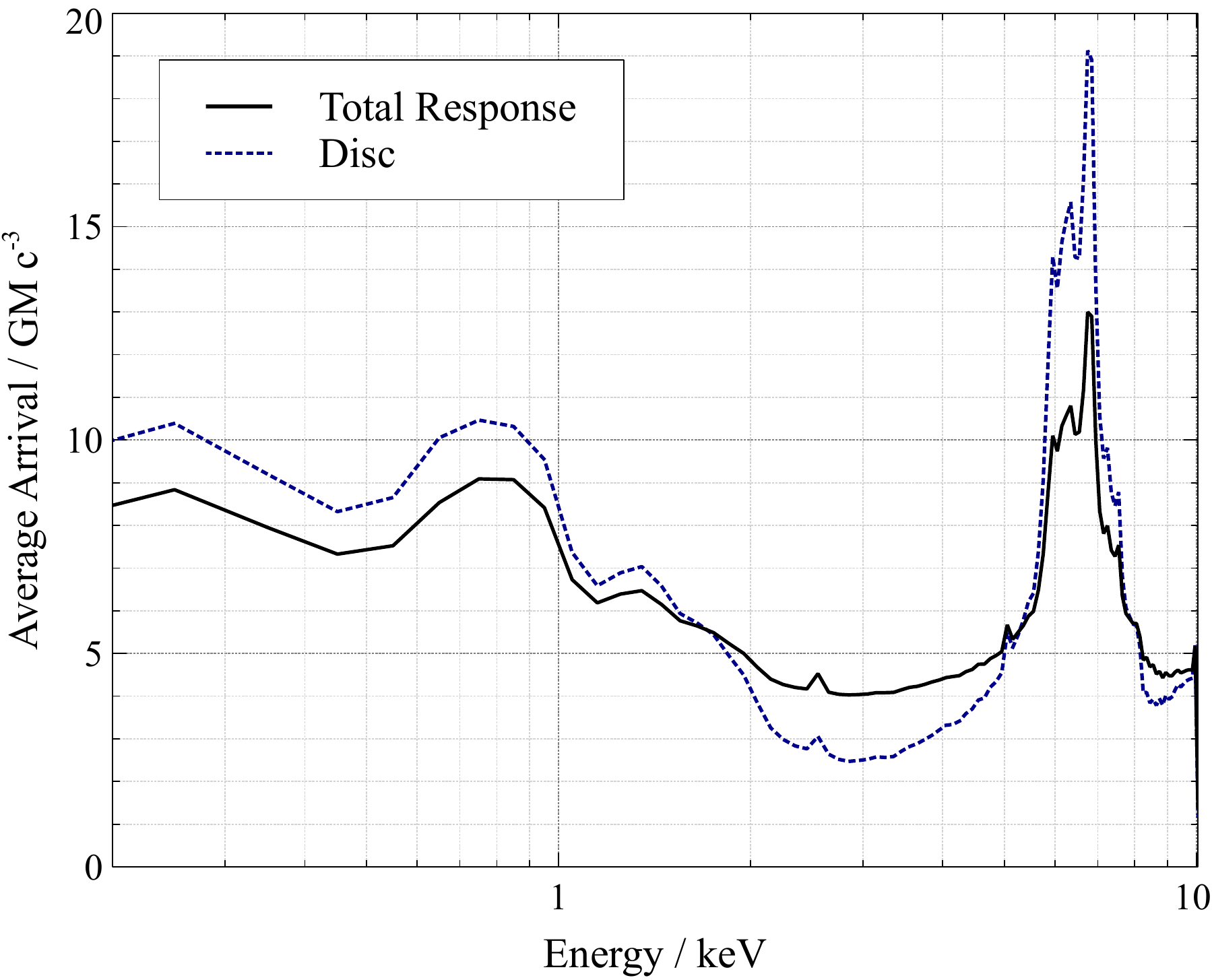}
\label{avg_arrival_propin.fig:c}
}
\subfigure[$a = 0.998$] {
\includegraphics[width=85mm]{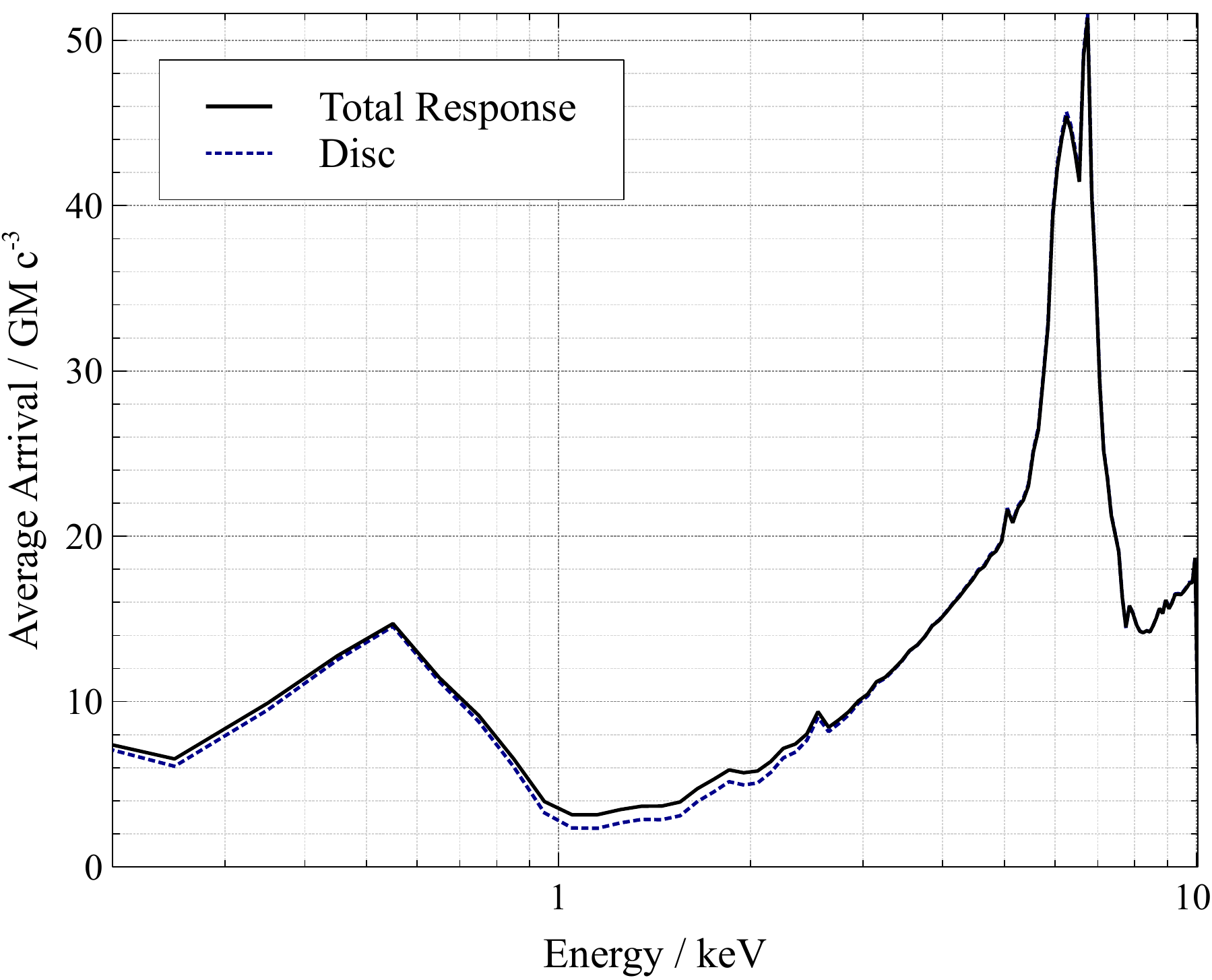}
\label{avg_arrival_propin.fig:0.1c}
}
\caption[]{The average response time as a function of photon energy (including both the reverberating and continuum emission components) comparing the total response to just that from the disc with no response from the plunging region.}
\label{avg_arrival.fig}
\end{figure*}

The average response times of successive photon energies for the non-spinning and maximally-spinning black hole cases are shown in Fig.~\ref{avg_arrival.fig}. As with the X-ray spectrum, the addition of emission from the plunging region to just the emission from the disc (as well as the directly-observed continuum) has only a small effect on the lag-energy spectrum. In the maximal spin case, a slight delay of $0.8\,GMc^{-3}$ is noted between energies of 1 and 2\keV. Since the true zero time is unknown and only the relative response times of different energies to correlated variations can be measured, this would appear as a very slight shortening of the iron K lag, as well as of the soft lag due to the reverberation in the soft X-ray emission lines, with respect to the 1-2\keV\ band that is taken as the proxy for the directly-observed continuum emission (it is the most continuum-dominated band). $GMc^{-3}$ is the light travel time over one gravitational radius for the black hole mass in question.

The addition of the reprocessing in the plunging region around a non-spinning black hole shortens the reverberation time across all energy bands. For a non-spinning black hole, the innermost stable orbit at a radius of 6\rg\ results in there being a significant region around the black hole in which there is no stably orbiting disc, hence there is a sizeable light travel time from an X-ray source on the polar axis to the ISCO. Accounting for reverberation much closer to the X-ray source shortens the response time of the iron K$\alpha$ and soft X-ray lines, though delays the response of the 2-4\keV\ band since this now contains a significant fraction of redshifted iron K$\alpha$ emission from the plunging region rather than being dominated by directly observed continuum emission. Again, however, only the relative response times of successive energy bands can be measured and the change to the lag-energy spectrum simply appears to a rescaling of the lag times around the average arrival time, which could also be accounted for by a smaller black hole mass (reducing the conversion from measured time to gravitational radii) or by an X-ray source at a lower height, closer to the reprocessor.

\subsubsection{Energy-Resolved Time Response}

\begin{figure*}
\centering
\subfigure[1-2\keV] {
\includegraphics[width=55mm]{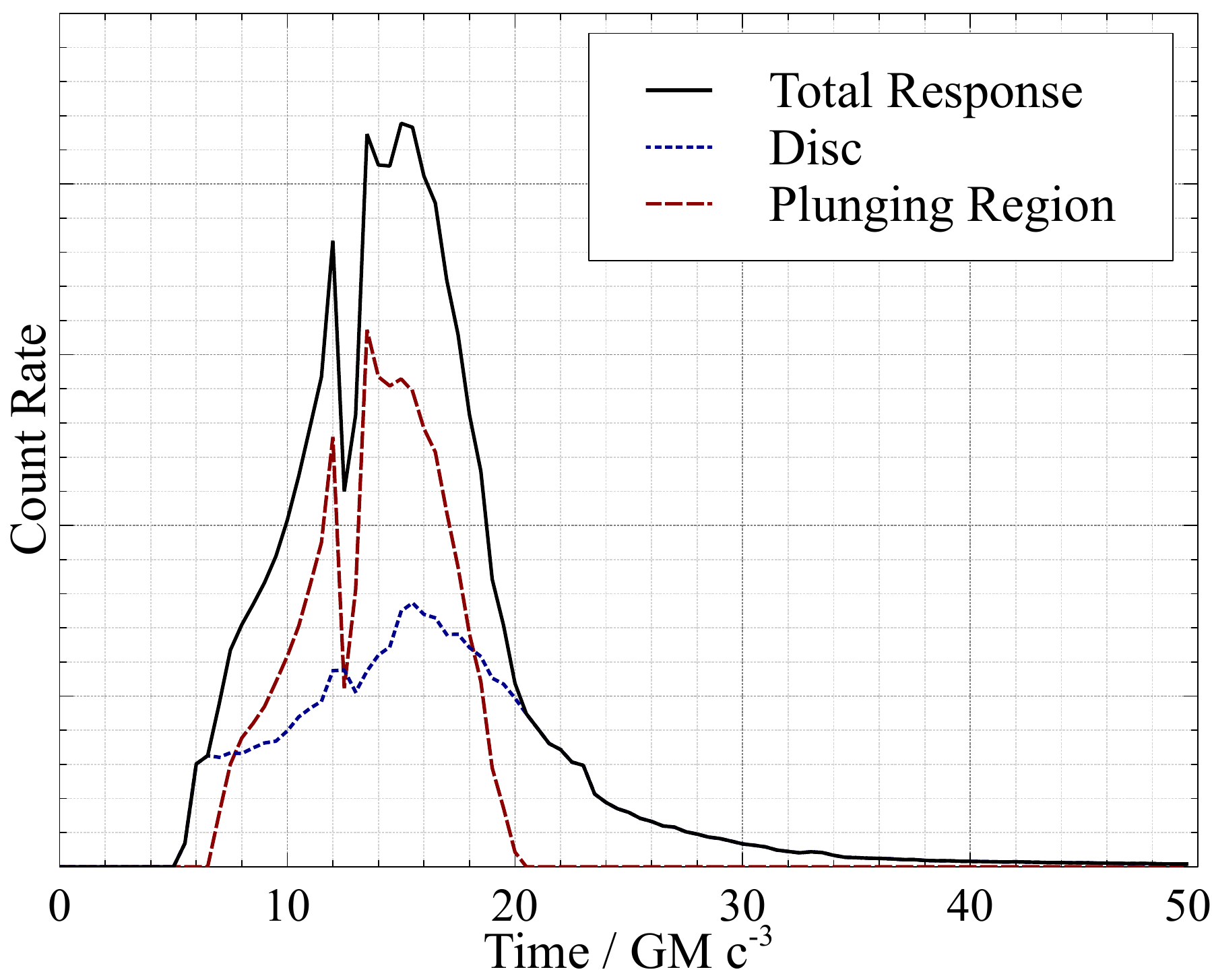}
\label{avg_arrival_propin.fig:c}
}
\subfigure[2-4\keV] {
\includegraphics[width=55mm]{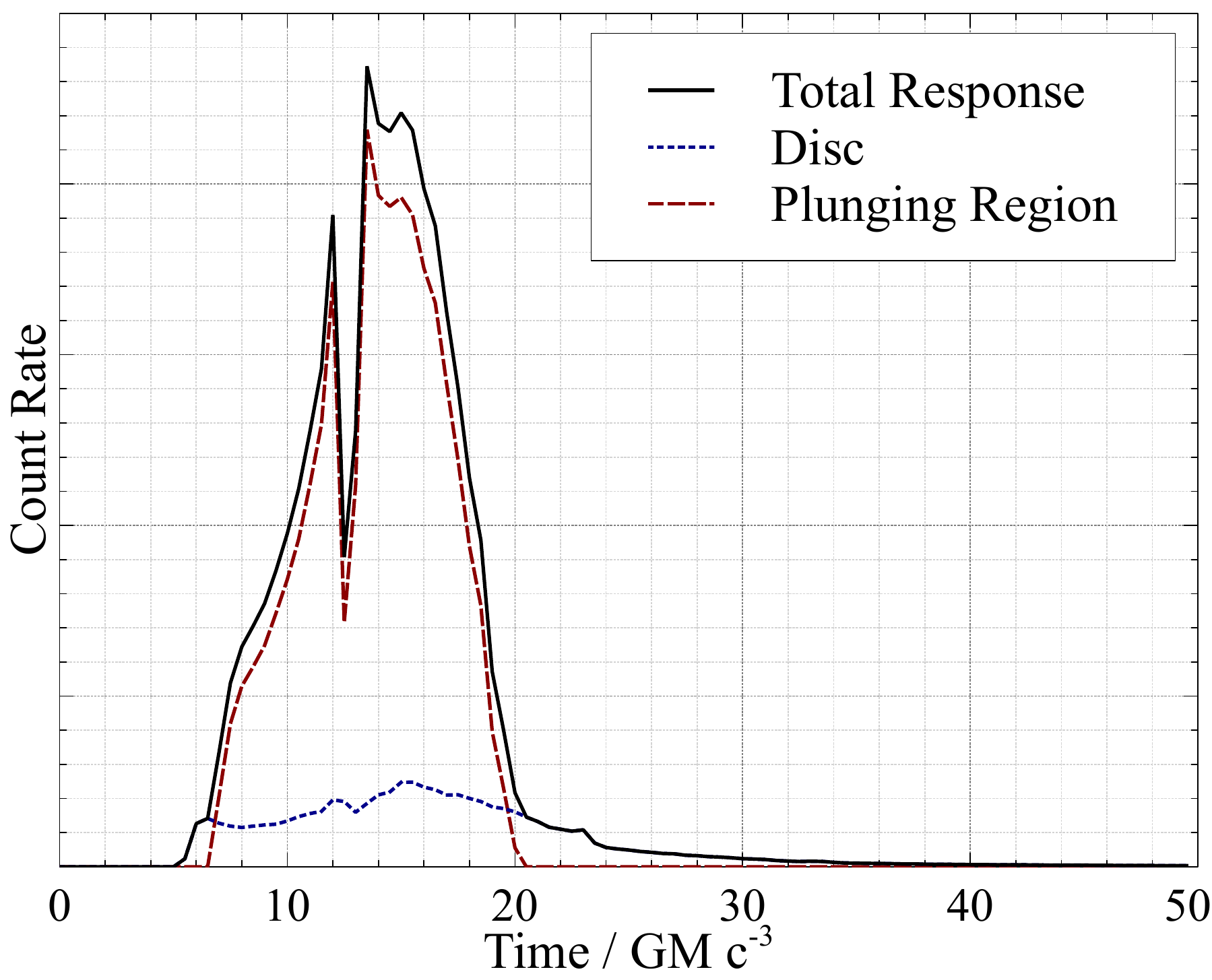}
\label{avg_arrival_propin.fig:0.01c}
}
\subfigure[4-7\keV] {
\includegraphics[width=55mm]{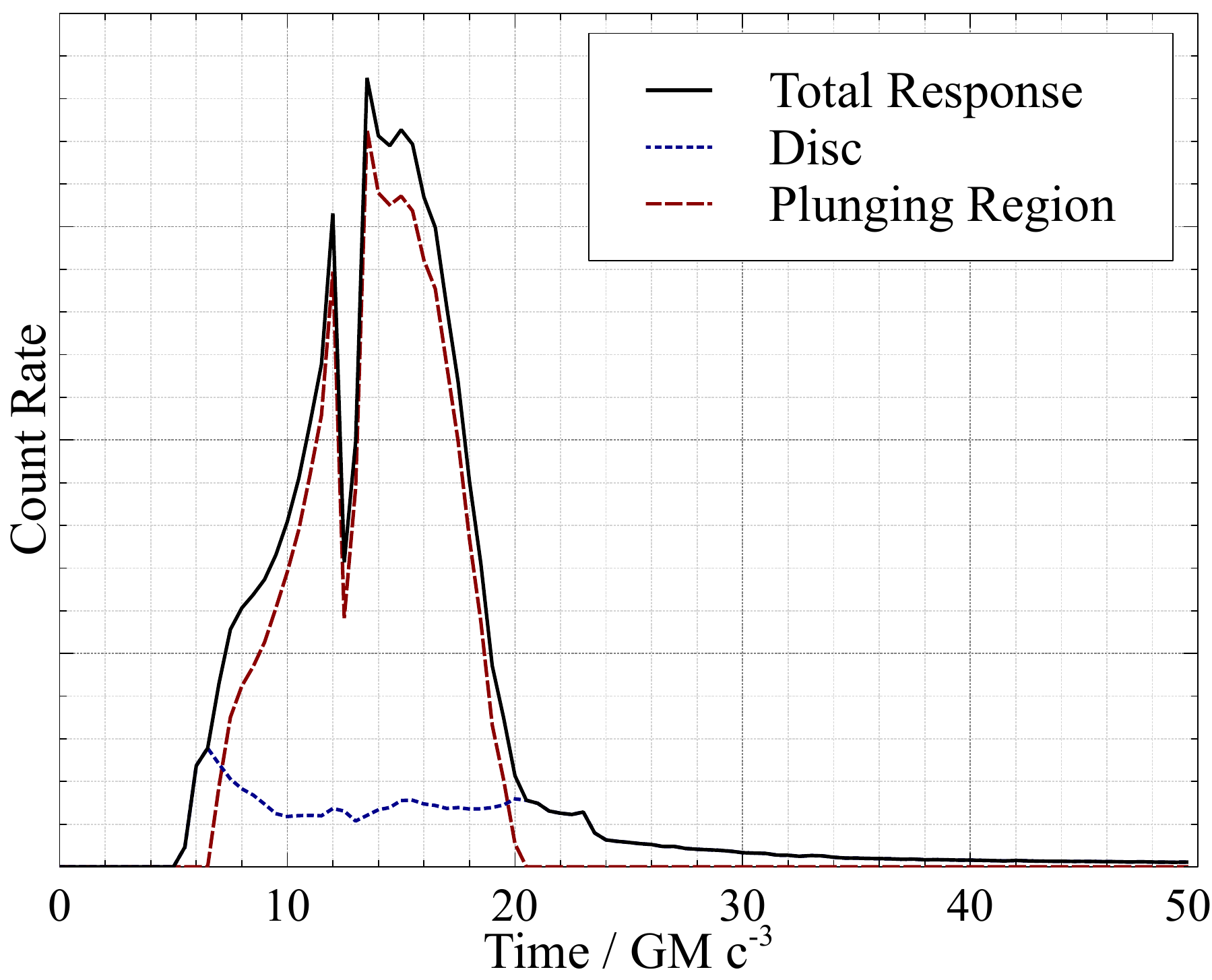}
}
\subfigure {
\includegraphics[width=55mm]{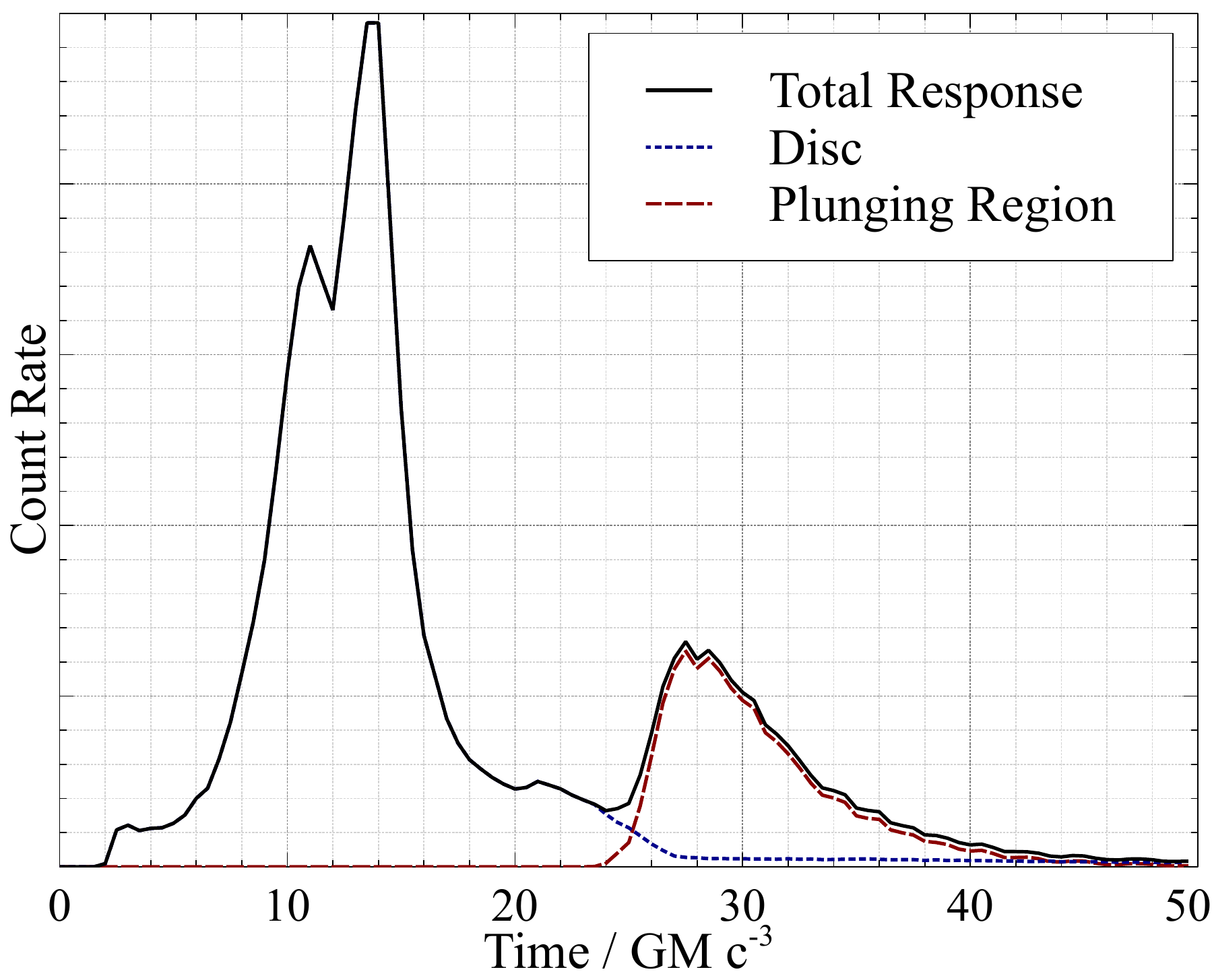}
\label{avg_arrival_propin.fig:0.1c}
}
\subfigure {
\includegraphics[width=55mm]{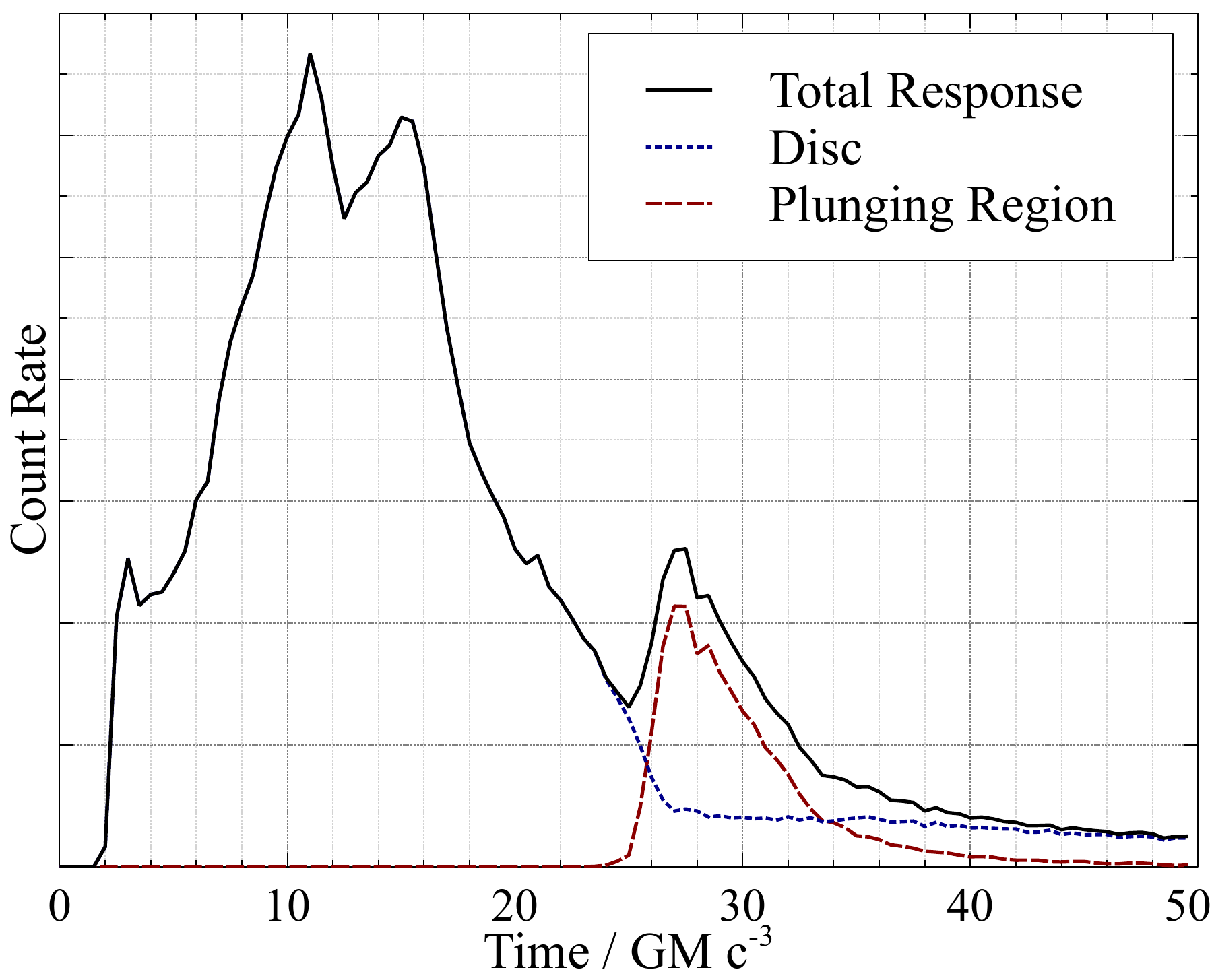}
}
\subfigure {
\includegraphics[width=55mm]{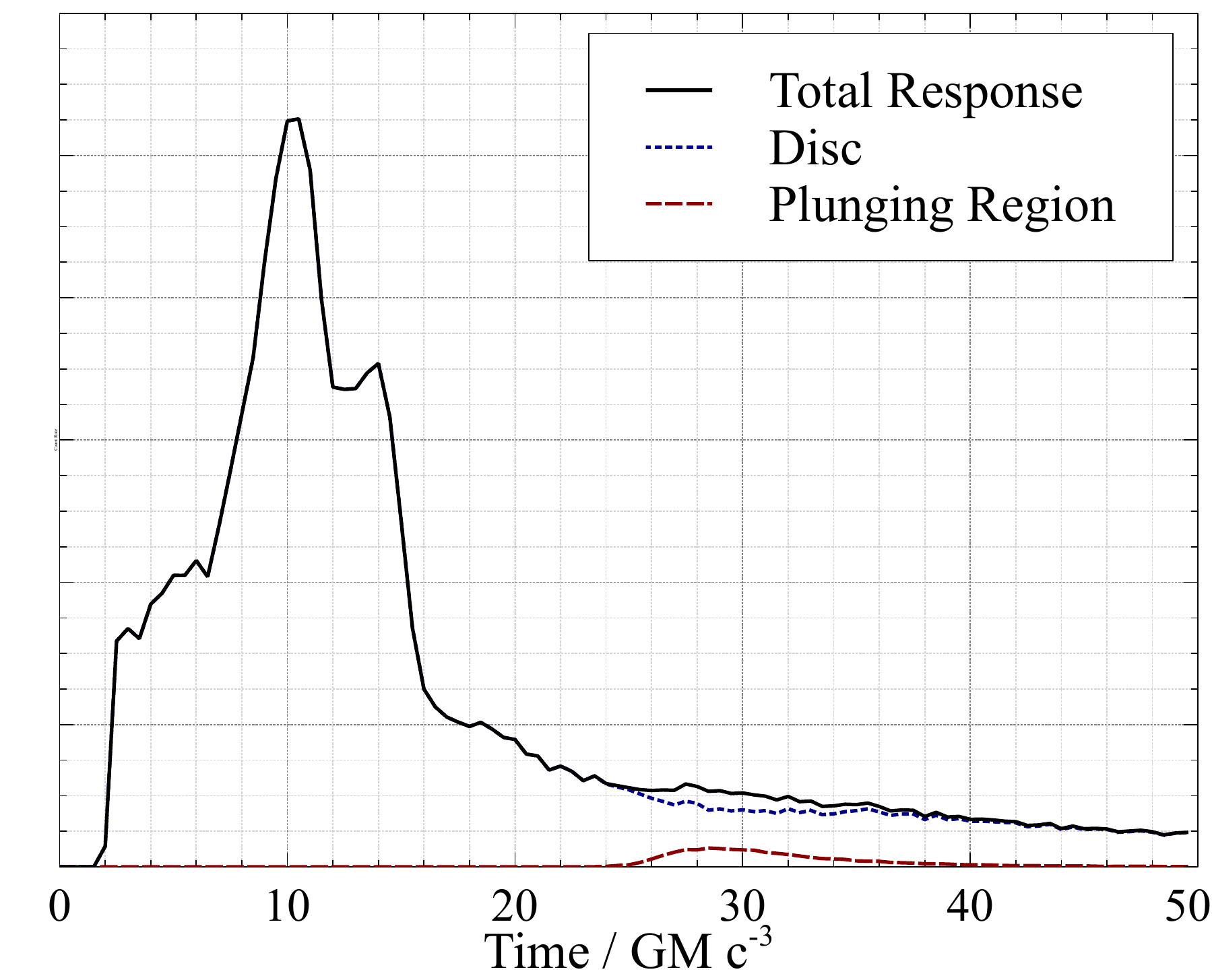}
}
\caption[]{The response functions showing the received photon count rate as a function of time since a flash of continuum emission from the point source in the 1-2\keV\ (\textit{left}), 2-4\keV\ (\textit{middle}) and 4-7\keV\ (\textit{right}) energy bands. The top panels are for a non-spinning black hole and the bottom panels are for a maximally spinning black hole with $a=0.998$. The total response is shown alongside the contributions from the stable-orbiting disc and the plunging region.}
\label{energy_response.fig}
\end{figure*}

In order to understand how the effect on the lag-energy spectra is so minimal despite the addition of delayed emission from the plunging region that is seen in the two-dimensional response functions, we examine the response function as a time series in select energy bands (Fig.~\ref{energy_response.fig}). The response is shown as a function of time for the 1-2\keV\ band (the most redshifted part of the iron K$\alpha$ emission line seen from the plunging region around a maximally spinning black hole), the 2-4\keV\ band (the redshifted wing of the iron line from the disc around a maximally spinning black hole or the plunging region around a non-spinning black hole) and the 4-7\keV\ band (dominated by the core of the line).

In the case of a non-spinning black hole, reverberation from the plunging region peaks more sharply than that from the stably orbiting portion of the accretion disc. The plunging region response is bright, since it is closer to the primary X-ray source than the disc, and peaks sharply because it occupies a relatively small range in radii. The peak of the plunging region emission is centred at roughly the same time as the disc emission, hence there is little change to the average arrival time in each energy band (the lag-energy spectrum).

These energy-resolved response functions show how the reverberation from the plunging region can be distinguished from the primary continuum. In the case of a point source, the response for a single variation is a delta function at $t=0$. The plunging region response, however, is delayed with respect to the continuum by $14\,GMc^{-3}$ (the average arrival time of the response function) for a source at height $h=5$\rg\ and the response function has a full width of $14\,GMc^{-3}$. Due to the smooth, power-law-like shape of the spectral component arising from the plunging region, the shape of the response remains approximately constant across all energy bands (indeed the average arrival time varies only by $0.25\,GMc^{-3}$ between 0.1 and 10\keV). The constant shape of the response means that the plunging region response cannot only be distinguished from the primary continuum by its delayed arrival, but can be distinguished from the reverberation response from the disc component of the accretion flow which varies across different energy bands as narrow emission lines are subject to a range of red- and blueshifts across different parts of the disc. The plunging region component will show strongly correlated variability across all energies.

Turning to the case of the maximally spinning black hole, the response from the plunging region is now seen to be delayed by, on average, $15\,GMc^{-3}$ with respect to the disc response. The plunging region response is most prominent in the 1-2\keV\ energy band in which predominantly extremely redshifted iron K$\alpha$ photons represent 20 per cent of the total reprocessed photon count. Some of the delayed photons from the plunging region are seen in the 2-4\keV\ band while the 4-7\keV\ band is almost entirely composed of line photons from the disc.

When the black hole is maximally spinning, the plunging region produces a second peak in the response function rather than smoothly blending into the tail of the disc response. The rapid rise in ionisation means that the K$\alpha$ lines from neutral and highly ionised iron are emitted from nearby in the disc. Since the highly ionised material is on plunging orbits closer to the black hole, the 6.97\keV\ photons are subject to more extreme redshifts than the 6.4\keV\ photons arising from circular orbits further out. The responses of these two lines therefore lie close together in energy and hence both contribute to the 1-2\keV\ band.

\begin{figure}
\centering
\includegraphics[width=85mm,trim=0 15mm 0 0]{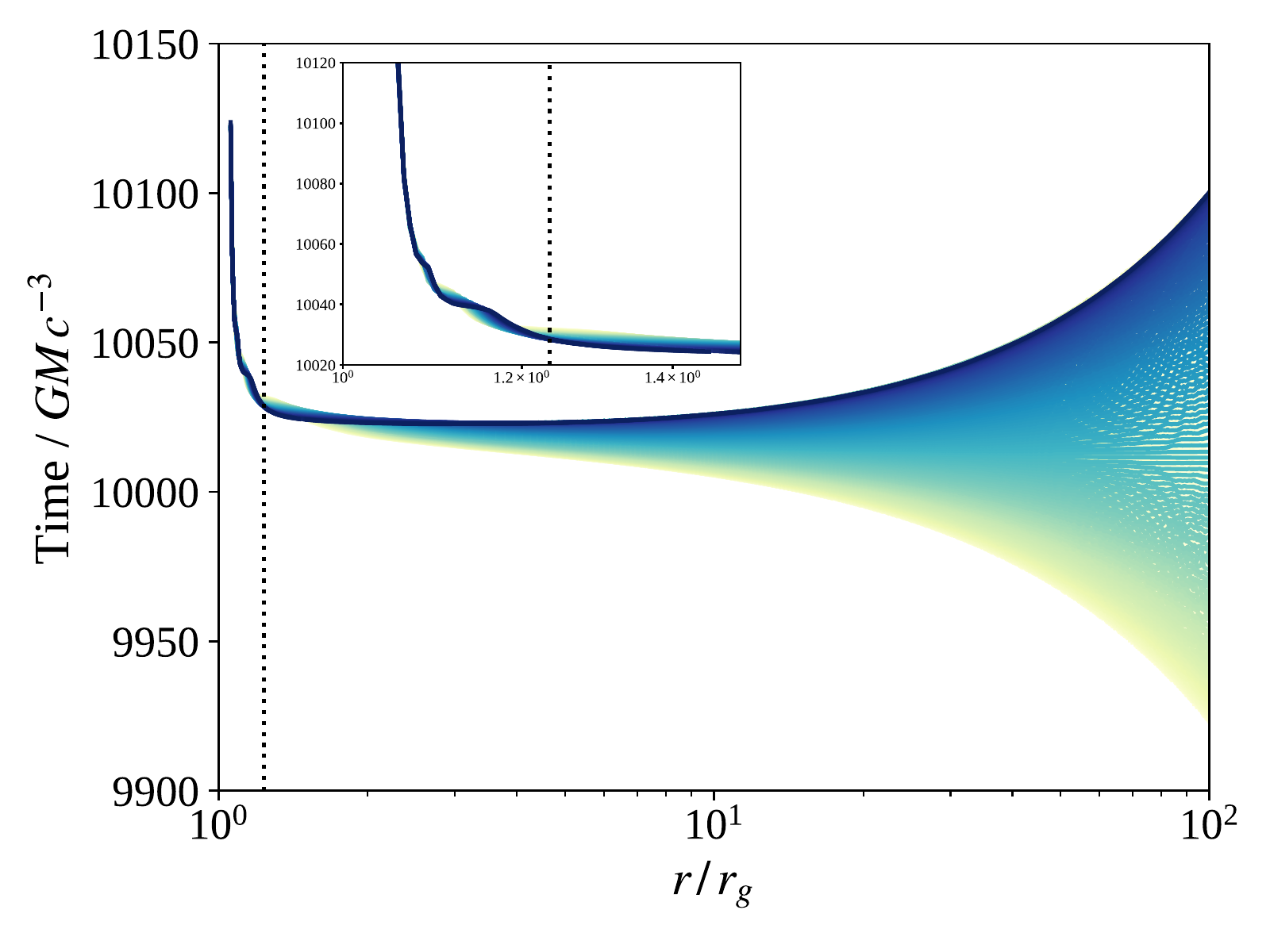}
\caption[]{The light travel time as a function of radius from the accretion flow around a maximally spinning black hole to an observer at 10,000\rg\ at an inclination of 60\,deg. Colours correspond to the light travel times from different azimuths, with lighter shades showing the front side of the disc at an angle $\varphi=0$ with respect to the line of sight, and darker shades showing the back side of the disc. The dotted line shows the location of the ISCO.}
\label{travel_time.fig}
\end{figure}

Moreover, the plunging region around a spinning black hole lies sufficiently deep within the potential well that the photons emitted from this region experience severe travel time delays as they pass close to the event horizon (and indeed inside the photon orbit). The light travel time to the observer as a function of radius on the disc is shown in Fig.~\ref{travel_time.fig}. It can be seen that within the ISCO, the travel time to the observer increases steeply. In these innermost regions, the variation in light travel time is dominated by the Shapiro delay through the curved spacetime close to the black hole. This causes the apparent bunching together of photons not only in energy, but in time, as the more highly redshifted photons appear later than those less highly redshifted, leading to the second peak in the response from the plunging region. 

The secondary peak is dominated by photons emitted from the approaching side of the plunging region such that the emission is beamed in the direction of the observer (though note that it is still redshifted, lying so deep in the potential well). The early part of this peak is emitted from the between 0 and 90\,deg to the line of sight on the approaching side, while the later part comes from between 90 and 180\,deg to the line of sight.

It is the sharp rise in ionisation, coupled with the variation in Doppler shifts between circular and plunging orbits that produces a detectable signature of the plunging region.

\subsubsection{Detecting Variations in the Response Function}
The response function is not typically measured in the time domain; this would be the response of each energy band as a function of time after a single delta-function flash of emission from the primary X-ray source. The observed light curves in different energy bands, $L_E(t)$, are the convolution of the response functions, $T_E(t)$, with the underlying driving light curve describing the stochastic variability of the corona, $L(t)$.
\begin{equation}
L_E(t) = L(t) \otimes T(E,t)
\end{equation}

The average response function over all variations is analysed by applying timing analysis techniques to the light curves in the Fourier domain. To measure the time lag between correlated variability in two light curves, the cross spectrum is computed from the Fourier transforms of those two light curves, noting that the Fourier transform of a light curve can be written as the product of the amplitude and phase of different frequency Fourier components, $\tilde{L}(\omega) = |\tilde{L}(\omega)|e^{-i\varphi}$.
\begin{equation}
\tilde{C}(\omega) = \tilde{L}_1^* \tilde{L}_2 = |\tilde{L}_1(\omega)||\tilde{L}_2(\omega)|e^{i(\varphi_1 - \varphi_2)}
\end{equation}
The argument of the cross spectrum corresponds to the phase lag (and hence the time lag, $\varphi = \omega\tau$) at a given Fourier frequency between the two light curves.

Using the convolution theorem, this can be written in terms of the transfer functions (the Fourier transforms of the impulse response functions) to compute the cross spectrum between two energy bands:
\begin{equation}
\label{crossspec.equ}
\tilde{C} = \tilde{L}^*_E \tilde{L}_\mathrm{ref} = \left|\tilde{L}\right|^2 \tilde{T}^*_E \tilde{T}_\mathrm{ref}
\end{equation}
$|\tilde{L}|^2$ is the power spectrum of the driving light curve, typically seen to follow $f^{-2}$ in frequency, while the term $\tilde{T}^*_E \tilde{T}_\mathrm{ref}$ is the `effective transfer function' that encodes the reverberation response of the accretion flow (the \textit{transfer function} is the Fourier transform of the time domain response function). In general, it is not possible to select from the observation a reference band that contains only continuum emission, hence the magnitude of effective transfer function will be the product of the power spectra of the response functions of the two bands and its phase will be the phase difference between the response of the two bands. In order to explore the structure of a single band's transfer function, we begin by taking the reference band response to be a delta function at $t=0$ (\textit{i.e.} just the continuum response) and we need only consider the Fourier transform of the response function of a single band.

\begin{figure*}
\centering
\subfigure {
\includegraphics[width=85mm]{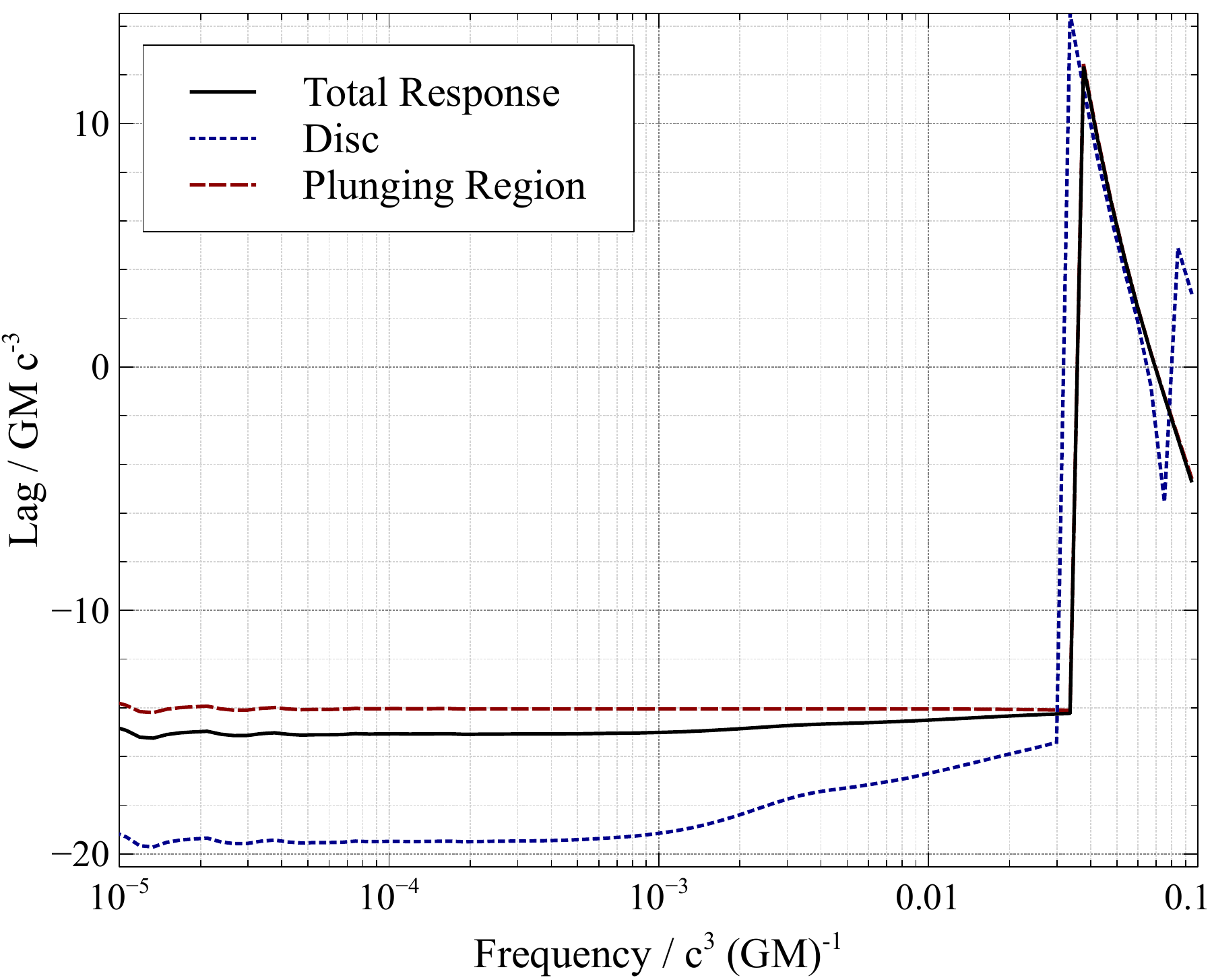}
\label{avg_arrival_propin.fig:c}
}
\subfigure {
\includegraphics[width=85mm]{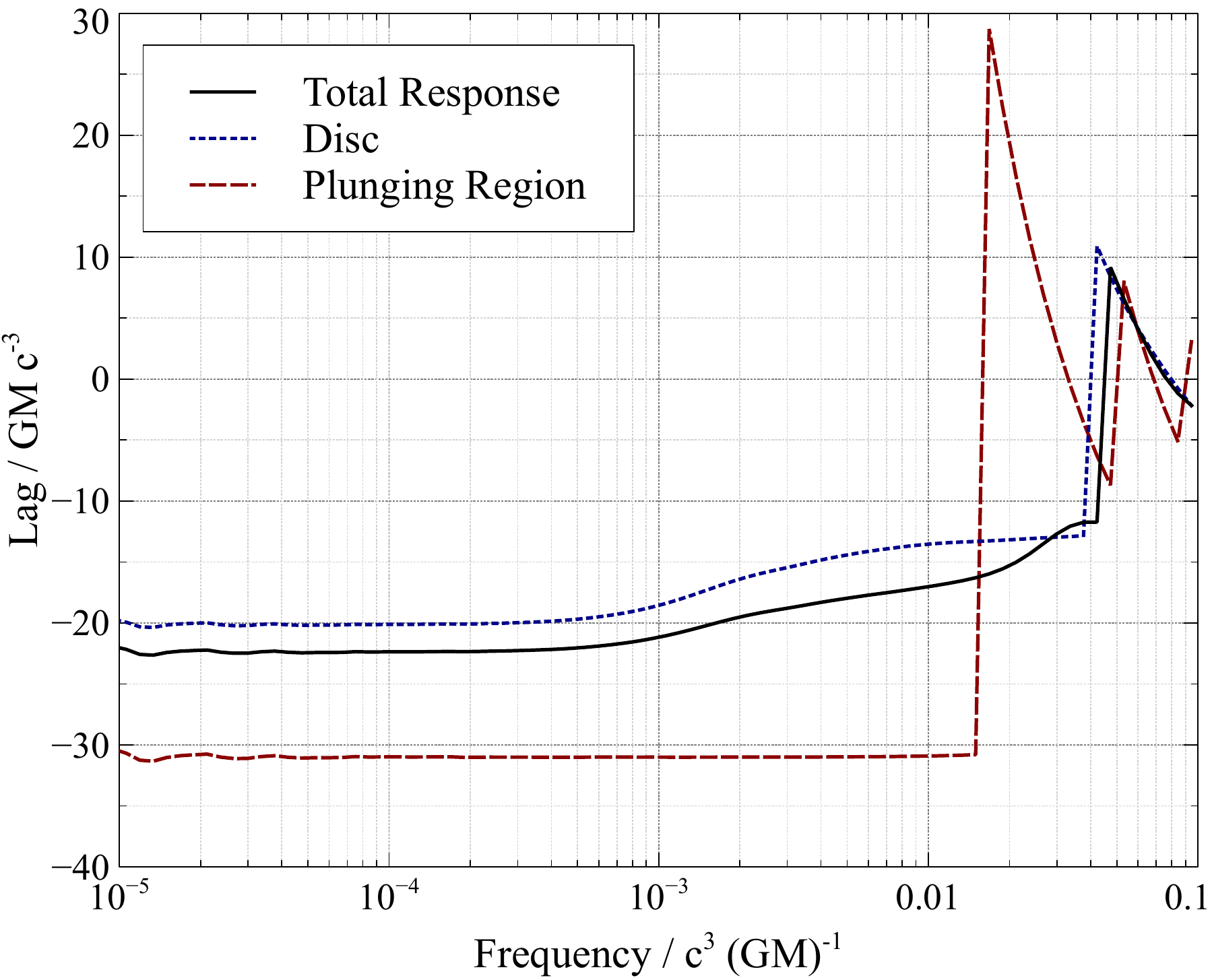}
\label{avg_arrival_propin.fig:0.1c}
}
\caption[]{The single band lag-frequency spectrum of the 2-4\keV\ response for a non-spinning black hole (\textit{left}) and the 1-2\keV\ energy band for a maximally spinning black hole (\textit{right}).}
\label{lagfreq.fig}
\end{figure*}

The time lag of the reverberation response as a function of the different frequency Fourier components making up the stochastic variability of the light curve (the \textit{lag-frequency} spectrum) for the energy bands exhibiting the strongest contribution from the plunging region (2-4\keV\ for spin $a=0$ and 1-2\keV\ for $a=0.998$) is shown Fig.~\ref{lagfreq.fig}. The total response function is compared to the response functions of both the disc and plunging regions of the accretion flow. In particular, comparing the total response to the disc response shows how the presence of the plunging region can be inferred from X-rays that are reprocessed from it.

For the accretion flow around the non-spinning black hole, the addition of the plunging region leads to a wholesale decrease in reverberation timescales across all frequencies compared to the response from the just the accretion disc. Since the primary X-ray source is taken to be an infinitesimal point source, the response function has relatively sharp features, which causes the lag time to be `phase-wrapped' at high frequency as discussed by \citet{lag_spectra_paper,cackett_ngc4151}. For a given time lag $\tau$, at frequencies greater than $1/2\tau$, the waveform is shifted more than half a cycle, hence it is not possible to determine whether it is shifted forwards or backwards in time causing the wrap as the phase shifts from $-\pi$ to $\pi$.

An increase in time lag that is due to either the X-ray source being placed at a greater height or due to an increase in black hole mass would cause the phase wrap to shift to a lower frequency \citep{cackett_ngc4151}. The decrease in lag due to the emission from the plunging region, however, does not lead to such a shift in the phase wrap, rather changing the gradient of the lag-frequency spectrum between frequencies of $10^{-3}\,c^3 (GM)^{-1}$ and the phase-wrap point at $3\times 10^{-2}\,c^3 (GM)^{-1}$ for a point source at $h=5$\rg. $c^3 (GM)^{-1}$ is the frequency unit corresponding to the light crossing time over 1\rg\ and can be scaled to the black hole mass of interest).

The sharp peak in the response of the plunging region increases the cross power relative to the case of reverberation from only the disc, though a change in normalisation of the cross spectrum alone would be insufficient to infer the presence of the plunging region (the normalisation of the underlying power spectrum is not known \textit{a priori}). The long light travel times from a centrally located point source to the ISCO act as a low pass filter in the reverberation response, leading to a cut-off in cross power at $0.05\,c^3 GM^{-1}$ for the disc response causing a drop in the power spectrum by a factor of 5 above this frequency when no emission from the plunging region is seen. No such cut-off in the cross power is seen when the plunging region is present in the reverberation response function.

The effect on the lag-frequency spectrum is more subtle for a maximally spinning black hole. In this case, there is little change in the lag time at low frequencies, however, a change is seen in the lag-frequency profile at frequencies between $10^{-3}\,c^3 GM^{-1}$ at the point at which the phase begins to wrap. The additional late-time response from redshifted iron K$\alpha$ photons in the 1-2\keV\ energy band leads to an increase in the time lag of 25 per cent at $10^{-2}\,c^3 GM^{-1}$.

\subsubsection{Observable lag-frequency spectra}

These single-band lag-frequency spectrum can only be measured directly if a reference band containing only continuum emission is available. In real accreting black hole systems, each energy band will have some contribution from both continuum and reprocessed emission and it is necessary to explore the structure of lags within the cross-spectrum between two observable energy bands.

Full spectral response functions were computed by adding the continuum emission to the reverberation response functions calculated above. The continuum emission from a point source arrives instantaneously since there is only a single light path from the source to the observer \citep{lag_spectra_paper}.

In order to maximise the X-ray count rate (and hence signal-to-noise), X-ray reverberation is typically observed between the reflection-dominated 0.3-1\keV\ and continuum-dominated 1-4\keV\ band, or between the 1-4\keV\ and the 4-7\keV\ iron K band. The signal from within the plunging region is manifested in either the 2-4\keV\ or 1-2\keV\ energy band, so to maximise signal-to-noise we compare this to the 0.3-1.0\keV\ band.

\begin{figure}
\centering
\includegraphics[width=85mm]{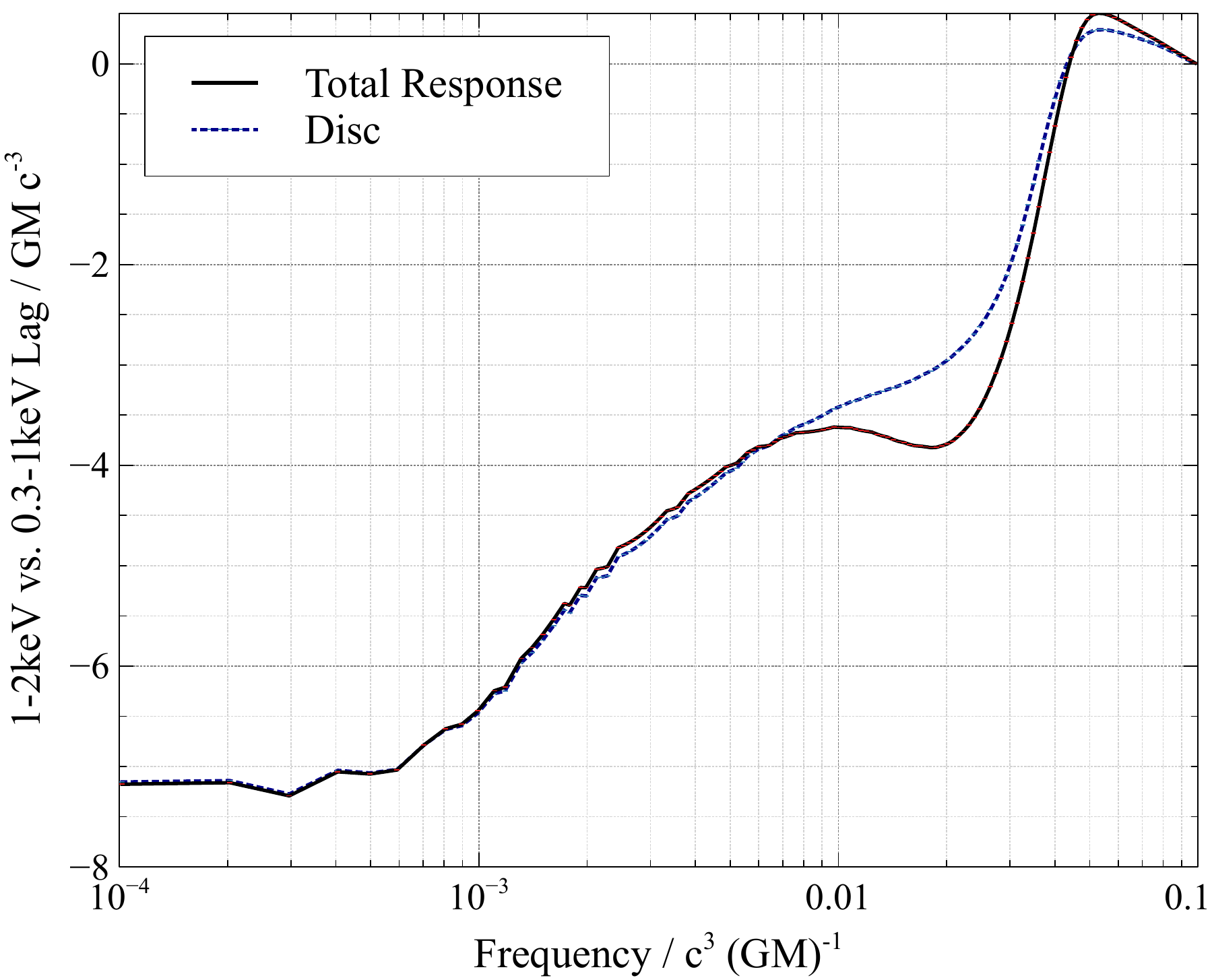}
\caption[]{The lag-frequency spectrum (including both the continuum and reverberation responses) between the the 0.3-1\keV\ and 1-2\keV\ energy band for a maximally spinning black hole comparing the total response (including emission reverberating within the plunging region) to the response containing reprocessed emission just from the stably orbiting disc.}
\label{lagfreq_2band.fig}
\end{figure}

Fig.~\ref{lagfreq_2band.fig} shows the lag-frequency spectrum between the 0.3-1\keV\ and 1-2\keV\ energy bands for an $a=0.998$ black hole, including the response from the plunging region. While the 0.3-1\keV\ band is dominated by reprocessed emission, the secondary peak from the plunging region is not seen in this band's response function. The highest energy line in this band is the iron L line at 0.7\keV. The equivalent of the iron~K 1-2\keV\ response from the plunging region for the iron L line lies below 0.2\keV, outside this band pass. Computing the cross spectrum with the 0.3-1\keV\ band acts like a matched filter, picking out the disc response which is approximately consistent between the 1-2\keV\ and 0.3-1\keV\ bands, leaving behind the signal due to the extra response, the delayed secondary peak from the plunging region, in the 1-2\keV\ band.

In this two-band lag-frequency spectrum, the manifestation of the plunging region in the high frequency tail is enhanced as a peak-like feature altering the lag in Fourier frequency components between 0.01 and $0.03\,c^{3}(GM)^{-1}$ by 20 per cent. The presence of the plunging region can be detected if the lag at $0.02\,c^{3}(GM)^{-1}$ can be measured to an accuracy better than 20 per cent in frequency bins sufficiently narrow to describe the peak feature.

\section{Dynamics of the Plunging Region}
Having discussed how X-ray emission from the plunging region may be detected through X-ray reverberation, we now explore what may be inferred about the dynamics of the plunging region. In particular, if it can be determined that material in this region is indeed on plunging orbits, rather than maintaining circular orbits, it would verify a prediction of General Relativity that there exists a last stable orbit some finite distance outside the event horizon of a black hole.

We seek to compare X-ray reverberation from material on plunging orbits between the ISCO and the event horizon with reverberation from material that remains on circular orbits through the ISCO. In order to do this, material within the ISCO was artificially assigned a velocity vector with components $[u^\mu] = \dot{t}(1, 0, 0, \Omega)$ with $\Omega$ given by Equation~\ref{angvel.equ}. Even though these orbits are not stable, this function remains continuous up to the event horizon and enables the redshift of rays either received from the source or emitted towards the observer to be calculated as if material had remained on circular orbits. Care must be taken with this prescription, however, since within the radius of the photon circular orbit, this orbit will become hyperluminal, hence the material must still depart the circular orbit at or before this radius. We therefore consider two cases; one where the disc transitions to a plunging orbit at the marginally bound orbit, $r_\mathrm{MB}$ (the innermost orbit at which \textit{bound} orbits, not necessarily stable, circular orbits, can exist) and one where the disc transitions to a plunging orbit at the photon orbit.

The propagation of rays was taken to follow the standard prescription in General Relativity and the Kerr metric is used to describe the geometry of the spacetime. While this does not represent a self-consistent modified gravity picture of the plunging region, it enables the sensitivity of X-ray observations to the orbital dynamics to be assessed.

\begin{figure*}
\centering
\subfigure[1-2\keV\ Impulse response function] {
\includegraphics[width=85mm]{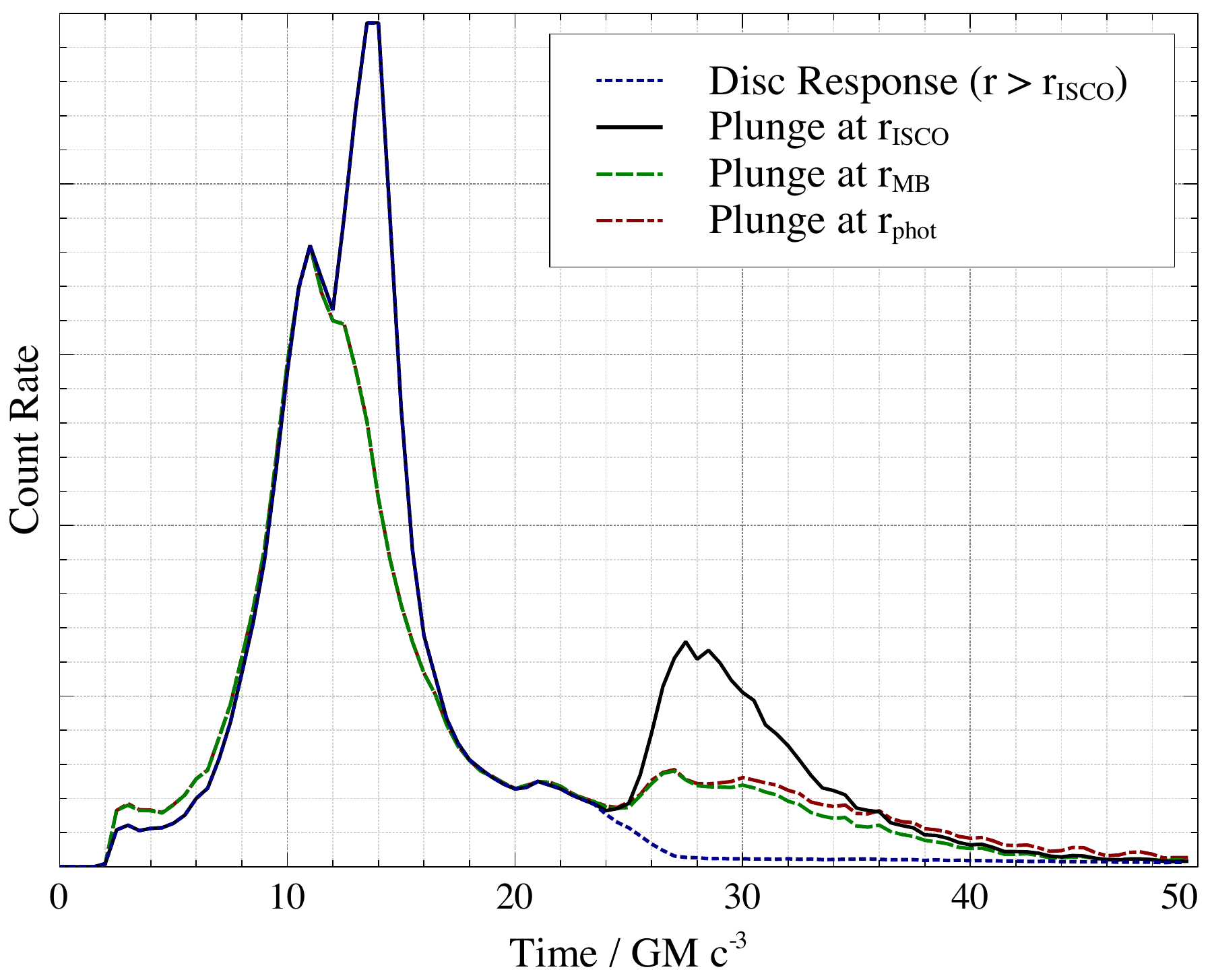}
\label{rplunge.fig:resp}
}
\subfigure[1-2\keV \textit{vs.} 0.3-1\keV\ lag-frequency spectrum] {
\includegraphics[width=85mm]{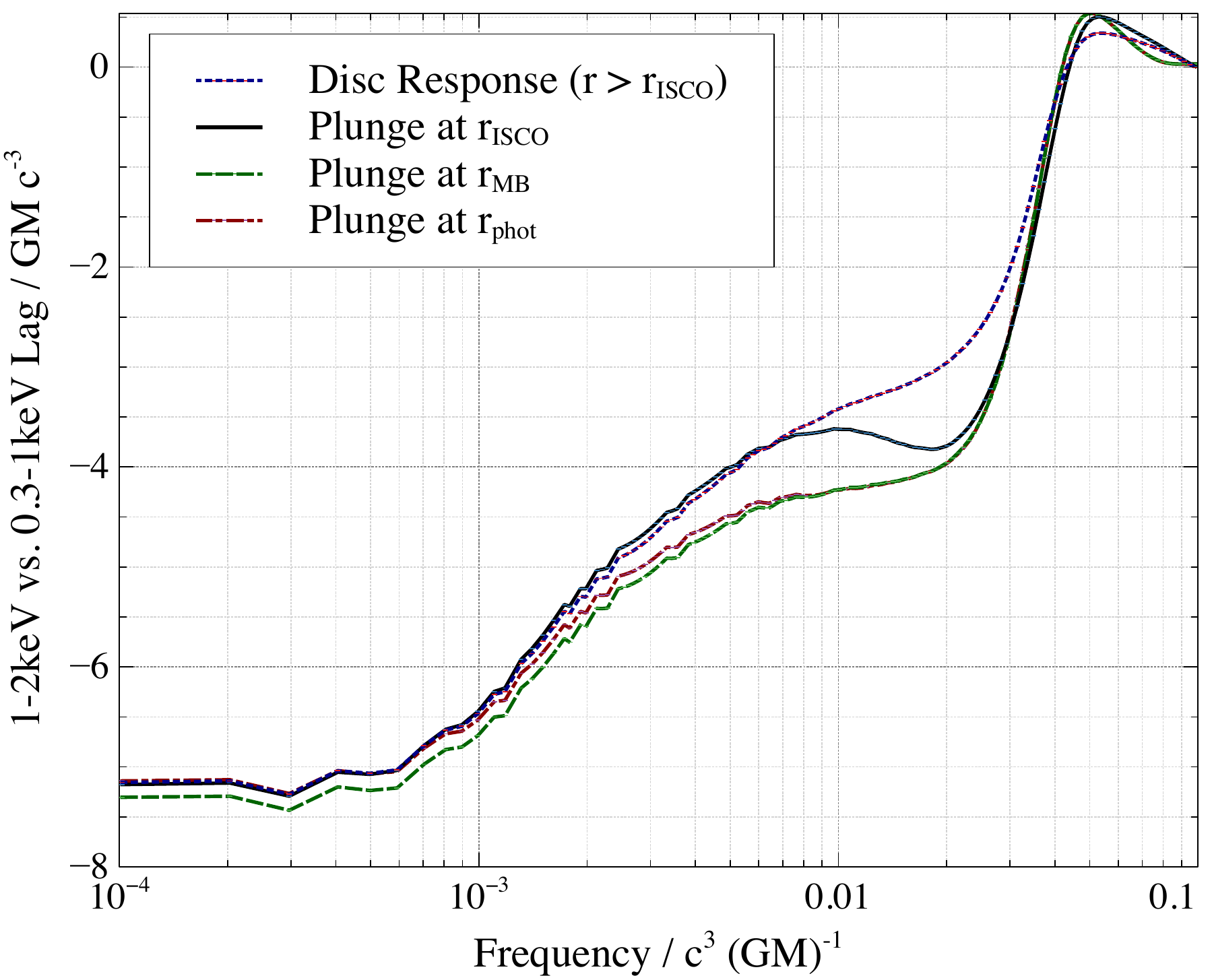}
\label{rplunge.fig:lagfreq}
}
\caption[]{\subref{rplunge.fig:resp} Impulse response functions and \subref{rplunge.fig:lagfreq} 2-band lag-frequency spectra for accretion flows in which circular orbits are maintained beyond the ISCO to test the precision with which the existence of the ISCO dynamics of material in the plunging can be probed by X-ray reverberation. Cases where material maintains circular orbits beyond the ISCO as far as either the marginally bound orbit (1.09\rg) or photon orbit (1.07\rg) are compared to the case where the plunging region begins at the ISCO (1.235\rg) and also the case where reverberation is only seen from the stably orbiting disc, not the plunging region.}
\label{rplunge.fig}
\end{figure*}

Fig.~\ref{rplunge.fig} shows the response functions in addition to the two-band lag-frequency spectra for accretion discs that maintain circular orbits beyond the ISCO around a maximally rotating Kerr black hole ($a=0.998$). In this case, the marginally bound orbit is at 1.09\rg\ while the (prograde) photon orbit is at 1.07\rg.

When circular orbits are maintained beyond the ISCO, the secondary peak in the 1-2\keV\ response function is diminished, though late-time response is still seen above that expected from a disc that truncates at the ISCO. The secondary peak resulted from high-ionisation iron K$\alpha$ line photons that are strongly redshifted into the 1-2\keV\ band and are delayed in their propagation as they travel to the observer from locations deep in the gravitational potential.

When material remains on a circular orbit to the smaller radius of either the marginally bound orbit or the photon orbit, the density drop occurs at smaller radius. A smaller fraction of the disc reaches such high ionisations and is on the plunging orbits that produce the greatest redshifts. This reduces the number of line photons that are shifted down to the 1-2\keV\ band in the early part of the secondary peak (due to the light travel time delays, the early parts of this peak correspond to the outer parts of the plunging region and the innermost parts are seen later). A significant part of the response inside the ISCO now extends above 2\keV\ and even up to 6\keV\ at early times, reducing the jump in count rate seen in the 1-2\keV\ band by a factor of 0.5. The innermost parts of the accretion flow (from which the time delay is the greatest) still follows plunging orbits and produces highly redshifted line photons and the late-time response tracks that seen when the plunging orbits begin at the ISCO.

As before, the delayed response from inside the ISCO is manifested in the high frequency tail of the lag-frequency spectrum; specifically the shape of the spectrum as the lags decay to zero. The location of the plunging radius and the dynamics of the material within the ISCO affects the shape of the lag-frequency spectrum between the 0.3-1\keV\ and 1-2\keV\ bands between variability frequencies of $2\times 10^{-3}$ and $3\times10^{-2}\,c^3 (GM)^{-1}$. While the lag at these end frequencies is not altered, being able to distinguish the lag between 0.01 and $0.02\,c^3 (GM)^{-1}$ to an accuracy of $0.5\,GM c^{-3}$ in the 2-band lag-frequency spectrum will allow for the presence of the plunging region to be distinguished from circular orbits within the ISCO.

\section{Discussion}
Through general relativistic ray tracing simulations of X-ray reverberation from the accretion flow, we are able to find the signatures of X-rays that are reprocessed by material inside in innermost stable circular orbit, offering a probe of material in its final moments as it plunges into a black hole.

The ability to probe the dynamics of the innermost regions of the accretion flow tests a key prediction of General Relativity in the strong-field regime, namely that there exists a radius, outside the event horizon, within which it is not possible to maintain a stable circular orbit. Thus, an accretion disc around a black hole will be truncated at this radius. The innermost extent of an accretion disc around a black hole is typically measured from the extremal redshift detected in the iron K$\alpha$ emission line, with more strongly redshifted line emission being produced the deeper the accretion disc extends into the gravitational potential. The line flux, however, decreases at lower energies producing a `wing' on the emission line profile that blends into the underlying X-ray continuum. This means that spectroscopy does not provide a firm detection of the innermost stable orbit and the transition to plunging orbits, but rather measures the extent of detectable emission from the accretion disc.

The accretion disc is typically assumed to extend down to the innermost stable orbit. Once the innermost extent of the disc has been measured, it is used to infer the spin of the black hole. If, however, the dynamics of the plunging region can be probed directly and the innermost stable circular orbit measured for known black hole spin, it will be possible to test the prediction of the ISCO radius made by General Relativity.

In the simulations presented here, we search for signatures of the most abrupt transition at the ISCO; that is from circular orbits to a plunging orbit assuming that the accreting material experiences zero torque once it crosses the ISCO and retains any angular momentum it possesses at the ISCO. This condition is justified \textit{a posteriori}, noting that the increase in radial velocity that follows the transition to the plunging orbit results in a drop in density. In the standard Shakura-Sunyaev picture of the accretion disc, angular momentum is transferred outwards by viscous stresses within the disc that will be greatly reduced once the density drops.

If the dynamics of material within the plunging region can be measured, this so-called \textit{zero-torque boundary condition} can be tested. It is likely that the highly ionised accretion disc is threaded by magnetic fields that in turn energise a corona that produces X-ray continuum emission, or, in the case of radio-loud AGN and micro-quasars, are responsible for the launching the jet. If magnetic field lines lie between regions of the stably-orbiting disc and the plunging region, torques will be applied, transferring angular momentum between the plunging material and more slowly orbiting material further out, altering the dynamics of both the plunging region and disc \citep{agol_krolik}. Magnetohydrodynamic (MHD) simulations of the inner regions of accretion flows around black holes suggest that magnetic torques lead to a drop in density of the accretion flow just outside the ISCO and a more gradual drop in density through the plunging region than is expected in the zero-torque case \citep{zhu+2012}. Probes of the dynamics of the plunging region and inner parts of the disc therefore also present tests of the physics of the accretion flow.

\subsection{Detectability of the plunging region}

The appearance of the plunging region in X-ray observations depends strongly upon the spin of the black hole. If the black hole is non-rotating, the innermost stable orbit is a significant distance outside the event horizon and there exists a substantial portion of the accretion flow on plunging orbits that will be strongly irradiated by a centrally located X-ray source. From mass conservation through the plunging region, it follows that this portion of the flow remains optically thick down to the event horizon and, as such, as much as 30 per cent of the observed reprocessed X-ray emission comes from within this region. X-ray emission from the plunging region is most easily detected in the redshifted wing of the iron K$\alpha$ fluorescence line. The most extremely redshifted photons from the receding side of the accretion disc are seen around 3.5\keV\ whereas line photons from the plunging region are observed as low as 1\keV\ with a significant contribution in the 2-4\keV\ band when $a=0$.

The drop in density as material crosses the ISCO, combined with strong illumination by the primary X-ray source in the inner regions, results in a rapid rise in ionisation in the plunging region. The high ionisation state of the plasma along with the extreme redshifts and blueshifts from plunging orbits close to the black hole mean that the time-averaged spectrum of the emission from the plunging region is smoothed out to mimic additional continuum emission. This emission is, however, delayed with respect to the primary continuum and the addition of light paths closer to the black hole means that the reverberation response from the strongly irradiated plunging region is more sharply peaked than that from the disc.

In the case of a maximally-rotating black hole, the ISCO is much closer to the event horizon, thus the plunging region makes up a substantially smaller area of the accretion flow. Only around 2 per cent of the reprocessed emission from the accretion flow around a maximally rotating black hole arises from the plunging region. Since the plunging region sits much deeper in the gravitational potential around a maximally spinning black hole, only redshifted photons are seen. The additional reverberation from the plunging region is seen most strongly as a delayed secondary peak in the response function in the 1-2\keV\ energy band that is visible due to the drop in density and subsequent jump in ionisation of material inside the ISCO.

X-ray reverberation from the plunging region does not significantly alter the average response time of each energy band to a variation in continuum luminosity (\textit{i.e.} the lag-energy spectrum), underscoring the importance of not just measuring the lag-energy spectrum, but the shape of the response function as a function of X-ray energy. Since the X-ray source varies stochastically, it is not possible to directly measure the response function (unless there were a single bright flare, shorter than the reverberation timescale). The response function is convolved with the stochastic variability of the primary X-ray source and is encoded in the time lags as a function of Fourier frequency between successive energy bands. In particular, the emission of the plunging region may be detected from the shape of the time lags as a function of Fourier frequency in the high frequency tail of the lag-frequency spectrum as the lags tend towards zero. 

In order to detect the component of the emission from the plunging region from the shape of the reverberation response function, it is necessary to measure the lag to an accuracy of 20 per cent at frequencies between  $0.01$ and $0.03\,c^3(GM)^{-1}$ in at least 3 frequency bins in a narrow-band light curve spanning 1-2\keV\ (compared to a 0.3-1\keV\ reference band for maximal signal-to-noise). These frequencies are scaled for the mass of the black hole in question and correspond to between 2 and $6\times 10^{-3}$\Hz\ for a $10^6$\Msun\ supermassive black hole or 200-600\Hz\ for a $10$\Msun\ stellar mass black hole in an X-ray binary, when the X-ray source is a few \rg\ from the singularity. In order to distinguish plunging orbits from the accreting material remaining on circular orbits within the ISCO, and hence ascertain the existence of the innermost stable orbit, 12 per cent accuracy at these frequencies is required.

If instead of a sharp transition to plunging orbits at the ISCO, magnetic torques mediate a less abrupt drop in density that begins outside the ISCO, the ionisation structure that leads to the observable signatures of the ISCO will be altered. Approximating the density drop seen across the ISCO in the MHD simulations of \citet{penna+2010,zhu+2012} by a broken power law to describe the altered density structure of the accretion flow (although not modelling the change in the velocities of the plunging orbits), we find that the less abrupt density drop, coupled with the density starting to fall at larger radius, results in the secondary peak in the 1-2\keV\ response function for a maximally spinning black hole increasing in amplitude and beginning earlier in time. The higher ionisation plasma contributing to this emission peak is now just outside the ISCO, where the density begins to drop. The emission is now less strongly redshifted and experiences less time delay close to the black hole. The corresponding peak in the high frequency tail of the lag-frequency spectrum increases by around 25 per cent and decays more slowly on the low frequency side. Measurement of the time lag to 12 per cent accuracy at these frequencies will not only confirm the presence of the ISCO and plunging region, but will probe the density structure and the nature of magnetic torques across the ISCO.

\subsection{Dependence on model parameters}

The detection of material on plunging orbits within the ISCO around rapidly spinning black holes is facilitated by the drop in density across the ISCO. This leads to a steep rise in ionisation in the plunging region, distinguishing the atomic features in the rest frame spectrum that are emitted from the plunging region from those emitted by material on stable circular orbits. This will, of course, be sensitive to the precise level of ionisation in the accretion flow around the black hole in question. In the simulations presented here, the shape of the ionisation profile is calculated self-consistently from the model density profile of the disc and its irradiation by a point source. The normalisation of the ionisation parameter is, however, a free parameter, here set at the ISCO. Shown here are the results for an ionisation parameter $\xi_0 = 1000$\ergcmps\ at the ISCO.

Decreasing the absolute disc ionisation, setting $\xi_0 = 100$\ergcmps, results in stronger fluorescence from neutral iron (a sharp narrow line at 6.4\keV\ in the rest frame) from the disc and stronger neutral and ionised line emission from throughout the plunging region. In the case of a maximally spinning black hole, the secondary peak in the 1-2\keV\ response is increased from 33 to 38 per cent of the flux from the primary peak (the disc response), marginally increasing its detectability (producing a 23 per cent shift in the lag time in the highest frequency Fourier components compared with 20 per cent). On the other hand, increasing the ionisation state of the disc, setting $\xi_0 = 10000$\ergcmps, causes the secondary peak to be diminished to only 8 per cent the magnitude of the primary peak. In this case, only weak, broadened lines from highly ionised ions are produced just outside the ISCO which become weaker as the number of completely ionised ions that do not produce fluorescence line emission increases inside the ISCO. As the secondary peak blends into the tail of the disc response, a shift in the lag time of only 6 per cent in the high-frequency Fourier components would be seen. For highly ionised accretion discs where strong, narrow emission lines are not produced just outside the ISCO, it becomes much more difficult to detect the presence of the plunging region.

Additional dissipation within the plunging region could steepen the ionisation gradient. If the material overflowing the ISCO is clumpy and slightly misaligned with the black hole spin, precessing streams could collide and experience shock heating. Maintaining fairly low ionisation just outside the ISCO and it rising rapidly as it begins to plunge is necessary for the secondary emission peak to be detected, but additional dissipation steepening the ionisation gradient will alter the amplitude and shape of this feature.

Many of the AGN in which strong reverberation signatures from the accretion disc have been observed are best fit by a spectrum with average ionisation parameters less than a few 100\ergcmps \citep{zoghbi+09, gallo+13, mrk335_corona_paper, jiang_iras}, although some exhibit more highly ionised discs \citep[\textit{e.g.}][]{kara_ark564}. For $\xi_0 = 1000$\ergcmps, the mean ionisation parameter across the disc, weighted by the flux observed at infinity as a function of radius when illuminated by a point source at 2\rg, is 118\ergcmps, suggesting that this model is a realistic representation of the ionisation profile of the AGN with the strongest detections of X-ray reverberation. For an isotropically emitting point source 2\rg\ above the singularity of a $10^6$\Msun\ black hole, ionising the innermost stable orbit of a disc with density $10^{19}$\pcmcu\ to $\xi_0 = 1000$\ergcmps\ implies a total coronal luminosity of $L\sim 10^{43}$\ergps (or 10 per cent of the Eddington limit). Note that due to strong light bending towards the black hole, and blueshifting of emission between the corona and inner regions of the disc, the innermost regions of the disc receive approximately 400 times the ionising flux they would in Euclidean space.

It should be noted that when emission lines from a disc with an ionisation gradient, as simulated here, are fit using a spectral model with constant ionisation over the whole disc (as commonly employed when fitting the observed spectra of AGN), systematic errors can be introduced in the parameters of the relativistic blurring, including overestimating the steepness of the emissivity profile of the inner disc \citep{svoboda+12} and underestimating the spin of the black hole from the location of the innermost stable orbit \citep{kammoun+2019}. These errors are introduced as the ionisation gradient weakens the redshifted wing of neutral iron fluorescence line at 6.4\keV\ from the inner disc relative to the core of the line from large radii and shifts emission at small radii to the higher ionisation lines at 6.67 and 6.97\keV. We further find that the weakening of the line from the inner disc can lead to the height of the corona above the disc being overestimated. The weakening of the line from the inner disc is accounted for in a constant ionisation model by raising the height of the primary X-ray source to reduce the inner disc illumination. When $\xi_0 = 1000$\ergcmps and the ionisation gradient corresponds to the illumination pattern of the disc, the height of the source can be overestimated by around 10 per cent, while for a highly ionised disc with $\xi_0 = 10,000$\ergcmps, the height can be overestimated by as much as a factor of two. For the purposes of this study however, the important factor is that the X-ray source is sufficiently low and compact to provide substantial illumination of the relatively small plunging region. This systematic error will reduce the intrinsic scale heights of coron\ae\ relative to the measured values, so will not substantially affect the conclusions about detectability of the plunging region.

Also important is the location of the primary X-ray source, particularly in the case of a maximally-spinning black hole where the small radius of the ISCO means that the plunging region subtends only a small solid angle at the source. Increasing the source height above a maximally-spinning black hole from $h=2$\rg\ to $h=3$\rg\ results in the reprocessed X-ray count rate from the plunging region dropping from a 2 to 0.7 per cent contribution. As a result, the signature in the lag-frequency spectrum of the 1-2\keV\ band drops from 20 per cent to 9 per cent and when the source height is increased to 5\rg, the feature is undetectable. Analysis of the X-ray spectrum and reverberation time lags from the accretion discs around a number of supermassive black holes reveals that in many cases, there is substantial X-ray emission from within 2-3\rg\ of the black hole \citep[\textit{e.g.}][]{understanding_emis_paper, lag_spectra_paper, parker_mrk335, mrk335_corona_paper, kara_global, jiang_iras}. Moreover, the close proximity of the plunging region to the black hole means it also receives significant illumination by reflected emission returning to the accretion flow due to strong light bending to be reflected a second time. This returning radiation enhances reflection features from the disc \citep{ross_fabian_ballantyne} and will likely enhance the detectability of the plunging region, and will be considered thoroughly in a forthcoming work.

While we consider the case that the accretion flow is illuminated by a compact point source located on the polar axis, it is possible that a significant contribution to the X-ray continuum comes from a component of the corona extending over the inner parts of the accretion disc \citep{understanding_emis_paper, mrk335_corona_paper}. The extended parts of the corona will vary in luminosity more slowly than the inner regions and \citet{propagating_lag_paper} show that when considering just the highest frequency Fourier components in which the signatures of the plunging region are seen, the extended corona behaves similarly to a compact point source. There is evidence that the highest frequency Fourier components are dominated by a collimated core within the corona that explains the profile of the observed high frequency lag-energy spectra \citep{1zw1_corona_paper}.

While the surface density, $\Sigma$, is well defined by mass conservation if the velocity is known through the plunging region, one of the greatest sources of uncertainty in this model is the height of the flow as a function of radius. For a given surface density profile, the height will determine the volume density which in turn determines the ionisation profile and the spectrum of the reprocessed emission. The height profile will depend on the detailed balance of the gas pressure, radiation pressure and any magnetic pressure within the plasma. In these models, we assume a constant ratio $h/r$ within the plunging region, extrapolating from the turnover in height close to the ISCO in the analytic solution of \citet{novthorne}. If, instead, the height of the flow remains constant through the plunging region, the density would drop more rapidly, enhancing the increase in ionisation towards the event horizon. Alternatively, if the height of the accretion flow dropped more steeply towards the event horizon, the density decrease and ionisation gradient would be less through the plunging region.

\subsection{Prospects with future missions}
At present, the state of the art in detecting X-ray reverberation around supermassive black holes are long observations with \textit{XMM-Newton} \citep{xmm}, using light curves recorded using the EPIC pn camera \citep{xmm_strueder}. The detectability of time lags depends on both the total count rate from the source and the magnitude of the X-ray variability, \textit{i.e.} the variable count rate \citep{kara_global}. During a 2\Ms\ observation of the narrow line Seyfert 1 galaxy IRAS\,13224$-$3809, the deepest observation to date of a source exhibiting some of the strongest signatures of X-ray reverberation \citep{alston_iras1}, it is possible to measure the lag in the 1-2\keV\ energy band with respect to a reference band encompassing the full bandpass of the detector in Fourier frequencies between $10^{-3}$ and $10^{-2}$\Hz\ to an accuracy of around 30 per cent. X-ray timing measurements at high frequencies are limited by Poisson noise owing to the finite count rate in the observed light curve.

The collecting area of the next generation X-ray observatory, \textit{Athena} \citep{athena}, promises to enhance the precision with which X-ray reverberation may be measured around supermassive black holes. In the case of low frequency variability (at mHz frequencies) where few wave cycles are observed during an observation but many photons are collected per cycle, the signal-to-noise increases as $\sqrt{N}$ as the count rate, $N$, increases \citep{reverb_review}. \textit{Athena} is planned to have an effective area of $5\times 10^4$\cmsq\ between 1 and 2\keV\ and 2500\cmsq\ at 6\keV, offering factors of 16 and 3 increase over \textit{XMM-Newton}. It is therefore expected to decrease the uncertainty by a factor of 4 in the time lag measurement in the 1-2\keV\ energy band. \textit{Athena} will likely be able to detect the signatures of X-ray reverberation from the plunging region in the shape high frequency tails of energy-resolved lag-frequency spectra if the contribution of the plunging region can be incorporated into models that incorporate the finite spatial extent and geometry of the corona as well as the propagation of luminosity fluctuations through the X-ray source \citep{propagating_lag_paper}.

The proposed X-ray timing instrument \textit{STROBE-X} \citep{strobex} promises a greater increase in the sensitivity of X-ray reverberation measurements. \textit{STROBE-X} is proposed to carry two large collecting area non-imaging detectors, a 3\msq\ \textit{X-ray Concentrator Array (XRCA)} sensitive between 0.2 and 10\keV\ and a 10\msq silicon drift detector, the \textit{Large Area Detector (LAD)}, sensitive between 2 and 30\keV. In the case of stellar mass black holes accreting in X-ray binary systems, X-ray reverberation is detected in the 10-1000\Hz\ frequency range where many wave cycles are captured over the course of an observation but with few photons per cycle. In this regime, the signal-to-noise in the time lag measurement increases as $N$ rather than $\sqrt{N}$. Coupled with the increase by a factor of $\sim 100$ in the X-ray count rates that are observed from nearby X-ray binaries in our own Galaxy compared to from more distant AGN, the enhanced collecting are provides greater gains for stellar mass black holes. Bright X-ray binaries in our own Galaxy may prove to be the most lucrative targets to probe the extreme environment immediately outside the event horizon of a black hole.

\citet{kara_maxireverb} were able to detect X-ray reverberation from the accretion disc during the bright outburst of the black hole X-ray transient MAXI\,J1820+070 using \textit{NICER}. The high frequency tail of the reverberation response was seen at 50\Hz\ and the lag in the tail was recorded with statistical uncertainty around 50 per cent. At 1\,keV, the \textit{STROBE-X XRCA} offers approximately 20 times greater effective area, reducing the statistical uncertainty in such an observation to around 3 per cent, within the margin required for the detection of reverberation signatures from the plunging region. The magnitude of the lag and relatively low frequency at which reverberation was detected in this system may suggest an X-ray source located further from the black hole (reducing the illumination fo the plunging region) or spin below maximal. The decaying power spectrum of the variability at high frequencies would reduce the signal-to-noise by a factor of eight for the detection of reverberation at 200\Hz\ from a plunging region illuminated by a more compact corona.

\section{Conclusions}

General relativistic ray tracing simulations of X-ray reverberation from the accretion flows around black holes reveal how the presence of an innermost stable circular orbit (ISCO) and the plunging region within which material falls rapidly into the black hole, may be inferred from X-ray measurements. X-ray emission from the plunging region is distinguished from that arising from the accretion disc by (1) the strong redshifts seen from rapid orbits deep within the gravitational potential, (2) the time delays associated with light passage close to the black hole, and (3) the sharp rise in ionisation of strongly irradiated material as the accretion flow density drops crossing the ISCO, varying the emission lines seen from the plunging region with respect to the disc.

The detectable signature of the ISCO and plunging region depends upon the spin of the black hole. In the case of a non-spinning black hole, as much as 30 per cent of the observed reverberating emission can arise from within the plunging region. Reverberation from the plunging region is most readily detected in a sharp peak in redshifted iron K$\alpha$ photons in the 2-4\keV\ energy band that respond more rapidly than the outer parts of the disc but are delayed with respect to the primary continuum.

In the case of a maximally spinning ($a=0.998$) black hole, the plunging region accounts for only 2 per cent of the reprocessed emission when the corona is located at a low height above the disc. In this case, reverberation from the plunging region is seen as a secondary peak at late times, comprised of highly redshifted K$\alpha$ photons from ionised iron in the 1-2\keV\ energy band.

Reverberation from the plunging region does not significantly alter the average spectrum or the average response time as a function of X-ray energy but is detected in the shape of the reverberation response function. The emission from the plunging region can be detected from a change in the profile of the time lag \textit{vs.} Fourier frequency in the 2-4 and 1-2\keV\ energy bands in the highest frequency components of the variability where the reverberation lag decays to zero. Detection of the plunging region above the disc emission requires a time lag measurement to an accuracy of 20 per cent at frequencies above $10^{-2}\,c^3(GM)^{-1}$, scaled to the mass of the black hole, in at least three frequency bins between 0.01 and $0.03\,c^3(GM)^{-1}$. Improving the accuracy to 12 per cent at $0.02\,c^3(GM)^{-1}$ will enable constraints to be placed on the dynamics of the material in the plunging region and enable plunging orbits to be distinguished from material remaining on stable circular orbits in this region, hence confirming the existence of the plunging region and ISCO.

Measurement of the location of the ISCO and dynamics of the plunging regions around supermassive black holes will be feasible with the next generation large X-ray observatory, \textit{Athena}. With specialised, large collecting area X-ray timing missions such as the proposed \textit{STROBE-X} it will be possible to extend such X-ray reverberation studies from AGN to stellar mass black holes in X-ray binaries. Detection of the ISCO and measurement of the dynamics of the plunging region not only tests a key prediction of general relativity in the strong-field regime, but probes the physical processes that drive the accretion of gas onto black holes, powering some of the most extreme objects in the Universe.

\section*{Acknowledgements}
DRW is supported by NASA through Einstein Postdoctoral Fellowship grant number PF6-170160, awarded by the \textit{Chandra} X-ray Center, operated by the Smithsonian Astrophysical Observatory for NASA under contract NAS8-03060. CSR acknowledges the support of the UK Science and Technology Facilities Council (STFC) via grant ST/R000867/1. ACF acknowledges the support of the European Research Council Advanced Grant 340442. This work used the XStream computational resource, supported by the National Science Foundation Major Research Instrumentation program (ACI-1429830). We thank the anonymous referee for their feedback on the original manuscript.

\bibliographystyle{mnras}
\bibliography{agn}

\begin{thebibliography}{}
\makeatletter
\relax
\def\mn@urlcharsother{\let\do\@makeother \do\$\do\&\do\#\do\^\do\_\do\%\do\~}
\def\mn@doi{\begingroup\mn@urlcharsother \@ifnextchar [ {\mn@doi@}
  {\mn@doi@[]}}
\def\mn@doi@[#1]#2{\def\@tempa{#1}\ifx\@tempa\@empty \href
  {http://dx.doi.org/#2} {doi:#2}\else \href {http://dx.doi.org/#2} {#1}\fi
  \endgroup}
\def\mn@eprint#1#2{\mn@eprint@#1:#2::\@nil}
\def\mn@eprint@arXiv#1{\href {http://arxiv.org/abs/#1} {{\tt arXiv:#1}}}
\def\mn@eprint@dblp#1{\href {http://dblp.uni-trier.de/rec/bibtex/#1.xml}
  {dblp:#1}}
\def\mn@eprint@#1:#2:#3:#4\@nil{\def\@tempa {#1}\def\@tempb {#2}\def\@tempc
  {#3}\ifx \@tempc \@empty \let \@tempc \@tempb \let \@tempb \@tempa \fi \ifx
  \@tempb \@empty \def\@tempb {arXiv}\fi \@ifundefined
  {mn@eprint@\@tempb}{\@tempb:\@tempc}{\expandafter \expandafter \csname
  mn@eprint@\@tempb\endcsname \expandafter{\@tempc}}}

\bibitem[\protect\citeauthoryear{{Agol} \& {Krolik}}{{Agol} \&
  {Krolik}}{2000}]{agol_krolik}
{Agol} E.,  {Krolik} J.~H.,  2000, \mn@doi [\apj] {10.1086/308177}, \href
  {http://adsabs.harvard.edu/abs/2000ApJ...528..161A} {528, 161}

\bibitem[\protect\citeauthoryear{{Alston} et~al.,}{{Alston}
  et~al.}{2019}]{alston_iras1}
{Alston} W.~N.,  et~al., 2019, \mn@doi [\mnras] {10.1093/mnras/sty2527}, \href
  {https://ui.adsabs.harvard.edu/abs/2019MNRAS.482.2088A} {482, 2088}

\bibitem[\protect\citeauthoryear{{Balbus} \& {Hawley}}{{Balbus} \&
  {Hawley}}{1991}]{balbus+91}
{Balbus} S.~A.,  {Hawley} J.~F.,  1991, \mn@doi [\apj] {10.1086/170270}, \href
  {http://adsabs.harvard.edu/abs/1991ApJ...376..214B} {376, 214}

\bibitem[\protect\citeauthoryear{{Balbus} \& {Hawley}}{{Balbus} \&
  {Hawley}}{1998}]{balbus+98}
{Balbus} S.~A.,  {Hawley} J.~F.,  1998, \mn@doi [Reviews of Modern Physics]
  {10.1103/RevModPhys.70.1}, \href
  {http://adsabs.harvard.edu/abs/1998RvMP...70....1B} {70, 1}

\bibitem[\protect\citeauthoryear{{Barcons} et~al.,}{{Barcons}
  et~al.}{2017}]{athena}
{Barcons} X.,  et~al., 2017, \mn@doi [Astronomische Nachrichten]
  {10.1002/asna.201713323}, \href
  {https://ui.adsabs.harvard.edu/\#abs/2017AN....338..153B} {338, 153}

\bibitem[\protect\citeauthoryear{{Blandford} \& {Znajek}}{{Blandford} \&
  {Znajek}}{1977}]{blandford_znajek}
{Blandford} R.~D.,  {Znajek} R.~L.,  1977, \mnras, \href
  {http://adsabs.harvard.edu/abs/1977MNRAS.179..433B} {179, 433}

\bibitem[\protect\citeauthoryear{{Brenneman} \& {Reynolds}}{{Brenneman} \&
  {Reynolds}}{2006}]{brenneman_reynolds}
{Brenneman} L.~W.,  {Reynolds} C.~S.,  2006, \mn@doi [\apj] {10.1086/508146},
  \href {http://adsabs.harvard.edu/abs/2006ApJ...652.1028B} {652, 1028}

\bibitem[\protect\citeauthoryear{{Cackett}, {Zoghbi}, {Reynolds}, {Fabian},
  {Kara}, {Uttley}  \& {Wilkins}}{{Cackett} et~al.}{2014}]{cackett_ngc4151}
{Cackett} E.~M.,  {Zoghbi} A.,  {Reynolds} C.,  {Fabian} A.~C.,  {Kara} E.,
  {Uttley} P.,   {Wilkins} D.~R.,  2014, \mn@doi [\mnras]
  {10.1093/mnras/stt2424}, \href
  {http://adsabs.harvard.edu/abs/2014MNRAS.438.2980C} {438, 2980}

\bibitem[\protect\citeauthoryear{{Event Horizon Telescope Collaboration}
  et~al.,}{{Event Horizon Telescope Collaboration} et~al.}{2019}]{eht1}
{Event Horizon Telescope Collaboration} et~al., 2019, \mn@doi [\apjl]
  {10.3847/2041-8213/ab0ec7}, \href
  {https://ui.adsabs.harvard.edu/abs/2019ApJ...875L...1E} {875, L1}

\bibitem[\protect\citeauthoryear{{Fabian} et~al.,}{{Fabian}
  et~al.}{2009}]{fabian+09}
{Fabian} A.~C.,  et~al., 2009, \mn@doi [\nat] {10.1038/nature08007}, \href
  {http://adsabs.harvard.edu/abs/2009Natur.459..540F} {459, 540}

\bibitem[\protect\citeauthoryear{{Gallo} et~al.,}{{Gallo}
  et~al.}{2013}]{gallo+13}
{Gallo} L.~C.,  et~al., 2013, \mn@doi [\mnras] {10.1093/mnras/sts102}, \href
  {http://adsabs.harvard.edu/abs/2013MNRAS.428.1191G} {428, 1191}

\bibitem[\protect\citeauthoryear{{Garc\'{\i}a} \& {Kallman}}{{Garc\'{\i}a} \&
  {Kallman}}{2010}]{garcia+2010}
{Garc\'{\i}a} J.,  {Kallman} T.~R.,  2010, \mn@doi [\apj]
  {10.1088/0004-637X/718/2/695}, \href
  {http://adsabs.harvard.edu/abs/2010ApJ...718..695G} {718, 695}

\bibitem[\protect\citeauthoryear{{Garc\'{\i}a}, {Fabian}, {Kallman}, {Dauser},
  {Parker}, {McClintock}, {Steiner}  \& {Wilms}}{{Garc\'{\i}a}
  et~al.}{2016}]{xillver_density}
{Garc\'{\i}a} J.~A.,  {Fabian} A.~C.,  {Kallman} T.~R.,  {Dauser} T.,  {Parker}
  M.~L.,  {McClintock} J.~E.,  {Steiner} J.~F.,   {Wilms} J.,  2016, \mn@doi
  [\mnras] {10.1093/mnras/stw1696}, \href
  {https://ui.adsabs.harvard.edu/\#abs/2016MNRAS.462..751G} {462, 751}

\bibitem[\protect\citeauthoryear{{George} \& {Fabian}}{{George} \&
  {Fabian}}{1991}]{george_fabian}
{George} I.~M.,  {Fabian} A.~C.,  1991, \mnras, \href
  {http://ukads.nottingham.ac.uk/abs/1991MNRAS.249..352G} {249, 352}

\bibitem[\protect\citeauthoryear{{Jansen} et~al.,}{{Jansen} et~al.}{2001}]{xmm}
{Jansen} F.,  et~al., 2001, \mn@doi [\aap] {10.1051/0004-6361:20000036}, \href
  {http://adsabs.harvard.edu/abs/2001A\%26A...365L...1J} {365, L1}

\bibitem[\protect\citeauthoryear{{Jiang} et~al.,}{{Jiang}
  et~al.}{2018}]{jiang_iras}
{Jiang} J.,  et~al., 2018, \mn@doi [\mnras] {10.1093/mnras/sty836}, \href
  {https://ui.adsabs.harvard.edu/\#abs/2018MNRAS.477.3711J} {477, 3711}

\bibitem[\protect\citeauthoryear{{Kammoun}, {Dom{\v c}ek}, {Svoboda}, {Dov{\v
  c}iak}  \& {Matt}}{{Kammoun} et~al.}{2019}]{kammoun+2019}
{Kammoun} E.~S.,  {Dom{\v c}ek} V.,  {Svoboda} J.,  {Dov{\v c}iak} M.,   {Matt}
  G.,  2019, \mn@doi [\mnras] {10.1093/mnras/stz408}, \href
  {https://ui.adsabs.harvard.edu/abs/2019MNRAS.485..239K} {485, 239}

\bibitem[\protect\citeauthoryear{{Kara}, {Alston}, {Fabian}, {Cackett},
  {Uttley}, {Reynolds}  \& {Zoghbi}}{{Kara} et~al.}{2016}]{kara_global}
{Kara} E.,  {Alston} W.~N.,  {Fabian} A.~C.,  {Cackett} E.~M.,  {Uttley} P.,
  {Reynolds} C.~S.,   {Zoghbi} A.,  2016, \mn@doi [\mnras]
  {10.1093/mnras/stw1695}, \href
  {https://ui.adsabs.harvard.edu/\#abs/2016MNRAS.462..511K} {462, 511}

\bibitem[\protect\citeauthoryear{{Kara}, {Garc\'{\i}a}, {Lohfink}, {Fabian},
  {Reynolds}, {Tombesi}  \& {Wilkins}}{{Kara} et~al.}{2017}]{kara_ark564}
{Kara} E.,  {Garc\'{\i}a} J.~A.,  {Lohfink} A.,  {Fabian} A.~C.,  {Reynolds}
  C.~S.,  {Tombesi} F.,   {Wilkins} D.~R.,  2017, \mn@doi [\mnras]
  {10.1093/mnras/stx792}, \href
  {https://ui.adsabs.harvard.edu/abs/2017MNRAS.468.3489K} {468, 3489}

\bibitem[\protect\citeauthoryear{{Kara} et~al.,}{{Kara}
  et~al.}{2019}]{kara_maxireverb}
{Kara} E.,  et~al., 2019, \mn@doi [\nat] {10.1038/s41586-018-0803-x}, \href
  {https://ui.adsabs.harvard.edu/abs/2019Natur.565..198K} {565, 198}

\bibitem[\protect\citeauthoryear{{Kerr}}{{Kerr}}{1963}]{kerr}
{Kerr} R.,  1963, Phys. Rev. Let., 11, 237

\bibitem[\protect\citeauthoryear{{Krolik}, {Hawley}  \& {Hirose}}{{Krolik}
  et~al.}{2005}]{krolik+05}
{Krolik} J.~H.,  {Hawley} J.~F.,   {Hirose} S.,  2005, \mn@doi [\apj]
  {10.1086/427932}, \href {http://adsabs.harvard.edu/abs/2005ApJ...622.1008K}
  {622, 1008}

\bibitem[\protect\citeauthoryear{{Loisel} et~al.,}{{Loisel}
  et~al.}{2017}]{loisel+17}
{Loisel} G.~P.,  et~al., 2017, \mn@doi [\prl] {10.1103/PhysRevLett.119.075001},
  \href {https://ui.adsabs.harvard.edu/abs/2017PhRvL.119g5001L} {119, 075001}

\bibitem[\protect\citeauthoryear{{McClintock} et~al.,}{{McClintock}
  et~al.}{2011}]{mcclintock_spin}
{McClintock} J.~E.,  et~al., 2011, \mn@doi [Classical and Quantum Gravity]
  {10.1088/0264-9381/28/11/114009}, \href
  {https://ui.adsabs.harvard.edu/\#abs/2011CQGra..28k4009M} {28, 114009}

\bibitem[\protect\citeauthoryear{{Novikov} \& {Thorne}}{{Novikov} \&
  {Thorne}}{1973}]{novthorne}
{Novikov} I.~D.,  {Thorne} K.~S.,  1973, in {C.~Dewitt \& B.~S.~Dewitt} ed.,
  {Black Holes (Les Astres Occlus)}. pp 343--450

\bibitem[\protect\citeauthoryear{{Parker}, {Wilkins}, {Fabian}, {Grupe},
  {Dauser}, {Matt}  \& {Harrison}}{{Parker} et~al.}{2014}]{parker_mrk335}
{Parker} M.~L.,  {Wilkins} D.~R.,  {Fabian} A.~C.,  {Grupe} D.,  {Dauser} T.,
  {Matt} G.,   {Harrison} F.~A.,  2014, \mn@doi [\mnras]
  {10.1093/mnras/stu1246}, \href
  {http://adsabs.harvard.edu/abs/2014MNRAS.443.1723P} {443, 1723}

\bibitem[\protect\citeauthoryear{{Penna}, {McKinney}, {Narayan},
  {Tchekhovskoy}, {Shafee}  \& {McClintock}}{{Penna} et~al.}{2010}]{penna+2010}
{Penna} R.~F.,  {McKinney} J.~C.,  {Narayan} R.,  {Tchekhovskoy} A.,  {Shafee}
  R.,   {McClintock} J.~E.,  2010, \mn@doi [\mnras]
  {10.1111/j.1365-2966.2010.17170.x}, \href
  {https://ui.adsabs.harvard.edu/abs/2010MNRAS.408..752P} {408, 752}

\bibitem[\protect\citeauthoryear{{Pringle}}{{Pringle}}{1981}]{pringle_81}
{Pringle} J.~E.,  1981, \mn@doi [Annual Review of Astronomy and Astrophysics]
  {10.1146/annurev.aa.19.090181.001033}, \href
  {https://ui.adsabs.harvard.edu/\#abs/1981ARA\&A..19..137P} {19, 137}

\bibitem[\protect\citeauthoryear{{Ray} et~al.,}{{Ray} et~al.}{2018}]{strobex}
{Ray} P.~S.,  et~al., 2018, in {Society of Photo-Optical Instrumentation
  Engineers (SPIE) Conference Series}. p. 1069919, \mn@doi{10.1117/12.2312257}

\bibitem[\protect\citeauthoryear{{Reynolds}}{{Reynolds}}{2013}]{reynolds_spin}
{Reynolds} C.~S.,  2013, \mn@doi [Classical and Quantum Gravity]
  {10.1088/0264-9381/30/24/244004}, \href
  {https://ui.adsabs.harvard.edu/\#abs/2013CQGra..30x4004R} {30, 244004}

\bibitem[\protect\citeauthoryear{{Reynolds} \& {Begelman}}{{Reynolds} \&
  {Begelman}}{1997}]{reynolds+97}
{Reynolds} C.~S.,  {Begelman} M.~C.,  1997, \mn@doi [\apj] {10.1086/304703},
  \href {https://ui.adsabs.harvard.edu/\#abs/1997ApJ...488..109R} {488, 109}

\bibitem[\protect\citeauthoryear{{Reynolds} \& {Fabian}}{{Reynolds} \&
  {Fabian}}{2008}]{reynolds_fabian2008}
{Reynolds} C.~S.,  {Fabian} A.~C.,  2008, \mn@doi [\apj] {10.1086/527344},
  \href {https://ui.adsabs.harvard.edu/abs/2008ApJ...675.1048R} {675, 1048}

\bibitem[\protect\citeauthoryear{{Reynolds}, {Young}, {Begelman}  \&
  {Fabian}}{{Reynolds} et~al.}{1999}]{reynolds+99}
{Reynolds} C.~S.,  {Young} A.~J.,  {Begelman} M.~C.,   {Fabian} A.~C.,  1999,
  \mn@doi [\apj] {10.1086/306913}, \href
  {http://adsabs.harvard.edu/abs/1999ApJ...514..164R} {514, 164}

\bibitem[\protect\citeauthoryear{{Ross} \& {Fabian}}{{Ross} \&
  {Fabian}}{2005}]{ross_fabian}
{Ross} R.~R.,  {Fabian} A.~C.,  2005, \mn@doi [\mnras]
  {10.1111/j.1365-2966.2005.08797.x}, \href
  {http://ukads.nottingham.ac.uk/abs/2005MNRAS.358..211R} {358, 211}

\bibitem[\protect\citeauthoryear{{Ross}, {Fabian}  \& {Ballantyne}}{{Ross}
  et~al.}{2002}]{ross_fabian_ballantyne}
{Ross} R.~R.,  {Fabian} A.~C.,   {Ballantyne} D.~R.,  2002, \mn@doi [\mnras]
  {10.1046/j.1365-8711.2002.05758.x}, \href
  {http://adsabs.harvard.edu/abs/2002MNRAS.336..315R} {336, 315}

\bibitem[\protect\citeauthoryear{{Shakura} \& {Sunyaev}}{{Shakura} \&
  {Sunyaev}}{1973}]{shaksun}
{Shakura} N.~I.,  {Sunyaev} R.~A.,  1973, \aap, \href
  {http://adsabs.harvard.edu/abs/1973A\%26A....24..337S} {24, 337}

\bibitem[\protect\citeauthoryear{{Str{\"u}der} et~al.,}{{Str{\"u}der}
  et~al.}{2001}]{xmm_strueder}
{Str{\"u}der} L.,  et~al., 2001, \mn@doi [\aap] {10.1051/0004-6361:20000066},
  \href {http://adsabs.harvard.edu/abs/2001A\%26A...365L..18S} {365, L18}

\bibitem[\protect\citeauthoryear{{Svoboda}, {Dov{\v c}iak}, {Goosmann},
  {Jethwa}, {Karas}, {Miniutti}  \& {Guainazzi}}{{Svoboda}
  et~al.}{2012}]{svoboda+12}
{Svoboda} J.,  {Dov{\v c}iak} M.,  {Goosmann} R.~W.,  {Jethwa} P.,  {Karas} V.,
   {Miniutti} G.,   {Guainazzi} M.,  2012, \mn@doi [\aap]
  {10.1051/0004-6361/201219701}, \href
  {http://adsabs.harvard.edu/abs/2012A\%26A...545A.106S} {545, A106}

\bibitem[\protect\citeauthoryear{{Taylor} \& {Reynolds}}{{Taylor} \&
  {Reynolds}}{2018}]{taylor_reynolds}
{Taylor} C.,  {Reynolds} C.~S.,  2018, \mn@doi [\apj]
  {10.3847/1538-4357/aaad63}, \href
  {https://ui.adsabs.harvard.edu/\#abs/2018ApJ...855..120T} {855, 120}

\bibitem[\protect\citeauthoryear{{Uttley}, {Cackett}, {Fabian}, {Kara}  \&
  {Wilkins}}{{Uttley} et~al.}{2014}]{reverb_review}
{Uttley} P.,  {Cackett} E.~M.,  {Fabian} A.~C.,  {Kara} E.,   {Wilkins} D.~R.,
  2014, \aapr

\bibitem[\protect\citeauthoryear{{Wilkins} \& {Fabian}}{{Wilkins} \&
  {Fabian}}{2011}]{1h0707_emis_paper}
{Wilkins} D.~R.,  {Fabian} A.~C.,  2011, \mn@doi [\mnras]
  {10.1111/j.1365-2966.2011.18458.x}, \href
  {http://adsabs.harvard.edu/abs/2011MNRAS.414.1269W} {414, 1269}

\bibitem[\protect\citeauthoryear{{Wilkins} \& {Fabian}}{{Wilkins} \&
  {Fabian}}{2012}]{understanding_emis_paper}
{Wilkins} D.~R.,  {Fabian} A.~C.,  2012, \mn@doi [\mnras]
  {10.1111/j.1365-2966.2012.21308.x}, \href
  {http://adsabs.harvard.edu/abs/2012MNRAS.424.1284W} {424, 1284}

\bibitem[\protect\citeauthoryear{{Wilkins} \& {Fabian}}{{Wilkins} \&
  {Fabian}}{2013}]{lag_spectra_paper}
{Wilkins} D.~R.,  {Fabian} A.~C.,  2013, \mn@doi [\mnras]
  {10.1093/mnras/sts591}, \href
  {http://adsabs.harvard.edu/abs/2013MNRAS.430..247W} {430, 247}

\bibitem[\protect\citeauthoryear{{Wilkins} \& {Gallo}}{{Wilkins} \&
  {Gallo}}{2015}]{mrk335_corona_paper}
{Wilkins} D.~R.,  {Gallo} L.~C.,  2015, \mn@doi [\mnras]
  {10.1093/mnras/stv162}, 449, 129

\bibitem[\protect\citeauthoryear{{Wilkins}, {Cackett}, {Fabian}  \&
  {Reynolds}}{{Wilkins} et~al.}{2016}]{propagating_lag_paper}
{Wilkins} D.~R.,  {Cackett} E.~M.,  {Fabian} A.~C.,   {Reynolds} C.~S.,  2016,
  \mn@doi [\mnras] {10.1093/mnras/stw276}, \href
  {http://adsabs.harvard.edu/abs/2016MNRAS.458..200W} {458, 200}

\bibitem[\protect\citeauthoryear{{Wilkins}, {Gallo}, {Silva}, {Costantini},
  {Brandt}  \& {Kriss}}{{Wilkins} et~al.}{2017}]{1zw1_corona_paper}
{Wilkins} D.~R.,  {Gallo} L.~C.,  {Silva} C.~V.,  {Costantini} E.,  {Brandt}
  W.~N.,   {Kriss} G.~A.,  2017, \mn@doi [\mnras] {10.1093/mnras/stx1814},
  \href {https://ui.adsabs.harvard.edu/\#abs/2017MNRAS.471.4436W} {471, 4436}

\bibitem[\protect\citeauthoryear{{Zhu}, {Davis}, {Narayan}, {Kulkarni}, {Penna}
   \& {McClintock}}{{Zhu} et~al.}{2012}]{zhu+2012}
{Zhu} Y.,  {Davis} S.~W.,  {Narayan} R.,  {Kulkarni} A.~K.,  {Penna} R.~F.,
  {McClintock} J.~E.,  2012, \mn@doi [\mnras]
  {10.1111/j.1365-2966.2012.21181.x}, \href
  {https://ui.adsabs.harvard.edu/abs/2012MNRAS.424.2504Z} {424, 2504}

\bibitem[\protect\citeauthoryear{{Zoghbi}, {Fabian}, {Uttley}, {Miniutti},
  {Gallo}, {Reynolds}, {Miller}  \& {Ponti}}{{Zoghbi} et~al.}{2010}]{zoghbi+09}
{Zoghbi} A.,  {Fabian} A.~C.,  {Uttley} P.,  {Miniutti} G.,  {Gallo} L.~C.,
  {Reynolds} C.~S.,  {Miller} J.~M.,   {Ponti} G.,  2010, \mn@doi [\mnras]
  {10.1111/j.1365-2966.2009.15816.x}, \href
  {http://ukads.nottingham.ac.uk/abs/2010MNRAS.401.2419Z} {401, 2419}

\bibitem[\protect\citeauthoryear{{Zoghbi}, {Fabian}, {Reynolds}  \&
  {Cackett}}{{Zoghbi} et~al.}{2012}]{zoghbi+2012}
{Zoghbi} A.,  {Fabian} A.~C.,  {Reynolds} C.~S.,   {Cackett} E.~M.,  2012,
  \mn@doi [\mnras] {10.1111/j.1365-2966.2012.20587.x}, \href
  {http://adsabs.harvard.edu/abs/2012MNRAS.422..129Z} {422, 129}

\makeatother
\end{thebibliography}

\appendix
\section{Proper Area Calculation}
\label{area.app}
In order to construct a fully covariant prescription for the area of a patch of the accretion flow, we consider a triangular patch $\bigtriangleup ABC$ seen in the rest frame of an observer comoving with the accretion flow in which the spacetime is locally flat. In flat, Euclidean space, the vector area of $\bigtriangleup ABC$ may be calculated from the cross product of vectors $\overrightarrow{AB}$ and $\overrightarrow{AC}$.

The proper area of a patch on the disc defined by four points, ABCD, specified in Boyer-Lindquist co-ordinates, is computed by first dividing the patch into two triangles $\bigtriangleup ABC$ and $\bigtriangleup ACD$. The displacement vectors describing the sides $\overrightarrow{AB}$ and $\overrightarrow{AC}$ are constructed in Boyer-Lindquist co-ordinates as the difference between the position vectors of the two points.
\begin{equation}
\mathbf{v}_{AB} = \mathbf{x}_B - \mathbf{x}_A
\end{equation}

The length of each side of the triangle as measured by the comoving observer is then calculated by projecting the displacement 4-vectors onto the observer's tetrad of basis vectors, ${\mathbf{e}_{(i)}}$, representing a Cartesian co-ordinate system that is measured by that observer. The timelike basis vector is taken to be a unit vector parallel to the observer's 4-velocity and three spatial unit vectors are constructed to be orthogonal to this. Assuming the distances between the vertices are small, the differences in the timelike co-ordinate may be neglected and the proper displacement measured by the observer comes from the three spatial components of the 4-vector.
\begin{equation}
\vec{v}'_{AB} = \sum_{i=1}^3 (\mathbf{v}_{AB} \cdot \mathbf{e}_{(i)} )\,\vec{e}_{(i)}
\end{equation}
The area of the triangular patch is then computed from the 3-vector cross product
\begin{equation}
A_{\bigtriangleup ABC} = |\vec{v}'_{AB} \times \vec{v}'_{AC}|
\end{equation}
The area of the other half of the patch, $\bigtriangleup ACD$, is computed in the same manner.

The tetrad of basis vectors defining the observer's frame can be calculated for an observer in a circular orbit in the equatorial plane (\textit{i.e.} for the disc) using the prescription laid out in \citet{understanding_emis_paper}, or for the plunging region, a numerical Gram-Schmidt orthogonalisation procedure is used, outlined in Appendix~\ref{gs.app}. The transformation from Boyer-Lindquist co-ordinates to the comoving frame accounts for all relativistic effects on the area of the patch including the conversion from co-ordinate to proper distances in the curved spacetime close to the black hole and the length contraction of a patch as seen by a stationary observer due to its orbital motion.

\section{Numerical Tetrad Calculation}
\label{gs.app}
In ray tracing calculations, it is often necessary to construct the tetrad of basis vectors, $\left\{ \mathbf{e}'_{(a)} \right\}$, describing the locally flat instantaneous rest frame of an observer. While analytic solutions can be found for specific cases, for instance azimuthal or radial motion of the source, it can be desirable to compute the basis vectors numerically for arbitrary motion of the source, described by its 4-velocity $\mathbf{u}$, such as within the plunging region.

The spacetime in the observer's rest frame reduces to Minkowski space, described by the metric
\begin{align}
\mathbf{e}'_{(a)}\cdot\mathbf{e}'_{(b)} = \eta_{ab}
\end{align}
Where the Minkowski metric $\left[\eta_{ab}\right] = \mathrm{diag}\left(1, -1, -1, -1\right)$.

Since the observer is at rest in its own rest frame, the spatial components of the 4-velocity must be zero in that frame, hence the observer's timelike basis vector is parallel to the 4-velocity. $u^t$ is found by imposing the normalisation condition on the 4-velocity that $|\mathbf{u}| = c$ and working in natural units with $\mu = c = 1$ will result in a timelike unit vector.

\begin{align}
\mathbf{\hat{e}}_0 = \mathbf{u}
\end{align}

The three spacelike basis vectors, which are orthogonal to the timelike unit vector and to each other, can be computed using the Gram-Schmidt orthogonalisation procedure.

We begin with an initial guess of the basis vectors, $\left\{ \mathbf{v}_a \right\}$. These are initially taken to be in the $r$, $\theta$ and $\varphi$ directions in Boyer-Lindquist co-ordinates. Note that in order to prioritise obtaining a basis vector corresponding to the radial direction to align with the polar axis of the source frame, $\mathbf{v}_1$ is taken to be in the radial direction. While there are an infinite number of possible tetrads (corresponding to rotations of the co-ordinate axes of the source frame), this will result in a vector as close to radial as is possible to be orthogonal with the 4-velocity and then will construct the other two basis vectors to be orthogonal to this).

\begin{align*}
\mathbf{v}_1 = \left( 0, 1, 0, 0 \right) \\
\mathbf{v}_2 = \left( 0, 0, 1, 0 \right) \\
\mathbf{v}_3 = \left( 0, 0, 0, 1 \right) \\
\end{align*}

Having set the initial basis vector to be the 4-velocity, the Gram-Schmidt procedure then creates the orthogonal tetrad by taking each of these starting vectors and subtracting the projections of this vector in the directions of the previously found basis vectors, hence ensuring that each vector has no component in the direction of the others.

For $i=1,2,3$
\begin{align}
\mathbf{e}_i = \mathbf{v}_i - \sum_{j=0}^{i-1} \frac{ \mathbf{v}_i \cdot \mathbf{e}_j }{ \mathbf{e}_j \cdot \mathbf{e}_j } \mathbf{e}_j
\end{align}
Where the scalar product is computed using the metric $\mathbf{u}\cdot\mathbf{v} = g_{\mu\nu}u^\mu v^\nu$.

Finally, the resulting set of orthogonal vectors is normalised to give the tetrad of unit basis vectors.
\begin{align}
\mathbf{\hat{e}}_a = \frac{\mathbf{e}_a}{|\mathbf{e}_a|} = \frac{\mathbf{e}_a}{\sqrt{|\mathbf{e}_a\cdot\mathbf{e}_a|}}
\end{align}

\label{lastpage}

\end{document}